\pdfoutput=1

\documentclass[11pt,a4paper]{article}
\usepackage{jheppub}
\usepackage{mciteplus}
\usepackage{hyperref}
\usepackage{amssymb}
\usepackage{amsthm}
\usepackage{amstext}
\usepackage{amsmath}
\usepackage{amsfonts}
\usepackage{dsfont}
\usepackage{cancel}
\usepackage{wrapfig}
\usepackage{graphicx}
\usepackage{pstricks}
\usepackage{color}

\usepackage{enumitem}

\usepackage{xparse}

\newcommand{\Dilog}[0]{\operatorname{Li}_2}
\newcommand{\Trilog}[0]{\operatorname{Li}_3}

\newcommand{\Cut}[0]{\operatorname{Cut}}
\renewcommand\Re{\operatorname{\mathrm{Re}}}
\renewcommand\Im{\operatorname{\mathrm{Im}}}
\newcommand\PV{\operatorname{PV}}

\DeclareMathOperator{\arsinh}{arsinh}
\DeclareMathOperator{\arcosh}{arcosh}

\newcommand{\dS}[1]{\mathrm{dS}_3^{#1}}
\newcommand{\dH}[1]{\mathrm{H}^3_{#1}}
\newcommand{\RP}{\widetilde{\mathbb{RP}}^{1,2}}

\newcommand{\nn}{\nonumber \\}

\def\eps{\epsilon}
\def\Ord{{\cal O}}

\def\Fau{{\cal F}}

\def\d{{\rm d}}

\newcommand{\refE}[1]{eq.~(\ref{#1})}

\newcommand{\refF}[1]{figure~\ref{#1}}

\newcommand{\refS}[1]{section~\ref{#1}}
\newcommand{\refA}[1]{appendix~\ref{#1}}

\newcommand{\colvec}[1]{
\left( \hspace{-0.2mm} \begin{array}{c}#1\end{array} \hspace{-0.2mm} \right)
}

\NewDocumentCommand\Tensor{mmggg}{
#1 \otimes #2 \IfNoValueTF{#3}{}{\otimes #3} \IfNoValueTF{#4}{}{\otimes #4} \IfNoValueTF{#5}{}{\otimes #5}
}

\theoremstyle{plain}

\newenvironment{algorithm}[1][Algorithm]{\begin{trivlist}
\item[ \hskip \labelsep {\bfseries Algorithm (#1)}]\hfill}{\end{trivlist}}

\numberwithin{equation}{section}

\title{Position-space cuts for Wilson line correlators}
\author[a,b,c]{Eric Laenen,}
\author[a,d]{Kasper J. Larsen,}
\author[a,b]{Robbert Rietkerk}
\affiliation[a]{Nikhef, Theory Group, Science Park 105, 1098 XG Amsterdam, The Netherlands}
\affiliation[b]{ITFA, University of Amsterdam, Science Park 904,
                1018 XE Amsterdam, The Netherlands}
\affiliation[c]{ITF, Utrecht University, Leuvenlaan 4, 3584 CE Utrecht, The Netherlands}
\affiliation[d]{Institute for Theoretical Physics, ETH Z{\"u}rich, 8093 Z{\"u}rich, Switzerland}
\emailAdd{Eric.Laenen@nikhef.nl}
\emailAdd{Kasper.Larsen@phys.ethz.ch}
\emailAdd{Robbert.Rietkerk@nikhef.nl}

\abstract{
We further develop the formalism for taking position-space cuts
of eikonal diagrams introduced in ref.~\cite{Laenen:2014jga}. These cuts
are applied directly to the position-space representation of any such diagram and
compute its discontinuity to the leading order in the dimensional
regulator. We provide algorithms for computing the position-space cuts
and apply them to several two- and three-loop eikonal diagrams,
finding agreement with results previously obtained in the literature.
We discuss a non-trivial interplay between the cutting prescription and
non-Abelian exponentiation. We furthermore discuss the relation of
the imaginary part of the cusp anomalous dimension
to the static interquark potential.}


\begin{document}

\begin{flushright}\begin{tabular}{r}
NIKHEF/2015-014
\\ ITP-UU-15/05
\end{tabular}\end{flushright}
\vspace{-14.1mm}
\maketitle

\pagebreak

\section{Introduction}\label{sec:introduction}

The infrared singularities of gauge theory scattering amplitudes
play a fundamental role in particle physics for phenomenological
as well as more theoretical studies. Determining the long-distance
singularities is necessary for combining the real and virtual contributions
to the cross section, as the divergences of the separate contributions
only cancel once they are added. Infrared singularities moreover dictate
the structure of large logarithmic contributions to the cross section,
allowing such terms to be resummed---which is in many cases required
in order to obtain reliable perturbative predictions. Beyond their significance
to collider phenomenology, long-distance singularities are highly
interesting from a theoretical point of view. Among several properties,
they have a universal structure among different gauge theories;
moreover, their exponentiation properties \cite{Yennie:1961ad,Sterman:1981jc,Gatheral:1983cz,Frenkel:1984pz,Magnea:1990zb,Magnea:2000ss,Gardi:2010rn,Mitov:2010rp,Gardi:2011wa,Gardi:2011yz,Dukes:2013wa,Dukes:2013gea,Gardi:2013ita}
and their relation to the renormalization of Wilson line correlators
\cite{Polyakov:1980ca,Arefeva:1980zd,Dotsenko:1979wb,Brandt:1981kf,Korchemsky:1985xj,Korchemsky:1985xu,Korchemsky:1987wg}
allow their perturbative expansion to be explored to all orders,
a feat currently unattainable for complete scattering amplitudes.

The basic tool for computing the infrared singularities of any
scattering amplitude is provided by the eikonal approximation.
In this limit the momenta of the soft gauge bosons emitted between
the partons emerging from the hard interaction are neglected with
respect to the hard momenta $p_i$. As a result, each hard parton $i$
simply acts as a source of soft gluon radiation
and is accordingly replaced by a semi-infinite Wilson line
\begin{equation}
\Phi_{v_i} \equiv \mathcal{P}
\exp\left( ig \int_0^\infty \hspace{-0.6mm} \d t \hspace{0.9mm}
v_i \cdot A(tv_i) \right) \,,
\label{eq:def_of_Wilson_line}
\end{equation}
which extends from time $t=0$, when the hard scattering takes place,
to infinity along the classical trajectory of the hard parton,
traced out by its four-velocity $v_i^\mu$. The long-distance singularities
of the scattering amplitude of the hard partons are then encoded in
the eikonal amplitude
\begin{equation}
\mathcal{S}(\gamma_{ij}, \eps) \equiv
\langle 0 \hspace{0.1mm} | \hspace{0.1mm} \Phi_{v_1} \otimes \Phi_{v_2}
\otimes \cdots \otimes \Phi_{v_n} \hspace{0.1mm} | \hspace{0.1mm} 0 \rangle \,,
\label{eq:Wilson_line_correlator_def}
\end{equation}
which has the same soft singularities as the original amplitude,
but is much simpler to compute.
An important feature of the eikonal amplitude
(\ref{eq:Wilson_line_correlator_def}) is the fact that it depends on
the kinematics only through the angles $\gamma_{ij}$ between the
four-velocities (defined through $\cosh \gamma_{ij} \equiv |v_i \cdot v_j|$). Before
renormalization, the integrals involved in the loop-level contributions
to $\mathcal{S}$ are thus scale invariant and vanish identically.
This in turn allows the infrared singularities at any loop order
to be computed by studying the ultraviolet renormalization factor of the
Wilson line correlator (\ref{eq:Wilson_line_correlator_def})
\cite{Korchemsky:1985xj,Korchemsky:1987wg,Kidonakis:1998nf,Kidonakis:1997gm,Becher:2009cu,Gardi:2009qi}.
This renormalization factor forms a matrix in the space of color configurations
available for the scattering process at hand, referred to as the
soft anomalous dimension matrix. In processes involving only two
Wilson lines, this matrix reduces to the cusp anomalous dimension,
a quantity which has been computed in QCD up to three loops \cite{Korchemsky:1987wg,Kidonakis:2009ev,Grozin:2014hna}.
In $\mathcal{N}=4$ super Yang-Mills theory, the cusp anomalous
dimension is known to three loops \cite{Correa:2012nk}, and
partial results have been obtained at four loops \cite{Henn:2012qz,Henn:2013wfa}.
For multi-parton amplitudes, the soft anomalous dimension matrix
has been computed through two loops for massless \cite{Aybat:2006wq,Aybat:2006mz}
as well as massive \cite{Becher:2009kw,Ferroglia:2009ep,Ferroglia:2009ii,Mitov:2010xw,Kidonakis:2010dk}
Wilson lines. Recently, much progress has been made toward
the calculation of the soft anomalous dimension matrix at three loops
\cite{Gardi:2013saa,Falcioni:2014pka}.

In this paper we continue exploring a notion of cuts of eikonal diagrams
(i.e., the diagrams contributing to the eikonal amplitude) introduced in
ref.~\cite{Laenen:2014jga}. Applied to any eikonal diagram,
the cuts compute the discontinuities of the
diagram, in analogy with the Cutkosky rules for standard Feynman
diagrams. The discontinuities are in turn readily combined
to produce the imaginary part of the diagram, a direct computational
method of which is desirable in several contexts. Indeed,
collinear factorization theorems for non-inclusive observables
were pointed out in refs.~\cite{Catani:2011st,Forshaw:2012bi}
to be violated due to exchanges of Glauber-region (i.e., maximally
transverse) gluons. The resulting factorization-breaking terms are
purely imaginary and take the form of the non-Abelian analog
of the QED Coulomb phase. Therefore, by utilizing the all-order exponentiation
property of the eikonal amplitude, the latter could be obtained
directly by computing the imaginary part of the exponent. The
resulting non-Abelian Coulomb phase \cite{Catani:1984dp,Catani:1985xt}
may also aid studies of interference effects. The importance of
understanding the imaginary part of eikonal diagrams has also
recently been highlighted in studies regarding rapidity gaps
\cite{Forshaw:2006fk,Forshaw:2008cq}. Moreover, cuts of Wilson
line correlators are naturally relevant for cross section calculations
\cite{Korchemsky:1992xv,Gardi:2005yi}.

A cutting prescription for eikonal diagrams may also provide the first
step toward extending the modern unitarity method \cite{Bern:1994zx,Bern:1994cg,Bern:1996je,Britto:2004nc,Britto:2005ha,%
Anastasiou:2006jv,Forde:2007mi,Mastrolia:2009dr,Kosower:2011ty,CaronHuot:2012ab}
to eikonal amplitudes. The development of the unitarity method has
led to a dramatic improvement in the ability to compute
loop-level (non-eikonal) scattering amplitudes at high multiplicity.
In this approach, the loop amplitude
is decomposed into a linear basis of loop integrals which are computed
independently (for example, by means of Feynman parametrization, or
differential equations \cite{Gehrmann:1999as,Henn:2013pwa}).
The calculation of the loop amplitude is then
reduced to the problem of determining the integral coefficients. This
step is performed by applying to both sides of the basis decomposition
of the loop amplitude a number of cuts which have the effect of putting
the internal lines on shell. In basic unitarity (as opposed to
generalized unitarity), the cuts employed measure the discontinuity
of the amplitude in its various kinematical channels. Unitarity
has proven highly successful, notably in computing one-loop amplitudes
with many partons in the final state. It is therefore natural
to look for extensions of this method to other physical quantities
with a perturbative expansion.

It should be emphasized that Cutkosky rules for eikonal diagrams
have been introduced previously in the literature, as a cut
prescription applied directly to the momentum-space representation
of the diagrams \cite{Korchemsky:1987wg}. In contrast, the cuts introduced
in ref.~\cite{Laenen:2014jga} and further studied here are
applied to the position-space representation of the eikonal diagrams.
A notion of position-space cuts of non-eikonal diagrams exists in the
literature in the form of a cutting equation that follows from Veltman's
largest-time equation \cite{'tHooft:186259}. However, that notion is
conceptually different from the position-space cuts in this paper,
since the former has the effect of cutting a given diagram in two disconnected
subdiagrams while the latter does not. Moreover, in practice, the largest-time equation
is typically not applied directly, but rather serves to derive the
momentum-space Cutkosky rules, which in turn are used to obtain the
imaginary part of a diagram. As already observed in ref.~\cite{Laenen:2014jga},
position-space cuts provide a substantial simplification over momentum-space cuts in the
computation of imaginary parts of eikonal diagrams. There has been
recent interest in the literature in studying Wilson line correlators
in position space, in particular
refs.~\cite{Erdogan:2011yc,Erdogan:2013bga,Erdogan:2014gha} which
investigate the structure of infrared singularities and factorization in position space.
Moreover, position-space analogs of generalized unitarity cuts
of Wilson line correlators were recently introduced in ref.~\cite{Engelund:2015cfa}.

The structure of this paper is as follows.
In section~\ref{sec:causality_and_unitarity_of_WL}
we discuss the origin of the imaginary part of Wilson line
correlators from the point of view of causality as well as unitarity.
We then show how the imaginary part can be computed from the position-
and the momentum-space representations at one loop.
In section~\ref{sec:Im_of_L-loop_Wilson_lines} we review the
formula in ref.~\cite{Laenen:2014jga} for the imaginary part of $L$-loop eikonal diagrams
containing no internal (i.e., three- or four-gluon) vertices
to the leading order in the dimensional regulator $\epsilon$.
We furthermore discuss the relation of the discontinuities of
the diagrams to their imaginary part.
In section~\ref{sec:examples} we apply the formalism to compute
the imaginary part of a number of two- and three-loop diagrams
and discuss a non-trivial interplay between the cutting prescription and
non-Abelian exponentiation. In section~\ref{sec:Im_of_diagrams_with_internal_vertices}
we turn to formulas for the imaginary part of eikonal diagrams
with internal vertices and provide details on its computation.
We give our conclusions in section~\ref{sec:Conclusions}.
Appendix~\ref{App:ImG} explains our method for computing
the principal-value integrals involved in the cutting prescription.
In appendix~\ref{App:algorithm} we present our algorithm
for re-expressing multiple polylogarithms in terms of ones with
constant indices.

\section{Imaginary parts of eikonal diagrams and their physical origin}\label{sec:causality_and_unitarity_of_WL}

In this section we will discuss the origin of the imaginary part of
Wilson line correlators from the point of view of causality
as well as unitarity. These viewpoints are naturally provided
by the representation of the correlator in position and
momentum space, respectively. We will show how the
imaginary part can be computed directly from each of
the respective integral representations at one loop.

We adopt the convention that all velocities are outgoing, such that
the velocities associated with outgoing and incoming states respectively
have positive and negative time components. We will take the gauge group
to be $SU(N)$ and work in Feynman--'t Hooft gauge with $(+,-,-,-)$
spacetime signature. Ultraviolet divergences will be regulated by computing
all diagrams in $D=4-2\eps$ dimensions with $\eps > 0$.
To avoid complications arising from regulating collinear singularities,
we take all velocities to be time-like, $v_i^2 = 1$.

{\vskip 10mm}

\begin{figure}[!h]
\begin{center}
\includegraphics[angle=0, width=0.80\textwidth]{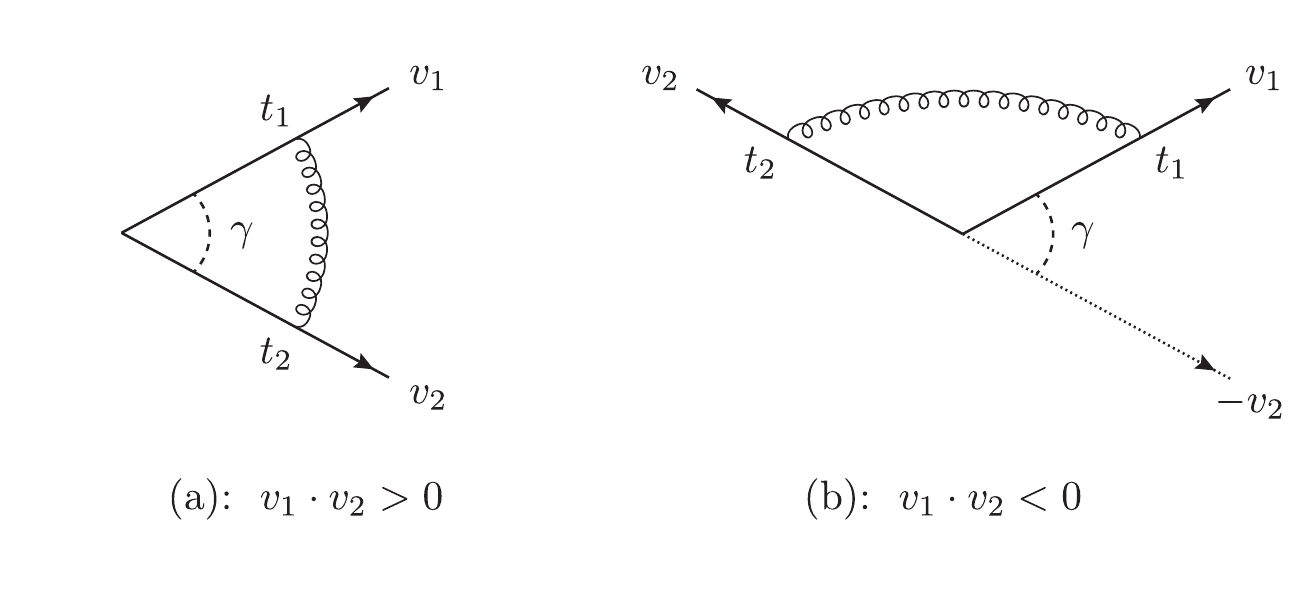}
{\vskip -5mm}
\caption{One-loop eikonal diagrams. In (a) both Wilson lines represent
final-state partons (as for example in $e^+ e^- \to q\bar{q}$).
Thus, the external velocities are in the region $v_1 \cdot v_2 > 0$,
and the partons can become lightlike separated.
In (b) one Wilson line represents a final-state parton, and one
represents an initial-state parton (as for example in deep inelastic scattering).
Thus, the external velocities are in the region $v_1 \cdot v_2 < 0$,
and the partons are never lightlike separated.}
\label{fig:one-loop_Wilson_line}
\end{center}
\end{figure}

We start our investigations by examining the simplest eikonal diagram,
the one-loop exchange illustrated in figure~\ref{fig:one-loop_Wilson_line}.
In both kinematic regions (a) and (b), the position-space
representation of the diagram is straightforwardly obtained
by direct perturbative expansion\footnote{See chapter 8
of ref.~\cite{sterman1993introduction} for the position-space
Feynman rules.} in $g$ of the correlator
(\ref{eq:Wilson_line_correlator_def}) and takes the form
\begin{equation}
F^{(1)} = C^{(1)} \mu^{2\epsilon} \int_0^\infty \d t_1
\int_0^\infty \d t_2 \frac{v_1 \cdot v_2}
{\big[ {-}(t_1 v_1 - t_2 v_2)^2 + i\eta \big]^{1-\epsilon}} \,,
\label{eq:one-loop_diagram_x-space}
\end{equation}
where the prefactor is defined as
$C^{(1)} \equiv g^2 C_F \frac{\Gamma(D/2 -1)}{4\pi^{D/2}}$,
with $C_F = \frac{N^2 -1}{2N}$ denoting the quadratic Casimir
of the fundamental representation.
Furthermore, $t_1, t_2$ have the dimension of time and denote the
positions of the attachment points of the soft-gluon propagator on
the Wilson lines spanned by the four-velocities $v_1$ and $v_2$.

The integrations in eq.~(\ref{eq:one-loop_diagram_x-space}) produce
an infrared divergence which can be extracted via the change of variables
$(t_1, t_2)= (\lambda x, \lambda(1-x))$ with $0\leq x \leq 1$,
where $\lambda$ has the dimension of length,
\begin{equation}
F^{(1)} = C^{(1)} \mu^{2\epsilon} \int_0^\infty
\frac{\d \lambda}{\lambda^{1-2\epsilon}}
\int_0^1 \d x \frac{v_1 \cdot v_2}{\big[{-}(x v_1 - (1-x) v_2)^2 + i\eta \big]^{1-\epsilon}} \,.
\label{eq:one-loop_diagram_x-space_lambda_s-channel}
\end{equation}
Indeed, the $\lambda$-integral is has an infrared divergence, owing to the
exchange of gluons of increasingly longer wavelength as
$\lambda \to \infty$. This divergence can be regularized
in a gauge invariant fashion by introducing an exponential damping
factor $e^{-\Lambda \lambda}$ with $\Lambda \ll 1$, whereby it
becomes
\begin{equation}
\mu^{2\epsilon} \int_0^\infty \frac{\d \lambda \hspace{0.7mm} e^{-\Lambda \lambda}}
{\lambda^{1-2\epsilon}} \hspace{1mm} = \hspace{1mm} \Gamma(2\epsilon)
\left( \frac{\mu}{\Lambda} \right)^{2\epsilon} \hspace{1mm} = \hspace{1mm} \frac{1}{2\epsilon}
\left( \frac{\mu}{\Lambda} \right)^{2\epsilon} + \mathcal{O}(\epsilon^0) \,.
\label{eq:exponential_damping_factor}
\end{equation}

The two diagrams in figure~\ref{fig:one-loop_Wilson_line}
have the same integrand; however, as the external kinematics
is taken from the distinct regions $v_1 \cdot v_2 > 0$ and $v_1 \cdot v_2 < 0$,
the integrations will produce distinct results. It is most convenient
to compute the diagram in figure~\ref{fig:one-loop_Wilson_line}(b) first
and obtain the result for figure~1(a) by analytic continuation as follows.
For the diagram in figure~1(b), we may define the deflection angle $\gamma > 0$
such that $\cosh \gamma = -v_1 \cdot v_2$, in terms of which
the diagram in figure~\ref{fig:one-loop_Wilson_line}(b) becomes,
to the leading order in $\epsilon$,
\begin{equation}
F^{(1)}_{\mathrm{1(b)}} = \frac{C^{(1)}}{2\epsilon}
\left( \frac{\mu}{\Lambda} \right)^{2\epsilon} \gamma \coth \gamma \,.
\label{eq:one-loop_diagram_spacelike_result}
\end{equation}
Likewise, for the diagram in figure~1(a), we may define the
cusp angle $\gamma > 0$ such that $\cosh \gamma = v_1 \cdot v_2$.
The integrated expression for this diagram can thus be obtained from
eq.~(\ref{eq:one-loop_diagram_spacelike_result}) by replacing
$\gamma^\mathrm{(b)} \to \pi i - \gamma^\mathrm{(b)} = \gamma^\mathrm{(a)}$,
\begin{equation}
F^{(1)}_{\mathrm{1(a)}} = \frac{C^{(1)}}{2\epsilon}
\left( \frac{\mu}{\Lambda} \right)^{2\epsilon} (\gamma - \pi i) \coth \gamma\,.
\label{eq:one-loop_diagram_timelike_result}
\end{equation}
We observe that the imaginary parts of the eikonal diagrams in
figs.~\ref{fig:one-loop_Wilson_line}(a)~and~\ref{fig:one-loop_Wilson_line}(b)
are respectively non-vanishing and vanishing.
Before turning to the question of how the imaginary parts
of the diagrams in
eqs.~(\ref{eq:one-loop_diagram_spacelike_result})--(\ref{eq:one-loop_diagram_timelike_result})
may be extracted from their integral representation
in eq.~(\ref{eq:one-loop_diagram_x-space_lambda_s-channel}),
let us consider their physical origin and interpretation.

\begin{figure}[t!]
\begin{center}
\includegraphics[angle=0, width=0.75\textwidth]{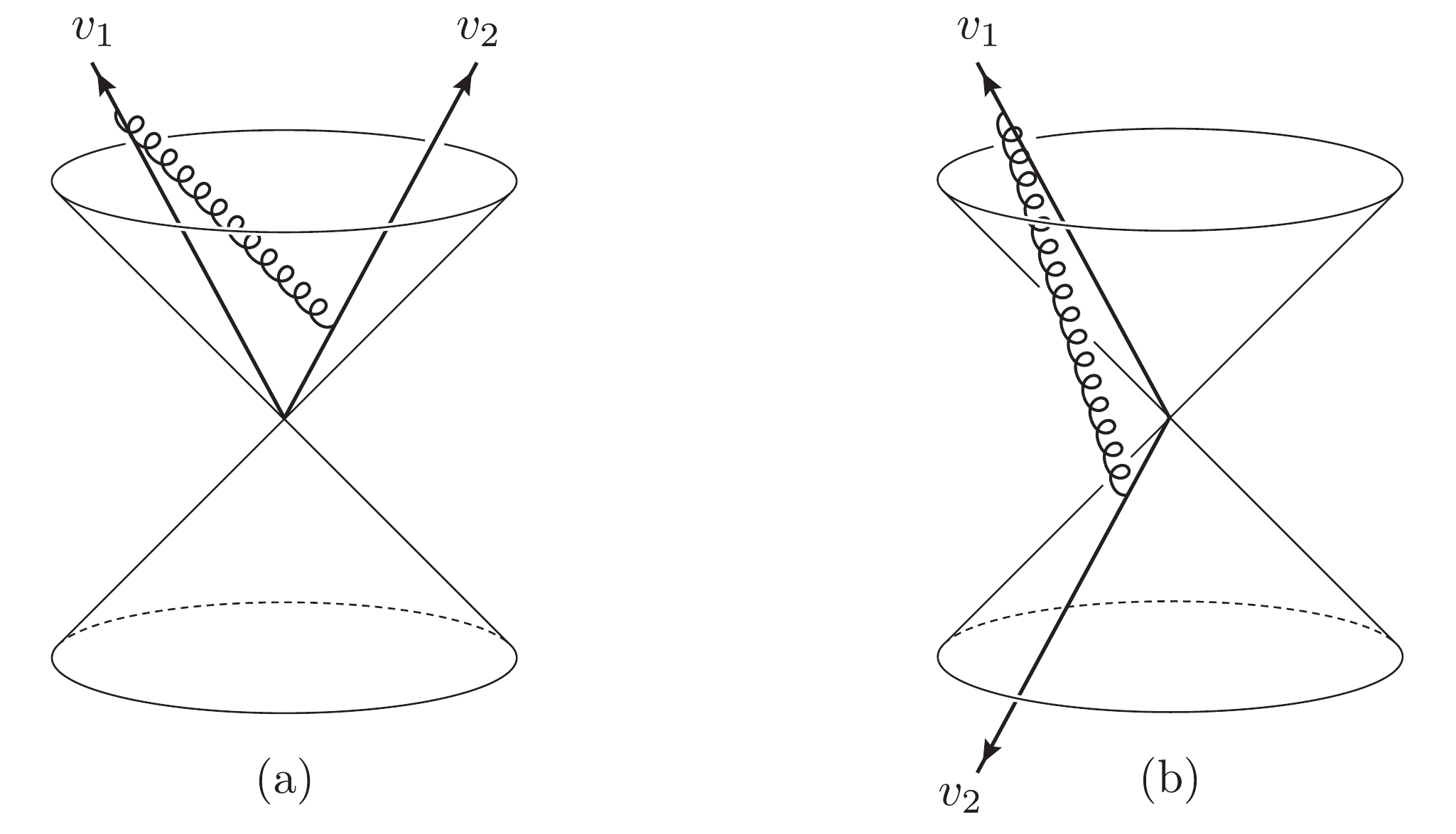}
{\vskip -2.5mm}
\caption{The one-loop eikonal diagrams of figure~\ref{fig:one-loop_Wilson_line} embedded in a space-time diagram. In (a) both Wilson lines are confined to the interior of the future light cone, describing two final state partons. In this case the partons can become lightlike separated, which is illustrated by the exchange of a lightlike gluon (i.e. a gluon that is aligned to the light cone). In (b) there is one incoming parton inside the past light cone and an outgoing parton inside the future light cone. These partons are never lightlike separated. Indeed, the gluon stretching between these two Wilson lines is necessarily time-like (i.e. off shell).}
\label{fig:spacetime_picture_one_loop_eikonal_diagrams}
\end{center}
\end{figure}

From the position-space representation
(\ref{eq:one-loop_diagram_x-space}) of the eikonal diagram,
the origin of the imaginary part may be understood from a
simple causality consideration as follows. As our focus is on
computing the imaginary part to the leading order in $\eps$,
the $\eps$ in the propagator
exponent may be dropped once the infrared divergence has been extracted.
After moreover stripping off real prefactors from eq.~(\ref{eq:one-loop_diagram_x-space}),
the integral takes the form,
\begin{equation}
\int_0^\infty \d t_1 \int_0^\infty \d t_2 \hspace{0.6mm} \frac{1}
{(t_1 v_1 - t_2 v_2)^2 - i\eta} \,.
\label{eq:one-loop_Wilson_lines_a_and_b}
\end{equation}
Now, for the kinematics corresponding to
the diagram in figure~\ref{fig:one-loop_Wilson_line}(a),
there are regions $\frac{t_1}{t_2} = e^{\pm \gamma}$ within the
integration domain where $(t_1 v_1 - t_2 v_2)^2 =0$. Here
the $-i\eta$ term becomes relevant and generates
an imaginary part. What is happening physically at
such times $t_1, t_2$ is that the two partons traveling
along $v_1$ and $v_2$ become lightlike separated.
This is illustrated in
figure~\ref{fig:spacetime_picture_one_loop_eikonal_diagrams}(a).
As a result, the phases of their states will change through
exchanges of lightlike gluons (or photons)---leading
to observable consequences that will be discussed
shortly. In contrast, for external kinematics corresponding to
the diagram in figure~\ref{fig:one-loop_Wilson_line}(b),
the integral in eq.~(\ref{eq:one-loop_Wilson_lines_a_and_b})
has a vanishing imaginary part: the denominator
$(t_1 v_1 - t_2 v_2)^2$ is strictly positive within the region
of integration, and the $-i\eta$ can therefore be dropped.
In this situation, the partons are never lightlike separated,
as illustrated in figure~\ref{fig:spacetime_picture_one_loop_eikonal_diagrams}(b),
and the phases of their states cannot change
through exchanges of lightlike massless gauge bosons.

These observations on the evolution of the phases of the hard-parton
states suggest that the imaginary part of the correlator of
two Wilson lines defines an interparton potential. Indeed,
in the non-relativistic limit, the final and initial
two-particle states are related in the interaction picture
through time evolution by $|f\rangle_I = e^{i \int_0^\infty
\d t \hspace{0.6mm} e^{-\Lambda t} V_I t} |i\rangle_I$ where $V_I$ denotes
the interaction potential. The relation of the correlator
to a non-relativistic potential can be made precise
in the situation where the pair of energetic particles carry
no color charges, as for example in the case of an $e^+ e^-$ pair.
In Abelian gauge theories, the correlator of two Wilson lines
can be written as the exponential of the sum of connected
diagrams \cite{Yennie:1961ad},
\begin{equation}
W \equiv \langle \Phi_{v_1} \Phi_{v_2} \rangle
= \exp \big( F^{(1)} + \mathcal{O}(g^4) \big) \,,
\label{eq:Abelian_exponentiation}
\end{equation}
where $F^{(1)}$ is the one-loop diagram in figure~\ref{fig:one-loop_Wilson_line},
and the additional diagrams contain a single lepton loop
connected to the Wilson lines by an arbitrary (even) number of
soft-photon exchanges. Using the result for the diagram
$F^{(1)}$ computed for time-like kinematics in
eq.~(\ref{eq:one-loop_diagram_timelike_result}) (with $C_F \to 1$
to recover the Abelian case), the anomalous dimension
of the Wilson-line correlator---i.e., the cusp anomalous
dimension---evaluates to
\begin{equation}
\Gamma_\mathrm{cusp} (\gamma) \equiv -\lim_{\eps \to 0}
\frac{\d \log W}{\d \log \mu} \hspace{0.7mm}=\hspace{0.7mm}
-\frac{g^2}{4\pi^2} (\gamma - \pi i) \coth \gamma \,.
\label{eq:QED_cusp_anomalous_dimension}
\end{equation}
The non-relativistic limit corresponds to the small-angle
regime $\gamma \approx 0$ where the two velocities $v_1$
and $v_2$ are nearly collinear, and the relative velocity
of the hard leptons thus small. Accordingly, expanding
eq.~(\ref{eq:QED_cusp_anomalous_dimension}) around
$\gamma = 0$ and taking the imaginary part, we find
\begin{equation}
\mathrm{Im} \hspace{0.7mm} \Gamma_\mathrm{cusp} (\gamma)
\hspace{0.7mm}=\hspace{0.7mm} \frac{g^2}{4\pi \gamma}
+ \mathcal{O}(\gamma^0) \,.
\end{equation}
We observe that the imaginary part of the cusp anomalous
dimension evaluated in time-like kinematics takes
the form of the non-relativistic Coulomb potential
(the appropriate dimension of energy is acquired after replacing
the angle $\gamma$ by the distance between the two fermions).

This relation does not extend to generic non-Abelian gauge theories,
as we will discuss shortly. It does, however, extend to the case of
conformal field theories, such as $\mathcal{N}=4$
super Yang-Mills theory, where the state-operator correspondence relates
Wilson-line operators in Minkowski space to states in
$\mathbb{R} \times \mathrm{AdS}_3$. In radial quantization,
a pair of Wilson lines intersecting at a cusp angle $\gamma$
with the resulting anomalous dimension $\Gamma_\mathrm{cusp} (\gamma)$
is mapped to a pair of static charges in $\mathrm{AdS}_3$ separated by
a distance of $\gamma$ with an electrostatic energy\footnote{The real
part of the cusp anomalous dimension gives rise to an imaginary part of
the electrostatic energy. As argued in ref.~\cite{Chien:2011wz},
the resulting non-unitary time evolution is accounted for by the real
radiation of soft and collinear gluons along the Wilson lines.} of
$\mathrm{Im} \hspace{0.6mm} \Gamma_\mathrm{cusp}$ \cite{Chien:2011wz}.
For small values of the cusp angle, the charges on $\mathrm{AdS}_3$ become
closer than the curvature scale, and the electrostatic energy
takes the form of the non-relativistic interquark potential in
flat space \cite{Correa:2012nk,Correa:2012hh}. (The non-relativistic
approximation becomes relevant here, as in the small-angle regime
$\gamma \approx 0$, the relative velocity of the hard partons is small,
as discussed above.)

However, for non-Abelian and non-conformal gauge theories such as QCD,
diagrams containing loop corrections to the soft propagators will
have a dependence on the beta function, thereby explicitly breaking
the scale invariance of the diagram. As a result, in QCD, the imaginary part
of the three-loop cusp anomalous dimension
$\Gamma^{(3)}_\mathrm{cusp}$ differs from the static
interquark potential by terms proportional to the beta function \cite{Grozin:2014hna}.
(This can be seen by comparing the $N_f^2$ contribution
to $\Gamma^{(3)}_\mathrm{cusp}$, given in eq.~(A.2) of
ref.~\cite{Beneke:1995pq}, against the $N_f^2$ term of the three-loop
static QCD potential\footnote{Note that in the literature on
the interquark potential, the loop order is often defined
as one less than the standard notion.}, given in eq.~(10) of
ref.~\cite{Schroder:1998vy}.)
\\
\\
Let us now turn to the question of how the imaginary part
of the eikonal diagrams in figure~\ref{fig:one-loop_Wilson_line}
may be obtained from their integral representation in
eq.~(\ref{eq:one-loop_diagram_x-space_lambda_s-channel}) where
the infrared divergence has been extracted. We will restrict
attention to the leading order in the dimensional regulator
$\eps$, and accordingly drop the $\eps$ in the propagator
exponent. We can then utilize the formula
\begin{equation}
\int_a^b \d x \hspace{0.5mm} \frac{f(x)}{D(x) \pm i\eta}
\hspace{0.8mm}=\hspace{0.8mm} \mathrm{PV} \int_a^b \d x \hspace{0.5mm} \frac{f(x)}{D(x)}
\hspace{0.6mm}\mp\hspace{0.6mm} \pi i \int_a^b \d x \hspace{0.5mm} f(x)
\hspace{0.3mm} \delta\big(D(x) \big) \,, \label{eq:PV-formula}
\end{equation}
where $\mathrm{PV}$ indicates that the Cauchy principal value
prescription is to be applied, and the integration bounds $a$ and $b$
are real numbers. The denominator $D(x)$ is a real-valued polynomial~in~$x$,
and the numerator $f(x)$ is an arbitrary real-valued function
with no poles or branch points inside the integration path.
As both integrals on the right-hand side of eq.~(\ref{eq:PV-formula})
are real, this formula achieves a decomposition into a purely
real and purely imaginary part.

\begin{figure}[!h]
\begin{center}
\includegraphics[angle=0, width=0.60\textwidth]{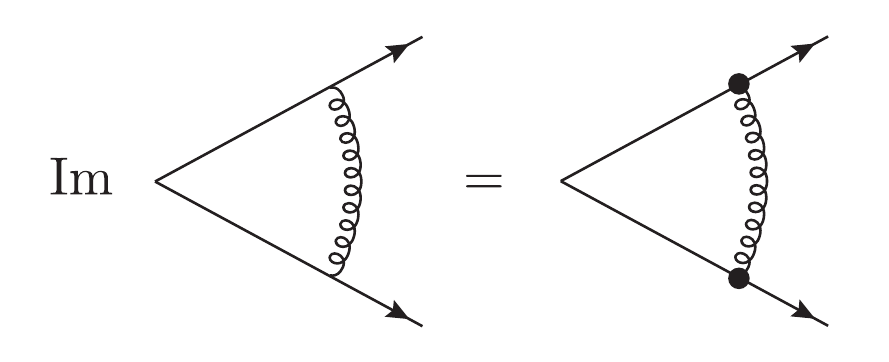}
\caption{Graphical representation of a cut of an eikonal diagram in its
position-space representation. The black dots represent the emission
and absorption of a lightlike gauge boson.}
\label{fig:x-space_cut}
\end{center}
\end{figure}

Accordingly, at one loop, we define the position-space cut prescription
\begin{equation}
\frac{1}{D(x) \pm i\eta} \hspace{1.3mm}\stackrel{\mathrm{cut}}{\longrightarrow}\hspace{1.3mm}
\mp \hspace{0.2mm} \pi i \hspace{0.3mm} \delta\big(D(x) \big) \,,
\label{eq:x-space_cut}
\end{equation}
in terms of which it is straightforward to obtain the
imaginary part of the diagrams in figure~\ref{fig:one-loop_Wilson_line}
to the leading order in $\eps$. For example, considering the
time-like kinematics situation in figure~\ref{fig:one-loop_Wilson_line}(a)
and applying the prescription (\ref{eq:x-space_cut})
to eq.~(\ref{eq:one-loop_diagram_x-space_lambda_s-channel})
with the $\eps$ in the propagator exponent set to zero, we find
\begin{equation}
\mathrm{Im} \hspace{0.7mm} F^{(1)}_{\mathrm{1(a)}}
= -\pi \hspace{0.3mm} C^{(1)} (v_1 \cdot v_2) \hspace{0.3mm}
\mu^{2\epsilon} \int_0^\infty \frac{\d \lambda \hspace{0.7mm} e^{-\Lambda \lambda}}
{\lambda^{1-2\epsilon}} \int_0^1 \d x \hspace{0.7mm}
\delta\big( \big( x v_1 - (1-x) v_2 \big)^2 \big) \,.
\label{eq:cut_one-loop_diagram_x-space_t-channel}
\end{equation}
We can integrate out the delta function by use of the identity
\begin{equation}
\int_0^1 \d x \hspace{0.7mm} \delta(Ax^2 + Bx + C) \hspace{0.7mm}=\hspace{0.7mm}
\frac{1}{\sqrt{\Delta}} \sum_{i=1,2} \theta (\rho_i) \hspace{0.2mm} \theta (1 - \rho_i) \,,
\label{eq:delta_of_quad_pol}
\end{equation}
where $\Delta \equiv B^2 - 4AC$ and $\rho_i$ respectively denote
the discriminant and roots of the polynomial. The roots
$\rho_i = \frac{1}{1 + e^{\pm \gamma}}$ of the delta function
argument in eq.~(\ref{eq:cut_one-loop_diagram_x-space_t-channel})
are manifestly located inside the domain of the $x$-integration.
The result of integrating out the delta function in
eq.~(\ref{eq:cut_one-loop_diagram_x-space_t-channel}) is
therefore
\begin{equation}
\mathrm{Im} \hspace{0.7mm} F^{(1)}_{\mathrm{1(a)}} = -\pi \frac{C^{(1)}}{2\epsilon}
\left( \frac{\mu}{\Lambda} \right)^{2\epsilon} \coth \gamma \,,
\end{equation}
in agreement with
eq.~(\ref{eq:one-loop_diagram_timelike_result}).
The calculation for the diagram in figure~\ref{fig:one-loop_Wilson_line}(b)
is completely analogous, except that in this case
$\int_0^1 \d x \hspace{0.7mm} \delta\big( \big(x v_1 - (1-x)v_2 \big)^2 \big) =0$,
as both roots $\rho_i = \frac{1}{1 - e^{\pm \gamma}}$
are located outside the domain of integration. We therefore find
a vanishing imaginary part, in agreement with
eq.~(\ref{eq:one-loop_diagram_spacelike_result}).
We conclude that in both cases (a) and (b), the cutting prescription
(\ref{eq:x-space_cut}) produces the correct imaginary part.
We introduce a graphical notation for the cutting prescription
(\ref{eq:x-space_cut}) in figure~\ref{fig:x-space_cut}.

It is natural to ask whether the imaginary part of eikonal diagrams
can also be obtained from their momentum-space representation%
\footnote{The momentum-space representation in eq.~(\ref{eq:one-loop_diagram_p-space})
is straightforwardly obtained from the eikonal Feynman rules.
By Schwinger parametrizing the eikonal propagators in
eq.~(\ref{eq:one-loop_diagram_p-space}) and performing
the resulting Fourier transform, one recovers the
position-space representation in eq.~(\ref{eq:one-loop_diagram_x-space}).}
\begin{equation}
F^{(1)} = i g^2 C_F \mu^{2\epsilon} \int \frac{\d^D k}{(2\pi)^D}
\frac{v_1 \cdot v_2}{(k^2 + i\eta) (v_1 \cdot k +i\eta) (v_2 \cdot k - i\eta)} \,.
\label{eq:one-loop_diagram_p-space}
\end{equation}
Such a cutting prescription was provided in ref.~\cite{Korchemsky:1987wg}.
Here it was shown that the imaginary part of the one-loop diagram
in eq.~(\ref{eq:one-loop_diagram_p-space}) may be obtained by
replacing the two eikonal propagators by delta functions,
\begin{equation}
\frac{1}{k \cdot v_i \pm i\eta} \hspace{0.9mm}\stackrel{\mathrm{cut}}{\longrightarrow}\hspace{0.9mm}
\mp \hspace{0.3mm} 2\pi i \hspace{0.3mm} \theta(v_i^0) \hspace{0.2mm} \delta (k \cdot v_i) \,.
\label{eq:eikonal_Cutkosky_rules}
\end{equation}
This prescription can be thought of as the eikonal limit of the
standard Cutkosky rules. It is illustrated in figure~\ref{fig:p-space_cut}
below. More explicitly, applying the prescription (\ref{eq:eikonal_Cutkosky_rules})
to eq.~(\ref{eq:one-loop_diagram_p-space}), the imaginary
part is determined as follows,
\begin{equation}
2i \hspace{0.4mm} \mathrm{Im} \hspace{0.6mm} F^{(1)} = (2\pi)^2 i g^2 C_F
\theta (v_1^0) \theta (v_2^0) (v_1 \cdot v_2) \mu^{2\epsilon}
\int \hspace{-0.5mm} \frac{\d^D k}{(2\pi)^D}
\frac{\delta(v_1 \cdot k) \delta(v_2 \cdot k)}{k^2 + i\eta} \,.
\label{eq:map_p-space_cut_to_x-space_1}
\end{equation}
This representation of the imaginary part of the one-loop
diagram motivates two remarks.

The first remark concerns
the region of momentum space which gives rise to the imaginary part.
Defining the light-cone variables $k^\pm \equiv \frac{1}{\sqrt{2}} (k^0 \pm k^3)$
and choosing the Lorentz frame in which the transverse components of
the velocities vanish, $v_{iT}=0$, the support of the delta
functions in eq.~(\ref{eq:map_p-space_cut_to_x-space_1})
is the region where the momentum of the exchanged gluon
is maximally transverse,
\begin{equation}
k_T \gg k^+ \sim k^- \approx 0 \,,
\end{equation}
which was identified in ref.~\cite{Korchemsky:1987wg} as
the Glauber region \cite{Collins:1983ju}.
This agrees with the discussion in section~\ref{sec:introduction}:
the imaginary part of eikonal diagrams arises from the exchanges
of Glauber-region gluons.

The second remark concerns the
physical interpretation of applying the momentum-space cuts
(\ref{eq:eikonal_Cutkosky_rules}). By writing the delta functions in
eq.~(\ref{eq:map_p-space_cut_to_x-space_1}) in terms of the
plane-wave representation
$\delta(A) = \frac{1}{2\pi} \int_{-\infty}^\infty \d u \hspace{0.6mm} e^{iuA}$
and performing the Fourier transform we find
\begin{equation}
2i \hspace{0.4mm} \mathrm{Im} \hspace{0.6mm} F^{(1)} = C^{(1)}
\theta (v_1^0) \theta (v_2^0) \mu^{2\epsilon} \int_{-\infty}^\infty \d t_1
\int_{-\infty}^\infty \d t_2 \frac{v_1 \cdot v_2}
{\big[ {-}(t_1 v_1 - t_2 v_2)^2 + i\eta \big]^{1-\epsilon}} \,.
\label{eq:map_p-space_cut_to_x-space_2}
\end{equation}
We observe that the resulting integration bounds
compared to those of the uncut diagram in eq.~(\ref{eq:one-loop_diagram_x-space})
are extended according to $\int_0^\infty \d t_i \longrightarrow \int_{-\infty}^\infty \d t_i$.
This state of affairs can be simply understood on physical grounds:
as the hard partons have been put on shell through the
cutting rule (\ref{eq:eikonal_Cutkosky_rules}), they are now
asymptotic states propagating from $t_i= -\infty$
to the interaction point.

\begin{figure}[!h]
\begin{center}
\includegraphics[angle=0, width=0.60\textwidth]{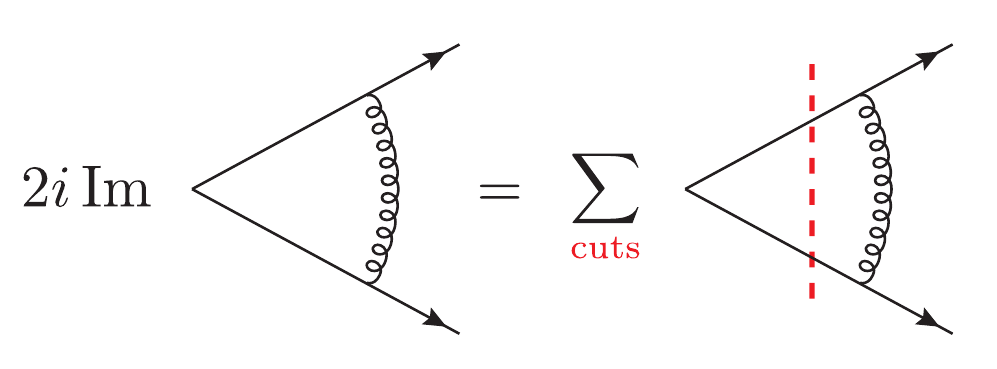}
\caption{Graphical representation of cuts of eikonal diagrams in momentum space.}
\label{fig:p-space_cut}
\end{center}
\end{figure}

We see that the position- and momentum-space representations
of eikonal diagrams offer complementary points of view
on the origin of their imaginary part. To summarize,
in the position-space representation, the imaginary part
is seen to arise from the exchanges of lightlike soft
gauge bosons whose emission and absorption change the
phases of the hard-parton states. In contrast, in
momentum space, the imaginary part (related to the branch
cut discontinuity through eq.~(\ref{eq:Im_and_discontinuities_relation}))
arises from the two hard partons going on shell and exchanging
Glauber gluons. Thus, the position- and momentum-space representations
explain the origin of the imaginary part from
the points of view of causality and unitarity, respectively.

\begin{figure}[!h]
\begin{center}
\includegraphics[angle=0, width=0.90\textwidth]{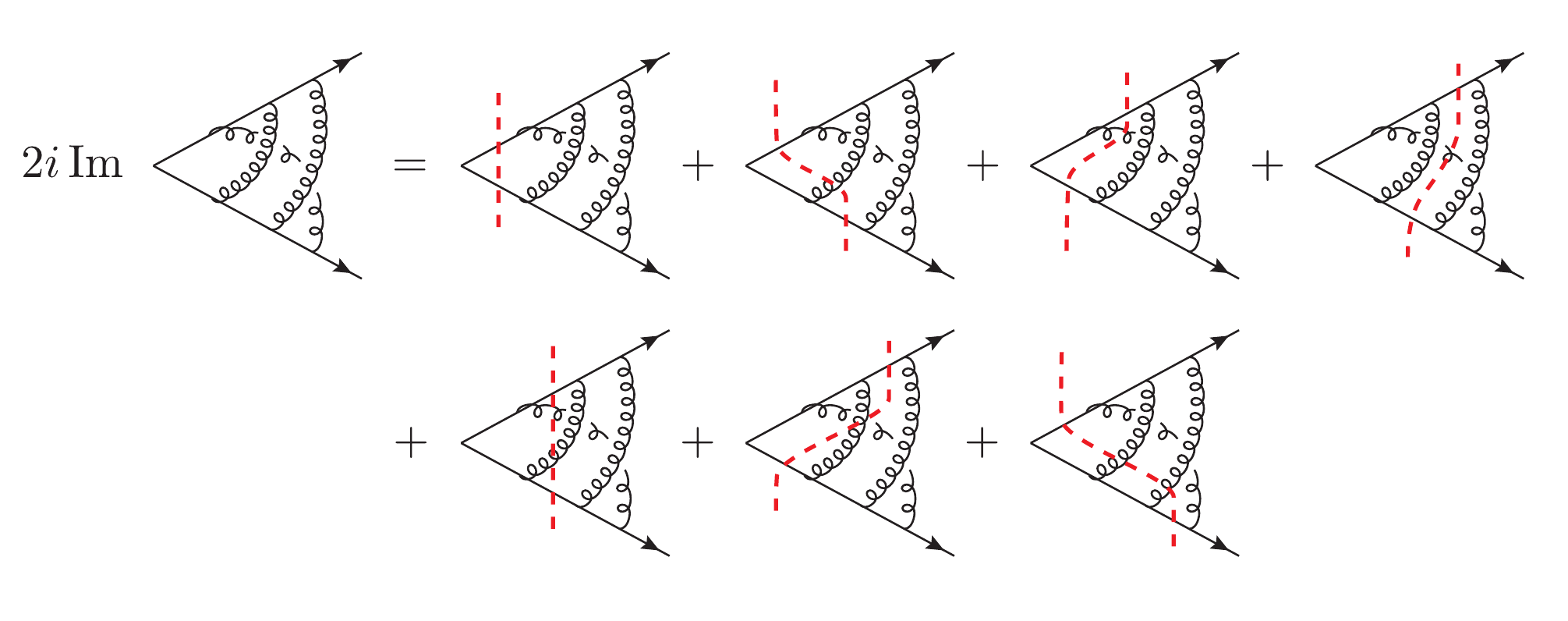}
{\vskip -5mm}
\caption{The non-vanishing momentum-space cuts of a non-planar three-loop ladder diagram.
The cuts require the evaluation of two-, three- and four-particle phase-space
integrals.}
\label{fig:Cut_non-planar_three-loop_ladder}
\end{center}
\end{figure}

The momentum-space cutting prescription in
eq.~(\ref{eq:eikonal_Cutkosky_rules}) has the conceptual advantage
of factoring eikonal diagrams into on-shell lower-loop and tree diagrams
which in turn can be computed as independent objects. However,
the resulting cut diagrams involve integrations over two-, three-,
four-, $\ldots$ particle phase space, as illustrated in
figure~\ref{fig:Cut_non-planar_three-loop_ladder}. In practice,
the evaluation of these phase-space integrals poses a substantial
computational challenge which limits the applicability
of the cut prescription (\ref{eq:eikonal_Cutkosky_rules})
for obtaining imaginary parts.

As we shall see in section~\ref{sec:Im_of_L-loop_Wilson_lines}, in position space,
eikonal diagrams without internal vertices take the form of iterated integrals.
In this representation, their imaginary parts can therefore be straightforwardly
obtained by applying the principal-value formula (\ref{eq:PV-formula}) recursively.

\section{Position-space cuts of eikonal diagrams without internal vertices}\label{sec:Im_of_L-loop_Wilson_lines}

For completeness, in this section we review the derivation
presented in ref.~\cite{Laenen:2014jga} of the imaginary part
of $L$-loop eikonal diagrams without internal (i.e., three- or four-gluon)
vertices to the leading order in the dimensional regulator $\epsilon$.
We will interchangeably refer to these diagrams as ladder-type diagrams.
The basic observation is that in position space
these diagrams are iterated integrals, and as a result their imaginary part
can be obtained by decomposing the real-line integrations into
principal-value and delta function contributions.

In position space, an arbitrary $L$-loop eikonal diagram without internal
vertices is composed of $L$ soft-gluon propagators, interchangeably referred
to here as rungs. Each rung extends between the Wilson lines
spanned by any two (possibly identical) external four-velocities
$v_1, \ldots, v_n$ where $1 \leq n \leq L+1$.
For the $j$th rung we will denote these four-velocities
by $v_{\ell_j}$ and $v_{r_j}$. We let $t_{i,k}$ denote the position
of the $k$th attachment on the Wilson line spanned by $v_i$, counting
from the hard interaction vertex and outwards, so that
$0\leq t_{i,1} < t_{i,2} < \cdots < t_{i,N_i -1} < t_{i,N_i}$, where
$N_i$ denotes the total number of soft-gluon attachments on the Wilson
line. In addition, for the $j$th rung, we let the variables
$m_j$ and $n_j$ record the soft-gluon attachment numbers on
the Wilson lines spanned by $v_{\ell_j}$ and $v_{r_j}$, respectively.
The $L$-loop eikonal diagram is then defined as the $2L$-fold iterated integral
\begin{equation}
F^{(L)} = C^{(L)} (g \mu^\epsilon)^{2L} \prod_{j=1}^L \int_0^\infty \d t_{\ell_j, m_j}
\d t_{r_j, n_j} \frac{(v_{\ell_j} \cdot v_{r_j}) \prod_{i=1}^n \prod_{k=0}^{N_i}
\theta (t_{i,k+1} - t_{i,k})}
{\big[ {-}(t_{\ell_j, m_j} v_{\ell_j} - t_{r_j, n_j} v_{r_j})^2
+ i\eta \big]^{1-\epsilon}} \,,
\label{eq:L-loop_ladder_diagram_def}
\end{equation}
where the kinematics-independent prefactor $C^{(L)}$ is determined
by the color structure of the diagram and where it is implied that
$t_{i,N_i +1} \equiv \infty$ and $t_{i,0} \equiv 0$.
Without loss of generality, we will assume that any rungs
with both endpoints attached to the same Wilson line
have been integrated out, and we suppress the resulting
pole factors in $\eps$. (The additional factors produced
by the integrations, involving epsilonic powers of
the remaining variables, will not be of importance here,
as our aim is to extract the imaginary part
of $F^{(L)}$ to the leading order in $\eps$.)

To extract the imaginary part of $F^{(L)}$ from the integral
representation in eq.~(\ref{eq:L-loop_ladder_diagram_def})
it turns out to be useful to perform a change of variables which leaves
each soft propagator dependent on a single variable.
To this end, we adopt a change of variables introduced
in ref.~\cite{Gardi:2011yz}. The idea is to first
express the attachment points of the $j$th rung in terms
of ``polar'' coordinates measuring the distance $\rho_j$ to the cusp
(in units of the infrared cutoff $1/\Lambda$) and
$x_j$ essentially measuring the emission
angle of the soft gluon to the Wilson line,\footnote{For a given rung,
the two endpoints may of course  be referred to interchangeably
as left or right. However, for practical calculations,
one particular choice may prove slightly more convenient.
We refer to section~\ref{sec:examples} for examples.}
\begin{equation}
\left( \hspace{-1.1mm} \begin{array}{c} t_{\ell_j, m_j} \\
t_{r_j, n_j} \end{array} \hspace{-1.1mm} \right)
\hspace{0.7mm}=\hspace{0.7mm} \rho_j \left( \hspace{-1.1mm}
\begin{array}{c} x_j \\ 1 - x_j \end{array} \hspace{-1.1mm} \right)
\hspace{6mm} \mathrm{where} \hspace{5mm}
\left\{ \hspace{-0.6mm} \begin{array}{l}
0 \leq \rho_j < \infty \\[0.1mm]
0 \leq x_j \leq 1 \,.
\end{array} \right.
\label{eq:radial_coordinates}
\end{equation}
After this change of variables, the diagram takes the form
\begin{equation}
F^{(L)} = C^{(L)} (g \mu^\epsilon)^{2L} \prod_{j=1}^L
\int_0^\infty \frac{\d \rho_j}{\rho_j^{1-2\epsilon}}
\int_0^1 \d x_j \hspace{0.6mm} P_{\ell_j r_j}^{[\epsilon]} (x_j)
\Theta (\boldsymbol{\rho},\boldsymbol{x}) \,,
\label{eq:L-loop_ladder_diagram_rep_1}
\end{equation}
where the soft propagators are defined as
\begin{equation}
P_{ij}^{[\epsilon]} (x) \equiv \frac{v_i \cdot v_j}
{\big[ {-}\big(x v_i - (1-x)v_j\big)^2 + i\eta \big]^{1-\epsilon}} \,,
\label{eq:propagator_notation}
\end{equation}
and where the nesting of the integrations is encoded in $\Theta$,
defined through
\begin{equation}
\Theta (\boldsymbol{\rho},\boldsymbol{x}) \equiv \prod_{i=1}^n \prod_{k=0}^{N_i}
\theta (t_{i,k+1} - t_{i,k}) \bigg|_{\tiny \left( \hspace{-0.5mm} \begin{array}{c} t_{\ell_j, m_j} \\
t_{r_j, n_j} \end{array} \hspace{-0.9mm} \right)
\hspace{0.7mm}=\hspace{0.7mm} \rho_j \left( \hspace{-0.3mm}
\begin{array}{c} x_j \\ 1 - x_j \end{array} \hspace{-0.7mm} \right)} \,.
\label{eq:nesting_function_def}
\end{equation}
We observe that the soft propagators' dependence
on the radial coordinates $\rho_j$ has scaled out in
eq.~(\ref{eq:L-loop_ladder_diagram_rep_1}), and that
each propagator now depends only on a single variable $x_j$.
This turns out to be particularly advantageous for the purpose
of extracting the imaginary part of the diagram, as this
circumvents the need to divide a higher-dimensional
domain of integration into subdomains characterized
by supporting a specific number of propagator roots.

Now we extract the overall infrared divergence of the
diagram by setting $\tau_1 \equiv \rho_1$ and then applying
the following sequence of $L-1$ substitutions
\begin{equation}
\left( \hspace{-1.1mm} \begin{array}{c} \tau_1 \\
\rho_2 \end{array} \hspace{-1.1mm} \right) =
\tau_2 \left( \hspace{-1.1mm} \begin{array}{c} y_1 \\
1 - y_1 \end{array} \hspace{-1.1mm} \right) \,,
\hspace{2mm} \ldots \,, \hspace{2mm}
\left( \hspace{-1.1mm} \begin{array}{c} \tau_{L-1} \\
\rho_L \end{array} \hspace{-1.1mm} \right) =
\tau_L \left( \hspace{-1.1mm} \begin{array}{c} y_{L-1} \\
1 - y_{L-1} \end{array} \hspace{-1.1mm} \right) \hspace{5mm} \mathrm{with} \hspace{4mm}
\left\{ \hspace{-0.7mm} \begin{array}{l}
0\leq \tau_j < \infty \\[0.1mm]
0\leq y_j \leq 1 \,,
\end{array} \right.
\label{eq:tau_y_variables}
\end{equation}
where the variables $\tau_j$ have the dimension of length and
the $y_j$ are dimensionless. The $L$-loop eikonal diagram then becomes
\begin{equation}
F^{(L)} = C^{(L)} \prod_{j=1}^L \int_0^1 \d x_j \hspace{0.5mm}
P_{\ell_j r_j}^{[\epsilon]} (x_j) \hspace{0.4mm} K(x_1,\ldots,x_L) \,,
\label{eq:L-loop_ladder_diagram_rep_2}
\end{equation}
where the infrared divergence of the diagram is now absorbed into the kernel
\begin{equation}
K(x_1,\ldots,x_L) = g^{2L} \hspace{0.6mm} \Gamma(2L\epsilon) \hspace{-0.2mm}
\left( \frac{\mu}{\Lambda} \right)^{2L\epsilon}
\prod_{j=1}^{L-1} \int_0^1 \d y_j \hspace{0.5mm}
y_j^{-1+2j\epsilon} (1- y_j)^{-1+2\epsilon}
\Theta \big(\{\boldsymbol{y}, \boldsymbol{x} \}\big) \,.
\label{eq:kernel}
\end{equation}
Here $\Theta \big(\{\boldsymbol{y}, \boldsymbol{x} \}\big)$
denotes the result of applying the substitutions
(\ref{eq:tau_y_variables}) to eq.~(\ref{eq:nesting_function_def}).
In analogy with section~\ref{sec:causality_and_unitarity_of_WL},
we have here regulated the infrared divergence in a gauge invariant way
through the exponential damping factor $e^{-\Lambda \tau_L}$
with $\Lambda \ll 1$. Eq.~(\ref{eq:kernel}) contains in addition
any potential ultraviolet subdivergences of the diagram (generated by the
nesting function $\Theta \big(\{\boldsymbol{y}, \boldsymbol{x} \}\big)$).

Having brought the $L$-loop eikonal diagram in the form
(\ref{eq:L-loop_ladder_diagram_rep_2}), we now turn to
extracting its imaginary part. Restricting our attention
to the leading order in the dimensional regulator $\eps$,
we will drop the dependence of the soft propagators on $\eps$,
\begin{equation}
F^{(L)} = C^{(L)} \prod_{j=1}^L \int_0^1 \d x_j \hspace{0.5mm}
P_{\ell_j r_j}^{[0]} (x_j) \hspace{0.4mm} K(x_1,\ldots,x_L)
\hspace{1mm} + \hspace{1mm} \mathcal{O}(\eps^{-d+1}) \,,
\label{eq:L-loop_ladder_diagram_rep_3}
\end{equation}
where $d$ denotes the degree of divergence of the diagram,
$F^{(L)} \sim \frac{1}{\eps^d} \times \mathrm{(finite)}$.

To compute the imaginary part of eq.~(\ref{eq:L-loop_ladder_diagram_rep_3}),
we start by observing that eq.~(\ref{eq:kernel}) is
manifestly purely real. As a result, the Feynman $i\eta$'s are
the only source of imaginary parts of
Eq.~(\ref{eq:L-loop_ladder_diagram_rep_3}). Each
of the $x_j$-integration paths can therefore be
decomposed into a principal-value part and
small semicircles around the propagator poles.
Given that the integrand takes purely imaginary values in the regions
close to the poles and is real-valued on the remaining domain of
integration, the resulting $2^L$ terms (which each involve
$L$ integrations) will be either purely real or purely imaginary.

To collect the imaginary contributions, we define
the cut propagator
\begin{align}
\label{eq:Dij}
\Delta_{ij}(x) \equiv
- \pi \, v_i \cdot v_j \, \delta\big( (x v_i-(1-x) v_j)^2 \big) \,,
\end{align}
and furthermore $p$-fold cutting operator
\begin{align}
\label{eq:Cut_x F}
\Cut_{x_{i_1},\dotsc,x_{i_p}} F^{(L)} &=
\prod_{j=1 \atop \!\!\! j \neq i_1,\dotsc,i_p \!\!\!\!\!\!}^{n} \PV \int_0^1 \d x_j \, P(x_j)
\prod_{k=1}^{p} \int_0^1 \d x_{i_k} \, \Delta(x_{i_k}) ~ K(x_1,\dotsc,x_L)\,.
\end{align}
The action of this operator is to replace the $p$ propagators
that depend on the specified variables by delta functions
and to place a principal-value prescription on the integrals
over the remaining variables. To simplify notation,
we here dropped the indices on the (cut) propagators:
$P(x_j) \equiv P_{\ell_j r_j}^{[0]} (x_j)$
and $\Delta (x_j) \equiv \Delta_{\ell_j r_j} (x_j)$.

The imaginary part of any $L$-loop eikonal diagram without
internal vertices can then be written, to the leading order in $\eps$,
\begin{equation}
\Im F^{(L)}=
\sum_{p=1 \atop p \hspace{0.7mm} \text{odd}}^L ~ \sum_{i_1,\dotsc,i_p = 1 \atop i_1 < \dotsm < i_p}^L i^{\,p-1} \Cut_{x_{i_1}, \ldots, x_{i_p}} F^{(L)} \,.
\label{eq:Master Im F}
\end{equation}
This is the central formula of our approach~\cite{Laenen:2014jga}.
The formula~(\ref{eq:Master Im F})
is illustrated schematically for a generic ladder diagram in
figure~\ref{fig:Im_of_three-loop_ladder} below.

We note that the decomposition of the line integrations
in eq.~(\ref{eq:L-loop_ladder_diagram_rep_3}) into
principal-value and delta function contributions
immediately shows that the imaginary part of
the integrated expression for the eikonal diagram
will have transcendentality weight one less than
the real part. This follows from the fact that the delta functions
will map the rational integrand to a rational expression
after being integrated out. Thus, compared to the real-part contribution
with $L$ principal-value integrals, the weight is dropped by one.

\begin{figure}[!h]
\begin{center}
\includegraphics[angle=0, width=1.0\textwidth]{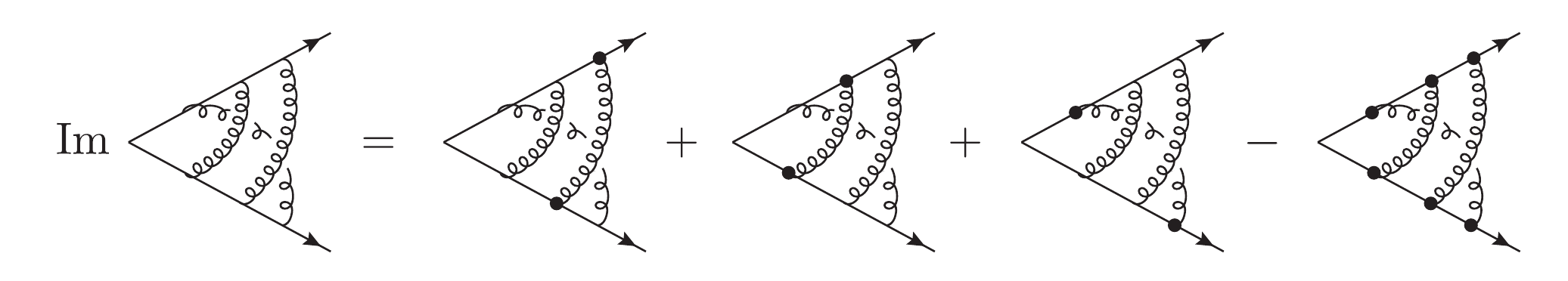}
{\vskip -3mm}
\caption{Schematic illustration of the formula~(\ref{eq:Master Im F})
for the imaginary part of an eikonal diagram without internal vertices.
The black dots at the endpoints of a soft-gluon propagator denote
that the propagator has been cut---that is, replaced by a delta function.
It is implied that the integrals over the attachment points of
uncut soft propagators are principal-value integrals. The relative
signs of the diagrams are determined by the factor $i^{p-1}$;
each individual diagram displayed here corresponds to the action
of the cutting operator (\ref{eq:Cut_x F}) on the eikonal diagram.}
\label{fig:Im_of_three-loop_ladder}
\end{center}
\end{figure}

It is natural to ask about the relation of the imaginary part
of the eikonal diagram to the discontinuities in its various
kinematic channels. This in turn leads us to ask for an
appropriate set of variables in terms of which to express
integrated results. A good choice of variables turns out
to be given by the exponentials of the cusp angles,
\begin{equation}
\chi_{ij} \equiv e^{-\gamma_{ij}} \,,
\end{equation}
with the cusp angles defined through $\cosh \gamma_{ij} = |v_i \cdot v_j|$.
Expressed in terms of the $\chi_{ij}$-variables, the
eikonal diagram has branch cuts located on the real line and
satisfies Schwarz reflection, $F^{(L)}(\overline{\chi_{ij}})
= \overline{F^{(L)}(\chi_{ij})}$. As a result, the discontinuities
of the diagram give rise to the imaginary part through the relation
\begin{equation}
2i \hspace{0.7mm} \mathrm{Im} \hspace{0.6mm}
F^{(L)} (\boldsymbol{\chi})
\hspace{0.7mm}=\hspace{0.7mm} \sum_{j=1}^L \theta(v_{\ell_j} \cdot v_{r_j}) \hspace{0.2mm}
\mathop{\mathrm{Disc}}_{\chi_{\ell_j r_j}} \hspace{0.3mm}
F^{(L)} (\boldsymbol{\chi}) \,.
\label{eq:Im_and_discontinuities_relation}
\end{equation}
Here, the step functions account for the fact that the
imaginary part has vanishing contributions from channels
with space-like kinematics $v_{\ell_j} \cdot v_{r_j} < 0$.
(This follows from the fact that propagators stretched
between mutually space-like eikonal lines have vanishing cuts,
as will be explained below eq.~(\ref{eq:propagator_roots}).)
We will see an explicit example of this in section~\ref{sec:Web121}
where we study a diagram that depends on two distinct
cusp angles in purely time-like as well as mixed
time- and space-like kinematics.
\\
\\
In section~\ref{sec:examples} we will work out examples of
how eq.~(\ref{eq:Master Im F}) is used in practice to compute
the imaginary part of ladder-type eikonal diagrams.
To this end it will be useful to record the following
partial-fractioned expressions, setting $\chi \equiv \chi_{ij}$,
\begin{align}
\label{eq:Pij(0)-partialfraction}
P_{ij}^{[0]}(x) &= \frac{R(\chi)}{2}\left(\frac{1}{x-\rho_1 +i\eta}-\frac{1}{x-\rho_2-i\eta}\right)
\nonumber \\[1mm]
\Delta_{ij}(x) &= - \pi \frac{R(\chi)}{2}\Big(\delta(x-\rho_1)+\delta(x-\rho_2)\Big) \,,
\end{align}
where the prefactor is the rational expression
$R(\chi) = \frac{1+\chi^2}{1-\chi^2} = \coth\gamma_{ij}$,
and the denominator roots are given by
\begin{equation}
(\rho_1, \rho_2) \hspace{0.9mm}=\hspace{0.9mm}
\left\{ \hspace{-0.7mm} \begin{array}{llllr}
\left( \frac{\chi}{\chi + 1} , \frac{1}{\chi + 1} \right) \hspace{5mm} &
\mathrm{for} \hspace{1mm} & v_i \cdot v_j
\hspace{-2.1mm}&\equiv&\hspace{-2.1mm} \cosh \gamma_{ij} > 0 \,\phantom{.} \\[2.5mm]
\left( \frac{\chi}{\chi -1} , \frac{1}{1 - \chi} \right) \hspace{5mm} &
\mathrm{for} \hspace{1mm} & v_i \cdot v_j
\hspace{-0.5mm}&\equiv&\hspace{-0.5mm} -\cosh \gamma_{ij} < 0 \,.
\end{array} \right.
\label{eq:propagator_roots}
\end{equation}
We note that in the upper case of eq.~(\ref{eq:propagator_roots}),
the roots satisfy $0<\rho_1<\tfrac{1}{2}<\rho_2<1$, whereas in the lower
case they satisfy $\rho_1 < 0 < 1 < \rho_2$. Since the delta functions
in eq.~(\ref{eq:Cut_x F}) are integrated over the interval $[0,1]$,
we may thus infer that the eikonal diagram will only have contributions
to its imaginary part from channels with time-like kinematics
$v_i \cdot v_j > 0$, as encoded in eq.~(\ref{eq:Im_and_discontinuities_relation}).
This is in agreement with the causality considerations
of section~\ref{sec:causality_and_unitarity_of_WL}.

In section~\ref{sec:examples} we will make extensive use of the fact
that the result for an eikonal diagram in time-like kinematics can
be immediately obtained from the space-like result
by analytic continuation of the cusp angle. To see this,
let us first recall that the soft propagator takes the same
form (\ref{eq:propagator_notation})~in space- and time-like
kinematics when expressed in terms of $v_i \cdot v_j$, owing to
our convention that all velocity vectors are outgoing.
However, once expressed in terms of the relative angle
$\gamma \equiv \gamma_{ij}$, it takes the respective forms
\begin{equation}
P_{ij}^{[\eps]}(x) \hspace{0.5mm}=\hspace{0.5mm}
\left\{ \hspace{-0.7mm} \begin{array}{llllr}
-\cosh \gamma \hspace{0.3mm} \big[{-}x^2 \hspace{-0.2mm}-\hspace{-0.2mm} (1-x)^2
\hspace{-0.3mm}-\hspace{-0.3mm} 2x(1-x) \cosh \gamma \big) \big]^{-1+\eps} \hspace{1mm} &
\mathrm{for} \hspace{-1mm} & v_i \cdot v_j
\hspace{-1.1mm}&\equiv&\hspace{-1.1mm} -\cosh \gamma < 0 \,\phantom{.} \\[2.5mm]
\cosh \gamma \hspace{0.3mm} \big[{-}x^2 \hspace{-0.2mm}-\hspace{-0.2mm} (1-x)^2
\hspace{-0.2mm}+\hspace{-0.2mm} 2x(1-x) \cosh \gamma + i\eta \big]^{-1+\eps} \hspace{1mm} &
\mathrm{for} \hspace{-1mm} & v_i \cdot v_j
\hspace{-1.1mm}&\equiv&\hspace{-1.1mm} \cosh \gamma > 0 \,,
\end{array} \right.
\end{equation}
where we dropped the $i\eta$ in the space-like case,
as the propagator roots are located outside the range $[0,1]$ of $x$.
Comparison of these expressions shows that we can
map space-like to time-like kinematics by means of the
analytic continuation
\begin{equation}
-\cosh \gamma \hspace{0.7mm}\longrightarrow\hspace{0.7mm} \cosh \gamma + i\eta \,,
\label{eq:analytic_continuation_gamma}
\end{equation}
or equivalently, in terms of $\chi \equiv e^{-\gamma}$,
\begin{equation}
\chi \hspace{0.7mm}\longrightarrow\hspace{0.7mm} -\frac{1}{\chi} - i\eta \,.
\end{equation}

\section{Examples}\label{sec:examples}

The aim of this section is to apply the formalism reviewed
in section~\ref{sec:Im_of_L-loop_Wilson_lines} to compute
the imaginary part of a number of ladder-type eikonal diagrams.
The main point to be addressed here concerns the evaluation of
the principal-value integrals involved in the $p$-fold cuts
in eq.~(\ref{eq:Cut_x F}).

\subsection{The non-planar two-loop ladder diagram}\label{sec:2loopNPladder}

As a first example we will consider the non-planar two-loop ladder diagram,
illustrated in \refF{fig:non-planar_two-loop_ladder} below.
This diagram contains no ultraviolet subdivergence and therefore
only has a simple pole in the dimensional regulator $\eps$.
In agreement with the observations at the end of
section~\ref{sec:Im_of_L-loop_Wilson_lines},
the diagram will only have an imaginary part for time-like kinematics
$v_1 \cdot v_2 > 0$. We therefore restrict our attention to this case.
Since the diagram contains only one cusp angle, we will
drop the subscripts for convenience and define $\cosh \gamma
\equiv v_1 \cdot v_2 $ as well as $\chi \equiv e^{-\gamma}$.

\begin{figure}[!h]
\begin{center}
\includegraphics[angle=0, width=0.35\textwidth]{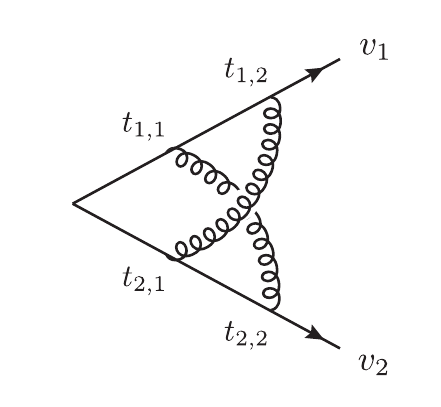}
{\vspace{-5mm}}
\caption{The non-planar two-loop ladder diagram.}
\label{fig:non-planar_two-loop_ladder}
\end{center}
\end{figure}

\noindent The non-planar two-loop ladder diagram has the position-space representation
\begin{align}
F^{(2)} &= C^{(2)} \mu^{4\epsilon}
\int_0^{\infty} \frac{\d t_{1,1} \,\d t_{1,2} \,\d t_{2,1} \,\d t_{2,2} \,\,\theta(t_{1,2}-t_{1,1})\,\theta(t_{2,2}-t_{2,1})\,(v_1 \cdot v_2)^2}
{\big[ {-} (t_{1,1} v_1 - t_{2,2} v_2)^2 +i\eta \big]^{1-\epsilon}\,
\big[ {-} (t_{1,2} v_1 - t_{2,1} v_2)^2 +i\eta \big]^{1-\epsilon}} \,,
\label{eq:NP_2-loop_ladder_eq_1}
\end{align}
where the prefactor is given by $C^{(2)} = - \frac{g^4 C_F}{2N}
\frac{\Gamma^2 (D/2 - 1)}{16\pi^D}$. To compute  the imaginary part
of this diagram, our first task is to write it in the form of
eq.~(\ref{eq:L-loop_ladder_diagram_rep_2}). This is achieved
through the change of variables in eq.~(\ref{eq:radial_coordinates}),
followed by that in eq.~(\ref{eq:tau_y_variables}),
\begin{align}
\colvec{ t_{1,1} \\ t_{2,2} }
=\lambda   \colvec{ x  \\ 1-x   } \,, \hspace{5mm}
\colvec{ t_{1,2} \\ t_{2,1} }
=\sigma  \colvec{ y  \\ 1-y  } \hspace{5mm}
\mbox{followed by} \hspace{3mm}
\colvec{ \lambda   \\ \sigma  }
=\beta \colvec{ t \\ 1-t } \,.
\label{eq:NP_2-loop_ladder_change_of_vars}
\end{align}
After these transformations the diagram takes the desired form,
\begin{align}
F^{(2)} &= C^{(2)}
\int_0^1 \d x
\int_0^1 \d y
~P_{12}^{[\eps]}(x) \, P_{12}^{[\eps]}(y) \, K(x, y) \,.
\label{eq:NP_2-loop_ladder_eq_2}
\end{align}
where the kernel $K(x, y)$, upon the additional
change of variable $u=\frac{t}{1-t}$, is given by
\begin{align}
K(x,y)&=
\mu^{4\epsilon}
\int_0^\infty \frac{\d\beta \hspace{0.7mm} e^{-\Lambda \beta}}{\beta^{1-4\eps}}
\int_0^\infty \d u
~ u^{-1+2\eps} (u+1)^{-4\eps}
\,\theta\left(\frac{y}{x}-u\right)
\,\theta\left(u-\frac{1-y}{1-x}\right) \,.
\label{eq:NP_2-loop_ladder_kernel_eq_1}
\end{align}
By comparing eqs.~(\ref{eq:NP_2-loop_ladder_eq_1})
and (\ref{eq:NP_2-loop_ladder_eq_2}), we see that the
effect of the first two transformations
of eq.~(\ref{eq:NP_2-loop_ladder_change_of_vars})
is to leave each soft propagator dependent on a single variable.
The effect of the last change of variable is to
extract the overall infrared divergence of the diagram.

To facilitate the computation of the cuts in eq.~(\ref{eq:Master Im F})
we will first evaluate the integral $K(x,y)$.
The $u$-integral in eq.~(\ref{eq:NP_2-loop_ladder_kernel_eq_1})
may be performed in terms of the ${}_2F_1$ hypergeometric function.
The primitive has the $\eps$-expansion
\begin{align}
f(u) &= \frac{u^{2\eps}}{2\eps} \, {}_2F_1(2\eps,4\eps;1+2\eps;-u)
= \frac{u^{2\eps}}{2\eps} \, \big( 1+\Ord(\eps^2) \big)
= \frac{1}{2\eps} + \log u + \Ord(\eps) \,,
\end{align}
and so $K(x,y)$ has the $\eps$-expansion
\begin{align}
K(x,y)&=
\Gamma(4\epsilon) \left(\frac{\mu}{\Lambda}\right)^{4\epsilon}
\theta(y-x)
\left[
f\left(\frac{y}{x}\right) - f\left(\frac{1-y}{1-x}\right)
\right]
\nn&=
\frac{1}{4\eps}\left(\frac{\mu}{\Lambda}\right)^{4\epsilon}
\theta(y-x) \left(\log \frac{y}{x} - \log \frac{1-y}{1-x}\right)
+\Ord(\eps^0) \,.
\label{eq:NP_2-loop_ladder_kernel_eq_2}
\end{align}
Substituting this result for $K(x,y)$ into eq.~(\ref{eq:NP_2-loop_ladder_eq_2}),
we can write the non-planar two-loop ladder diagram in the convenient form
\begin{align}
\label{eq:F2-v3}
F^{(2)} &= \frac{C^{(2)}}{4\eps}\left(\frac{\mu}{\Lambda}\right)^{4\epsilon} \Fau^{(2)} \,,
\end{align}
where $\Fau^{(2)}$ is finite, given to leading order in $\eps$ as,
\begin{align}
\label{eq:Fau2-v1}
\Fau^{(2)}&=
\int_0^1 \d x
\int_0^1 \d y
~P_{12}^{[0]}(x)\, P_{12}^{[0]}(y)\,
\Big( \theta(y-x) - \theta(x-y) \Big)
\log \frac{y}{x} \,.
\end{align}
Here we have dropped the dependence of the soft propagators
on $\eps$ and furthermore rewritten the integrand of $\Fau^{(2)}$
to make its symmetry under $x \longleftrightarrow y$ manifest.
(This was achieved by changing variables $(x,y) \mapsto (1-x,1-y)$
on the second term $\log\frac{1-y}{1-x}$ arising from
eq.~(\ref{eq:NP_2-loop_ladder_kernel_eq_2}).)

As the prefactor of $\Fau^{(2)}$ in eq.~(\ref{eq:F2-v3}) is real,
it factors out on both sides of eq.~(\ref{eq:Master Im F}), yielding the formula
\begin{align}
\label{eq:ImF2-v1}
\Im \Fau^{(2)} &=\Cut_{x} \Fau^{(2)} + \Cut_{y} \Fau^{(2)} \,,
\end{align}
where we recall that $\Cut_{x_i}$ is defined in eq.~(\ref{eq:Cut_x F})
and replaces the propagator depending on the specified
variable by a delta function and places a principal-value
prescription on the integral over the remaining variable.
The two cuts on the right-hand side are equal because the
integrand of eq.~(\ref{eq:Fau2-v1}) is symmetric under
the interchange of $x$ and $y$. (This also follows
from the $v_1 \longleftrightarrow v_2$ symmetry of the
original diagram.) Thus, it suffices to compute $\Cut_{x} \Fau^{(2)}$,
given by
\begin{align}
\label{eq:Cut-x-F2-v1}
\Cut_{x} \Fau^{(2)} &=
\int_0^1 \d x
\PV \int_0^1 \d y
~\Delta_{12}(x)\, P_{12}^{[0]}(y)\,
\Big( \theta(y-x) - \theta(x-y) \Big)
\log \frac{y}{x} \,.
\end{align}
Inserting the partial-fractioned expressions for
$\Delta_{12}(x)$ and $P_{12}^{[0]}(y)$ given in \refE{eq:Pij(0)-partialfraction}
and performing the trivial integral over $x$ produces
\begin{align}
\label{eq:Cut-x-F2-v2}
\Cut_{x} \Fau^{(2)} &=
-\frac{\pi}{4}R(\chi)^2
\sum_{k=1}^2
\PV\left(
\int_{\rho_k}^1 \d y
-\int_0^{\rho_k} \d y
\right)
\left(\tfrac{1}{y -\rho_1+i\eta}-\tfrac{1}{y -\rho_2-i\eta}\right)
\log \frac{y}{\rho_k} \,,
\end{align}
where the propagator roots $\rho_{1,2}$ are given
in the upper part of eq.~(\ref{eq:propagator_roots}).

We are now confronted with the task of evaluating
principal-value integrals. As such integrals
do not immediately take the form of iterated integrals,
our strategy for evaluation will be to write them
as differences of iterated integrals, which in turn
are readily expressible in terms of multiple polylogarithms.
The basic observation is that the principal-value
integral equals the corresponding full integral
minus the imaginary part of the latter,
cf. eq.~(\ref{eq:PV-formula}).

As a simple illustration, let us consider the
evaluation of the following principal-value integral,
\begin{align}
\label{eq:PVisREisFULL-IM}
\PV \int_{0}^{1}\frac{\d y}{y-\rho_1 + i\eta}
=\int_{0}^{1}\frac{\d y}{y-\rho_1 + i\eta} - i \Im \int_{0}^{1}\frac{\d y}{y-\rho_1 + i\eta} \,.
\end{align}
The full integral evaluates to $G(\rho_1;1)$ by definition,
cf. eqs.~(\ref{eq:G definition recursion})--(\ref{eq:G definition}).
Its imaginary part arises from the pole of
the integrand and is extracted by localizing
the integration variable,
\begin{align}
\Im \int_{0}^{1}\frac{\d y}{y-\rho_1+i\eta}
= -\pi \int_{0}^{1}\d y \,\delta(y-\rho_1)
= -\pi \,,
\end{align}
where we used in the last step that the pole is
located inside the range of integration, in agreement with
the discussion below eq.~(\ref{eq:propagator_roots}).
Thus we arrive at
\begin{align}
\label{eq:PVlog}
\PV \int_{0}^{1}\frac{\d y}{y-\rho_1 +i\eta} &=G(\rho_1;1) + \pi i \,.
\end{align}
In this simple example we could have computed the real part more directly,
\begin{align}
\PV \int_{0}^{1}\frac{\d y}{y-\rho_1}
&=\Re G(\rho_1;1)
=\Re \log\big(1-\tfrac{1}{\rho_1}\big)
=\log\big|1-\tfrac{1}{\rho_1}\big|
=\log\big(\tfrac{1}{\rho_1}-1\big)\,.
\end{align}
However, an extension of this direct approach to
higher-weight cases requires the use of a sequence
of functional identities which in practice is case-dependent
and thus not applicable in a systematic way. In contrast,
the above method relies only on a construction of the
imaginary part which can be derived systematically
as demonstrated in \refA{App:ImG}.

Returning to eq.~(\ref{eq:Cut-x-F2-v2}) and evaluating
the principal-value integrals following the steps outlined
above, the result for the cut is readily expressed in terms
of multiple polylogarithms,
\begin{align}
\label{eq:Cut-x-F2-v3}
\Cut_{x} \Fau^{(2)}
&= -\frac{\pi}{4}R(\chi)^2
\Big(
{-}2G(\rho_2,0;1)+2G(\rho_1,0;1)-2G(0,\rho_2;\rho_2)-2G(0,\rho_2;\rho_1)
\nn&\hspace{23mm}
+2G(0,\rho_1;\rho_2)+2G(0,\rho_1;\rho_1)+G(0;\rho_2)G(\rho_2;1)+G(0;\rho_1)G(\rho_2;1)
\nn&\hspace{23mm}
-G(0;\rho_2)G(\rho_1;1)-G(0;\rho_1)G(\rho_1;1)-2\pi i G(0;\tfrac{\rho_1}{\rho_2})
\Big) \,,
\end{align}
where we refer to eqs.~(\ref{eq:G definition recursion})--(\ref{eq:G definition})
for definitions. In this expression, the multiple polylogarithms depend on
the propagator roots $\rho_{k}(\chi)$ through both their indices
and their arguments. This expression can in turn be rewritten in
terms of polylogarithms with constant indices by exploiting
the Hopf algebra structure of multiple polylogarithms,
which encodes the plethora of functional identities
within this class of functions
\cite{Goncharov:2005sla,Goncharov:2010jf,Brown:2011ik,Duhr:2011zq,Duhr:2012fh,Anastasiou:2013srw}.
Utilizing this algebraic structure, we have implemented
the steps required to achieve the desired functional
form as a general algorithm. We refer to \refA{App:algorithm}
for further details. The algorithm leaves a simplified
form of eq.~(\ref{eq:Cut-x-F2-v3}) expressed in terms of
harmonic polylogarithms which, using \refE{eq:GtoClassic},
can be simplified further into classical polylogarithms,
\begin{align}
\label{eq:Cut-x-F2-v4}
\Cut_{x} \Fau^{(2)}
&= -\pi R(\chi)^2
\Big({-}G(0,1;\chi)+G(0,0;\chi)-G(0,-1;\chi)-\frac{1}{2}\zeta_2\Big) \nn
&= -\frac{\pi}{2}R(\chi)^2
\Big(\text{Li}_2(\chi^2)+\log^2\chi-\zeta_2\Big) \,.
\end{align}
With this result, we can now immediately obtain the imaginary part
of the two-loop ladder from \refE{eq:ImF2-v1}, recalling
that the two cuts are equal. We find
\begin{align}
\label{eq:ImF2-v2}
\Im \Fau^{(2)} &=
-\pi R(\chi)^2
\big(\text{Li}_2(\chi^2)+\log^2 \chi -\zeta_2\big)\,,
\end{align}
and we recall that multiplying the infrared pole
cf. \refE{eq:F2-v3} onto both sides of
eq.~(\ref{eq:ImF2-v2}) gives the imaginary part of
the original diagram $F^{(2)}$. This completes
the evaluation of the imaginary part of the
non-planar two-loop ladder diagram to the leading
order in $\eps$.
\\
\\
As a crosscheck of the result in eq.~(\ref{eq:ImF2-v2}),
we can alternatively compute the imaginary part of the non-planar
two-loop ladder by evaluating the diagram for space-like
kinematics $v_1 \cdot v_2 < 0$, in which case it will
be purely real (cf. the discussion at the end of
section~\ref{sec:Im_of_L-loop_Wilson_lines}), and subsequently
perform the analytic continuation to time-like kinematics.
We refer to the end of section~\ref{sec:Im_of_L-loop_Wilson_lines}
for a more detailed discussion of analytic continuations.

\begin{figure}[!h]
\begin{center}
\includegraphics[angle=0, width=0.75\textwidth]{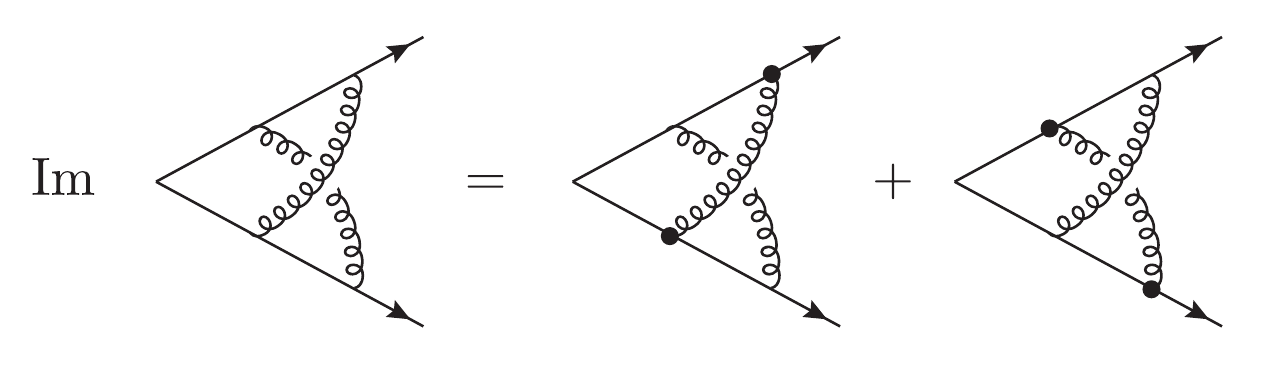}
{\vskip -3mm}
\caption{Graphical representation of the cutting prescription for the
non-planar two-loop ladder stated in eq.~(\ref{eq:ImF2-v1}).
The black dots at the endpoints of a soft-gluon propagator
indicate that the propagator has been cut; i.e., replaced
by a delta function. It is implied that the integrals over
the attachment points of uncut soft propagators are
principal-value integrals.}
\label{fig:Im_of_two-loop_ladder}
\end{center}
\end{figure}

To the leading order in $\eps$, the two-loop ladder is given by
eq.~(\ref{eq:Fau2-v1}), although we must bear in mind that for
space-like kinematics the propagator roots $\rho_k$ are given
by the lower case of eq.~(\ref{eq:propagator_roots}). Inserting
into eq.~(\ref{eq:Fau2-v1}) the expressions for $P^{[0]}_{12}$
given in eq.~(\ref{eq:Pij(0)-partialfraction}),
the diagram readily evaluates into multiple polylogarithms,
\begin{align}
\label{eq:F2-tch-v2}
\widetilde{\Fau}^{(2)} &=
\frac{R(\chi)^2}{2} \big(G(\rho_1,0,\rho_1;1)
-G(\rho_1,0,\rho_2;1)-G(\rho_2,0,\rho_1;1)
+G(\rho_2,0,\rho_2;1)\big) \,.
\end{align}
The tilde on the left-hand side serves to remind us that
the expression for the diagram on the right-hand side is valid for
space-like kinematics.
We can use the algorithm in appendix~\ref{App:algorithm}
to recast this representation in terms of polylogarithms
with constant indices. In fact, the two-loop ladder
diagram can be expressed in terms of classical polylogarithms,%
\footnote{Note that in the representation in the second line of
eq.~(\ref{eq:two-loop_ladder_space-like_result}),
the polylogarithms and logarithms have the respective
arguments $\chi^2$ and $\chi$. This form has the advantage
of being particularly compact as well as convenient
for the purpose of performing the analytic continuation
$\chi \rightarrow -1/\chi-i\eta$ which then maps
the arguments of the (poly)logarithms to the vicinity
of the branch cuts. (In general, computer algebra software,
such as \texttt{Mathematica} augmented with the package
\texttt{HPL} \cite{Maitre:2005uu,Maitre:2007kp},
does not detect the $2 \pi$ monodromy of $\log (\chi^2)$
after tracing out a complete circle around the branch point.)}
\begin{align}
\widetilde{\Fau}^{(2)} &=
R(\chi)^2\,
\big(2G(0,1,0;\chi)-2G(0,0,0;\chi)+2G(0,-1,0;\chi)- \zeta_2 G(0;\chi) -\zeta_3\big)
\nn&=
R(\chi)^2\,
\big(\text{Li}_3(\chi^2)-\log \chi \hspace{0.6mm}
\text{Li}_2(\chi^2)-\tfrac{1}{3}\log^3 \chi - \zeta_2 \log \chi -\zeta_3\big)\,.
\label{eq:two-loop_ladder_space-like_result}
\end{align}
We can now find the result for the two-loop ladder
diagram in time-like kinematics by performing
the analytic continuation $\chi \rightarrow -1/\chi-i\eta$
on eq.~(\ref{eq:two-loop_ladder_space-like_result}).
Under the analytic continuation, the rational function
$R(\chi)$ picks up a minus sign, while the polylogarithms
transform according to
\begin{align}
\log \chi &\rightarrow -\log \chi -\pi i \nn
\Dilog(\chi^2) &\rightarrow
-\Dilog(\chi^2) - 2\log^2 \chi + 2\zeta_2 - 2\pi i \log \chi \nn
\Trilog(\chi^2) &\rightarrow
\Trilog(\chi^2) + \tfrac{4}{3}\log^3 \chi
- 4 \zeta_2 \log \chi + 2 \pi i \log^2 \chi \,.
\end{align}
Applying these replacements to \refE{eq:two-loop_ladder_space-like_result}
we find the following result for the non-planar
two-loop ladder with time-like kinematics,
\begin{align}
\Fau^{(2)} &=
R(\chi)^2
\,\Big(\,
\text{Li}_3(\chi^2)
-\log \chi \hspace{0.7mm} \text{Li}_2(\chi^2)
-\tfrac{1}{3}\log^3 \chi
+5\zeta_2 \log \chi
-\zeta_3
\nn&\hspace{18mm}
-i\pi\big(
\text{Li}_2(\chi^2)
+\log^2 \chi
-\zeta_2\big)
\Big)\,.
\label{eq:two-loop_ladder_time-like_result}
\end{align}
We observe that the imaginary part of
eq.~(\ref{eq:two-loop_ladder_time-like_result})
agrees with the result found in eq.~(\ref{eq:ImF2-v2}),
as expected. We conclude that the cutting prescription
for the two-loop ladder stated in eq.~(\ref{eq:ImF2-v1})
produces the correct imaginary part. The cutting
prescription (\ref{eq:ImF2-v1}) is illustrated in
figure~\ref{fig:Im_of_two-loop_ladder}.

\subsection{Three-loop non-planar ladder diagram}\label{sec:3loopNPladder}

\begin{figure}[!h]
\begin{center}
\includegraphics[angle=0, width=0.35\textwidth]{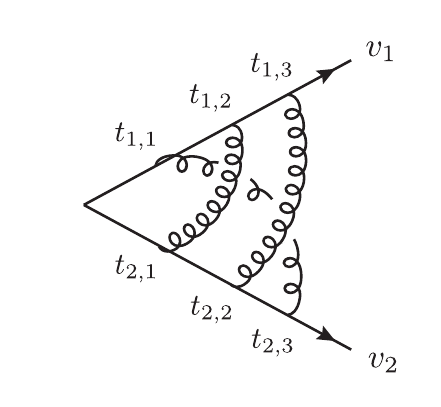}
{\vskip -4mm}
\caption{A three-loop non-planar ladder diagram.}
\label{fig:three-loop_non-planar_ladder}
\end{center}
\end{figure}

To demonstrate that the principal-value integrals
involved in the $p$-fold cuts in eq.~(\ref{eq:Cut_x F}) can indeed
be evaluated in non-trivial cases, we consider in this section
the three-loop ladder diagram illustrated in \refF{fig:three-loop_non-planar_ladder}.
This diagram also represents an example of an eikonal diagram
with multiple-cut contributions to its imaginary part (in the
case at hand, a triple cut). As in section~\ref{sec:2loopNPladder}, we take
$\cosh \gamma \equiv v_1 \cdot v_2 > 0$, in order to have
a non-vanishing imaginary part, and set $\chi \equiv e^{-\gamma}$.

\noindent The position-space representation of the diagram
in \refF{fig:three-loop_non-planar_ladder} takes the form
\begin{align}
\label{eq:F3-v1}
F^{(3)} &= C^{(3)} \mu^{6\epsilon}
\int_0^{\infty}
\Bigg(\prod_{i=1}^2\prod_{j=1}^3 \d t_{i,j}\Bigg)
\frac{\prod_{i,j=1}^2\theta(t_{i,j+1}-t_{i,j}) \,
(v_1 \cdot v_2)^3 }{
D(t_{1,1},t_{2,3}) \,
D(t_{1,2},t_{2,1}) \,
D(t_{1,3},t_{2,2})
} \,,
\end{align}
where $D(t_1,t_2) = \left[ -(t_1 v_1-t_2 v_2)^2+i\eta \right]^{1-\eps}$.
To compute the imaginary part of this diagram from
eq.~(\ref{eq:Master Im F}), our first task is to bring it into the form of
eq.~(\ref{eq:L-loop_ladder_diagram_rep_2}). This is achieved
through the changes of variables in eq.~(\ref{eq:radial_coordinates})
(with $t_{\ell_j, m_j} = t_{1,j}$), followed by
the sequence of substitutions in eq.~(\ref{eq:tau_y_variables}),
setting $(x_1, x_2, x_3) = (x,y,z)$ and $(y_1,y_2)=(t,u)$ for convenience.

After these transformations, the diagram takes the form
\begin{align}
\label{eq:F3-v2}
F^{(3)} &= C^{(3)}
\int_0^1 \d x \hspace{0.7mm} \d y \hspace{0.7mm} \d z
\, P_{12}^{[\eps]}(x) \, P_{12}^{[\eps]}(y) \, P_{12}^{[\eps]}(z) \, K(x,y,z) \,,
\end{align}
where the kernel is given by
\begin{align}
\label{eq:K(xyz)-v2}
K(x,y,z)&= \Gamma(6\eps) \left( \frac{\mu}{\Lambda} \right)^{6\eps}
\int_0^1 \d t ~ t^{-1+2\eps} (1-t)^{-1+2\eps}
\int_0^1 \d u ~ u^{-1+4\eps} (1-u)^{-1+2\eps}
\nn&\hspace{10mm}
\times\theta\Big(\tfrac{z}{y} - \tfrac{(1-t)\,u}{1-u}\Big) \,
\theta\Big(\tfrac{y}{x} - \tfrac{t}{1-t}\Big) \,
\theta\Big(\tfrac{t\,u}{1-u} - \tfrac{1-z}{1-x}\Big) \,
\theta\Big(\tfrac{1-z}{1-y} - \tfrac{(1-t)\,u}{1-u}\Big) \,.
\end{align}
The arguments of the step functions simplify after rescaling
the integration variables according to $\frac{t}{1-t} \mapsto t$
and $\frac{u}{1-u} \mapsto u$. As we are interested in computing
the imaginary part of $F^{(3)}$ only to the leading order in
$\eps$, we may set $\eps$ to zero in the $u$-integral. Performing
the $t$-integral in terms of hypergeometric functions and
subsequently expanding in $\eps$, we find the expression
\begin{align}
\label{eq:K(xyz)-v3}
K(x,y,z)&=
\frac{1}{6\eps}\left(\frac{\mu}{\Lambda}\right)^{6\epsilon} \bigg[~
\theta(y-z)\theta(z-x)\tfrac{1}{2} \log^2\bigg(\frac{1-x}{x}\frac{z}{1-z}\bigg)
+ (y \longleftrightarrow z)
~\bigg]
\nn&
\equiv \frac{1}{6\eps}\left(\frac{\mu}{\Lambda}\right)^{6\epsilon}  ~\mathcal{K}(x,y,z)
\,,
\end{align}
valid to the leading order in $\eps$. We observe that
eq.~(\ref{eq:K(xyz)-v3}) is symmetric under the interchange
of $y$ and $z$; hence the integrand of the full diagram in \refE{eq:F3-v2} is as well.
(In other words, interchanging the two parallel gluon lines
in figure~\ref{fig:three-loop_non-planar_ladder} leaves the diagram invariant.)
This observation implies that $\Cut_y F^{(3)} = \Cut_z F^{(3)}$
and thereby reduces the number of independent cuts to be computed.

\begin{figure}[!h]
\begin{center}
\includegraphics[angle=0, width=0.75\textwidth]{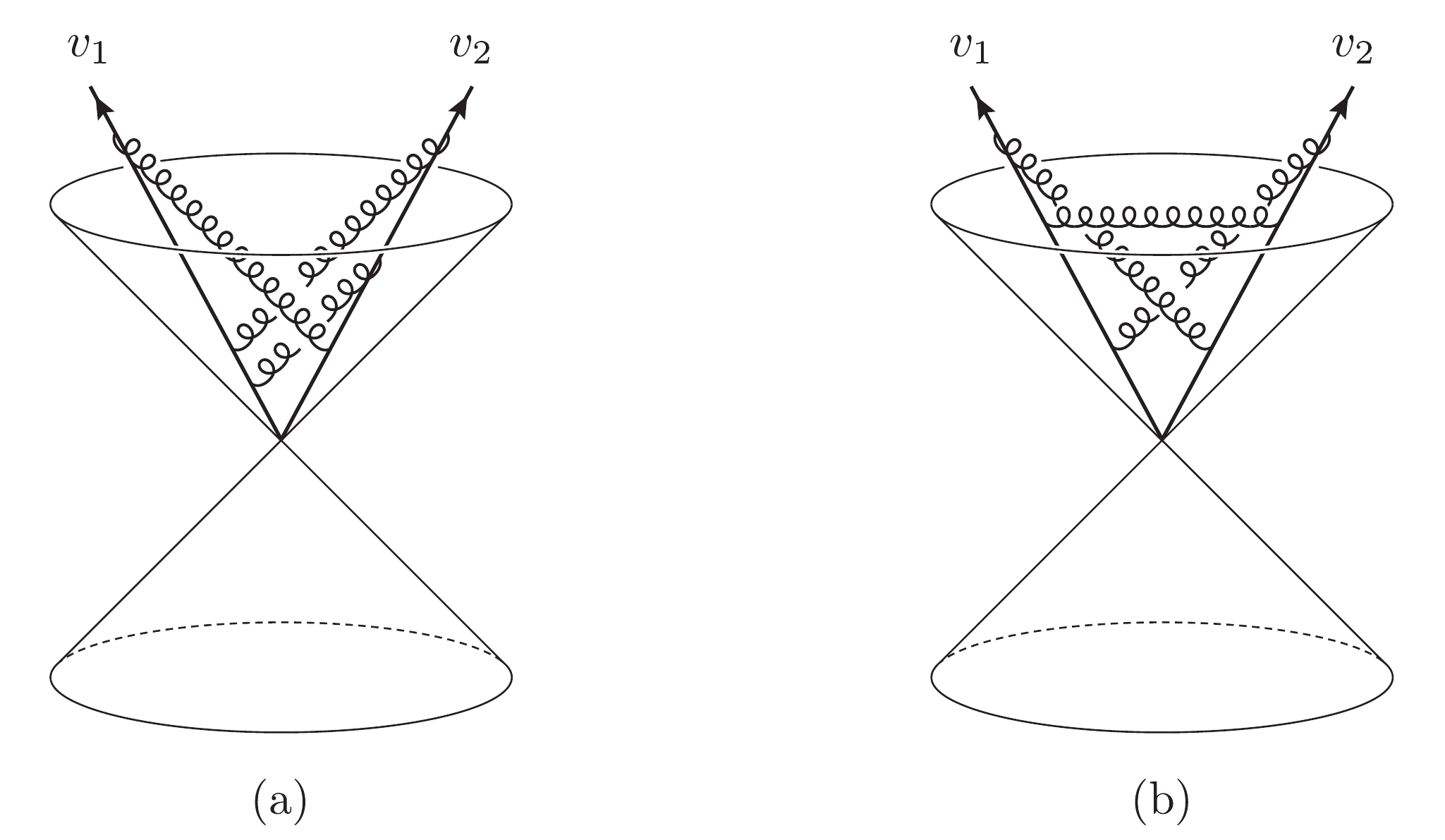}
{\vskip -4mm}
\caption{Triple cut of the two three-loop diagrams that have a single pole divergence. In (a) the non-vanishing triple cut of $F^{(3)}$ is illustrated by the exchange of three lightlike gluons. In contrast, (b) shows graphically that the triple cut of the maximally crossed diagram vanishes, because the three gluons cannot simultaneously be aligned with the light cone (i.e., go on shell).}
\label{fig:triple_cuts_three_loop_diagrams}
\end{center}
\end{figure}

Substituting eq.~(\ref{eq:K(xyz)-v3}) into eq.~(\ref{eq:F3-v2})
we can write the non-planar three-loop ladder diagram
in the convenient form
\begin{align}
\label{eq:F3-v3}
F^{(3)} &= \frac{C^{(3)}}{6\eps}\left(\frac{\mu}{\Lambda}\right)^{6\epsilon} \Fau^{(3)} \,,
\end{align}
where $\Fau^{(3)}$ is finite, given to leading order in $\eps$ as
\begin{align}
\label{eq:Fau3-v1}
\Fau^{(3)}&=
\int_0^1 \d x \hspace{0.7mm} \d y \hspace{0.7mm} \d z \hspace{0.7mm}
P_{12}^{[0]}(x)\, P_{12}^{[0]}(y)\, P_{12}^{[0]}(z)\, \mathcal{K}(x,y,z) \,.
\end{align}
As the prefactor of $\Fau^{(3)}$ in eq.~(\ref{eq:F3-v3}) is real,
it factors out on both sides of eq.~(\ref{eq:Master Im F}),
yielding the formula
\begin{align}
\label{eq:ImF3-v1}
\Im \Fau^{(3)} &=\Cut_{x} \Fau^{(3)} + \Cut_{y} \Fau^{(3)} + \Cut_{z} \Fau^{(3)} - \Cut_{x,y,z} \Fau^{(3)} \,.
\end{align}
The appearance of a multiple-cut contribution represents a new
feature for diagrams with more than two loops. Incidentally,
the diagram $F^{(3)}$ is the only three-loop ladder diagram
with an $\mathcal{O}(\frac{1}{\eps})$ divergence that has a non-vanishing triple cut. This point is illustrated in figure~\ref{fig:triple_cuts_three_loop_diagrams}.
The diagram $F^{(3)}$ therefore provides an excellent example to demonstrate a multiple-cut contribution.

Let us start by evaluating
this triple cut: as all three integrations are localized
by delta functions, the cut is immediately computed,
\begin{align}
\label{eq:Cut(xyz)F3-v1}
\Cut_{x,y,z} \Fau^{(3)} &=
\int_0^1 \d x \hspace{0.7mm} \d y \hspace{0.7mm} \d z \hspace{0.7mm}
\Delta_{12}(x)\,\Delta_{12}(y)\,\Delta_{12}(z)\,\mathcal{K}(x,y,z)
\nn&=
- \frac{\pi^3}{8} R(\chi)^3
\sum_{k,l,m=1}^{2} \mathcal{K}(\rho_k,\rho_l,\rho_m) \,.
\end{align}
In the second line we inserted the form of $\Delta_{12}$
given in \refE{eq:Pij(0)-partialfraction} and then integrated
out the delta functions. Now observe that
$\mathcal{K}(\rho_k,\rho_l,\rho_m)$, which is implicitly
defined in \refE{eq:K(xyz)-v3}, is non-zero
if and only if $(k,l,m)=(1,2,2)$. Indeed, in the first term
of \refE{eq:K(xyz)-v3}, the logarithm is non-zero only when
$k \neq m$, while the step functions dictate that
$\rho_k \leq \rho_m \leq \rho_l$. Since $\rho_1 < \rho_2$,
we must therefore have $k=1$ and $l=m=2$. An identical argument
applies to the second term of \refE{eq:K(xyz)-v3}.
We arrive at the simple result
\begin{align}
\Cut_{x,y,z} \Fau^{(3)}
\label{eq:Cut(xyz)F3-v2}
= - \frac{\pi^3}{2} R(\chi)^3 \, \log^2 \chi \,.
\end{align}
This completes the evaluation of the triple-cut contribution
to eq.~(\ref{eq:ImF3-v1}).

We now turn to the single cuts. We will focus on
the contribution $\Cut_x \Fau^{(3)}$,
the remaining cuts being computed completely analogously.
(Also recall from the discussion below eq.~(\ref{eq:K(xyz)-v3})
that $\Cut_y \Fau^{(3)} = \Cut_z \Fau^{(3)}$, so that
only one additional single cut needs to be computed.)
Using the partial-fractioned expressions for $\Delta_{12}$
and $P_{12}^{[0]}$ in eq.~(\ref{eq:Pij(0)-partialfraction})
and integrating out the delta function we find, dropping $\pm i\eta$
for notational convenience,
\begin{align}
\label{eq:Cut(x)F3-v1}
\Cut_{x} \Fau^{(3)} &=
\int_0^1 \d x
\PV \int_0^1 \d y
\PV \int_0^1 \d z
~\Delta_{12}(x) P_{12}^{[0]}(y) P_{12}^{[0]}(z) \,\mathcal{K}(x,y,z)
\nn&=
- \frac{\pi}{8} R(\chi)^3
\sum_{k=1}^{2}
\PV \int_{\rho_k}^1 \d y \,
\Big(\tfrac{1}{y-\rho_1} {-} \tfrac{1}{y-\rho_2}\Big)
\PV \int_{\rho_k}^y \d z  \,
\Big(\tfrac{1}{z-\rho_1} {-} \tfrac{1}{z-\rho_2}\Big)
\log^2\Big(\tfrac{1-\rho_k}{\rho_k}\tfrac{z}{1-z}\Big) \,,
\end{align}
exploiting the residual $y \longleftrightarrow z$ symmetry
of the integrand.

The principal-value integrals may be evaluated as
the corresponding full integrals minus the imaginary
part of the latter. Formulas of imaginary parts of
multiple polylogarithms are listed up to
weight four in \refA{App:ImG}.
The first PV integral gives rise to step functions involving
the variables $\rho_1, \rho_2$ and $y$. Splitting the $y$-integral
in the $k=1$ term into two integrals with the respective
domains $[\rho_1,\rho_2]$ and $[\rho_2,1]$ allows
all step functions to be resolved. Ultimately, the
cut is found to evaluate into the expression
\begin{align}
\label{eq:Cut(x)F3-v2}
\Cut_{x} \Fau^{(3)} &=
\frac{\pi}{2} R(\chi)^3 \Big[
{-}\tfrac{1}{3}\log^4 \chi
+\log^2 \chi \Big(H_2(\chi^2)-3\zeta_2\Big)
-2\log \chi \Big(H_3(\chi^2)-\zeta_3\Big)
\nn&\hspace{20mm}
+H_{2,2}(\chi^2)
+2H_4(\chi^2)
-\zeta_2 H_2(\chi^2)
-\tfrac{1}{4}\zeta_4
\Big] \,.
\end{align}
This result is expressed in terms of harmonic polylogarithms,
defined through \refE{eq:GtoHPL}. Similarly, the
$y$- and $z$-cuts are found to take the form
\begin{align}
\Cut_y \Fau^{(3)} &= \Cut_z \Fau^{(3)} =
\frac{\pi}{4} R(\chi)^3 \Big[
{-}\tfrac{1}{3}\log^4 \chi
+\log^2 \chi \Big(H_2(\chi^2)+3\zeta_2\Big)
\nn&\hspace{41mm}
+2H_{3,1}(\chi^2)
+H_{2,2}(\chi^2)
-\zeta_2 H_2(\chi^2)
+\tfrac{5}{4}\zeta_4
\Big] \,.
\end{align}
Combining all single and triple cuts according to \refE{eq:ImF3-v1} yields
the following imaginary part
\begin{align}
\label{eq:ImF3-v2}
\Im \Fau^{(3)} &= \pi R(\chi)^3 \Big[
{-}\tfrac{1}{3}\log^4 \chi
+\log^2 \chi \Big(H_2(\chi^2)+3\zeta_2\Big)
-\log \chi \Big(H_3(\chi^2)-\zeta_3\Big)
\nn&\hspace{19mm}
+H_{3,1}(\chi^2)
+H_{2,2}(\chi^2)
+H_4(\chi^2)
-\zeta_2 H_2(\chi^2)
+\tfrac{1}{2}\zeta_4
\Big]
\,.
\end{align}
\\
\\
As a crosscheck of this result, we can alternatively
compute the imaginary part of the three-loop ladder by
evaluating the diagram for space-like kinematics $v_1 \cdot v_2 < 0$,
in which case it will be purely real (cf. the discussion at the end
of section~\ref{sec:Im_of_L-loop_Wilson_lines}), and subsequently
perform the analytic continuation to time-like kinematics.

To the leading order in $\eps$, the three-loop ladder is given by
eq.~(\ref{eq:Fau3-v1}), although we must bear in mind that for
space-like kinematics the propagator roots $\rho_k$ are given
by the lower case of eq.~(\ref{eq:propagator_roots}). Inserting
into eq.~(\ref{eq:Fau3-v1}) the expressions for $P^{[0]}_{12}$
and $\mathcal{K}(x,y,z)$ given in eqs.~(\ref{eq:Pij(0)-partialfraction})
and (\ref{eq:K(xyz)-v3}), respectively, the diagram is directly
expressible in terms of multiple polylogarithms,
\begin{align}
\widetilde{\Fau}^{(3)}
&=
\frac{R(\chi)^3}{4} \sum_{i,j,k,l,m=0,1} (-1)^{i+j+k+l+m}
G(\rho_{i+1}, \rho_{j+1}, k, l, \rho_{m+1}; 1) \,,
\end{align}
where the tilde on the left-hand side indicates that the diagram is
computed for space-like kinematics.
We can use the algorithm in appendix~\ref{App:algorithm}
to recast this representation in terms of polylogarithms
with constant indices. In fact, the three-loop ladder
diagram can be expressed in terms of harmonic polylogarithms,
\begin{align}
\widetilde{\Fau}^{(3)}&=
\frac{R(\chi)^3}{4} \Big[
{-}\tfrac{4}{15}\log^5 \chi
+\tfrac{4}{3}\log^3 \chi \Big(H_2(\chi^2)-\zeta_2\Big)
-2\log^2 \chi \Big(H_3(\chi^2)-\zeta_3\Big)
\nn&\hspace{20mm}
+4\log \chi \Big(
H_{3,1}(\chi^2)
+H_{2,2}(\chi^2)
+H_4(\chi^2)
+\zeta_2H_2(\chi^2)
+\tfrac{3}{2}\zeta_4
\Big)
\nn&\hspace{20mm}
-6H_{4,1}(\chi^2)
-6H_{3,2}(\chi^2)
-4H_{2,3}(\chi^2)
-6H_5(\chi^2)
-2\zeta_2 H_3(\chi^2)
\nn&\hspace{20mm}
+4\zeta_3 H_2(\chi^2)
+3\zeta_5
+2\zeta_2 \zeta_3
\Big]
\,.
\label{eq:three-loop_ladder_space-like_result}
\end{align}
We have cross-checked this expression with previous
results in the literature, finding agreement.%
\footnote{More specifically, adding the diagram
in figure~\ref{fig:three-loop_non-planar_ladder}
to the maximally-crossed three-loop ladder, which we have computed by
the same methods, we find agreement with eq.~(A.1)
of ref.~\cite{Correa:2012nk} (in its published version;
or alternatively eq.~(73) of the corresponding arXiv e-print (v2)).
Furthermore, the two color structures of
the $(3,3)$ web are linear combinations of these two
diagrams, according to eq.~(4.26) in ref.~\cite{Falcioni:2014pka}.
Inserting our results for the two diagrams, we recover
the color structures in their eqs.~(4.29)~and~(4.33)
(where the basis functions are given explicitly in appendix~A),
thereby cross-checking our results for both diagrams individually,
and in particular our result in eq.~(4.37) for the
three-loop non-planar ladder.}

We can now find the result for the three-loop ladder
diagram in time-like kinematics by performing
the analytic continuation $\chi \rightarrow -1/\chi-i\eta$
on eq.~(\ref{eq:three-loop_ladder_space-like_result}).
Under the analytic continuation, the rational function
$R(\chi)$ picks up a minus sign, while polylogarithms
transform according to
\begin{align}
\log \chi &\rightarrow \log(-1/\chi-i\eta) = -\log \chi - \pi i \,,
\nn
H_{\vec{a}}(\chi^2) &\rightarrow H_{\vec{a}}(1/\chi^2+i\eta) \,.
\end{align}
Thus, all harmonic polylogarithms are evaluated slightly above
the branch cut $[1,\infty)$. They were subsequently expressed
in terms of $H_{\vec{a}}(\chi^2)$ and $\log \chi$ using
the {\tt Mathematica} package {\tt HPL} \cite{Maitre:2005uu,Maitre:2007kp}.
In this way, we find the following result for the
three-loop ladder with time-like kinematics,
\begin{align}
\Fau^{(3)}&=
-\frac{R(\chi)^3}{4} \bigg[
\tfrac{4}{15}\log^5 \chi
-\tfrac{4}{3}\log^3 \chi \Big(H_2(\chi^2)+11\zeta_2\Big)
+2\log^2 \chi \Big(H_3(\chi^2)-\zeta_3\Big)
\nn&\hspace{22mm}
-4\log \chi \Big(
H_{3,1}(\chi^2)
+H_{2,2}(\chi^2)
+H_4(\chi^2)
-5\zeta_2 H_2(\chi^2)
-\tfrac{27}{2}\zeta_4 \Big)
\nn&\hspace{22mm}
+6H_{4,1}(\chi^2)
+6H_{3,2}(\chi^2)
+4H_{2,3}(\chi^2)
+6H_5(\chi^2)
-10\zeta_2H_3(\chi^2)
-4\zeta_3H_2(\chi^2)
\nn&\hspace{22mm}
-3\zeta_5
+10\zeta_2\zeta_3
\nn&\hspace{22mm}
+4 \pi i \Big(
\tfrac{1}{3}\log^4 \chi
-\log^2 \chi \Big(H_2(\chi^2)+3\zeta_2\Big)
+\log \chi \Big(H_3(\chi^2)-\zeta_3\Big)
\nn&\hspace{34mm}
-H_{3,1}(\chi^2)
-H_{2,2}(\chi^2)
-H_4(\chi^2)
+\zeta_2H_2(\chi^2)
-\tfrac{1}{2}\zeta_4
\Big) \,
\bigg] \,.
\label{eq:three-loop_ladder_time-like_result}
\end{align}
We observe that the imaginary part of
eq.~(\ref{eq:three-loop_ladder_time-like_result})
agrees with the result found in \refE{eq:ImF3-v2},
as expected. We conclude that the cutting prescription
for the three-loop ladder stated in eq.~(\ref{eq:ImF3-v1})
produces the correct imaginary part. The cutting
prescription (\ref{eq:ImF3-v1}) is illustrated in
figure~\ref{fig:Im_of_three-loop_ladder}.

\subsection{Two-loop web with three Wilson lines}\label{sec:Web121}

The formalism of section~\ref{sec:Im_of_L-loop_Wilson_lines}
allows us to compute the imaginary part of eikonal diagrams
only to the leading order in $\eps$. This appears to
limit the applicability of the approach, but in practice a large
class of diagrams have only simple poles in $\eps$, and the
coefficient of the $\frac{1}{\eps}$~pole of the correlator
of Wilson lines defines physical observables of interest, such
as for example the cusp anomalous dimension (i.e.,
the anomalous dimension of the correlator of two Wilson lines,
cf. eq.~(\ref{eq:QED_cusp_anomalous_dimension})). As it turns out,
the cusp anomalous dimension can be expressed
entirely in terms of diagrams with simple poles in $\eps$
(once the diagrams are expressed in terms of the renormalized
coupling $g_R$). We observed this already for Abelian gauge theories
in eq.~(\ref{eq:Abelian_exponentiation}), but the statement
extends to the case of non-Abelian gauge theories as well.
This owes to the non-Abelian exponentiation theorem for two Wilson
lines~\cite{Gatheral:1983cz,Frenkel:1984pz} which states
that the two-line correlator can be written as the exponential
of the sum of \emph{webs}, defined as the subclass of diagrams
which are eikonal-line two-particle irreducible. The webs
appear in the exponent with their color prefactors
appropriately modified to account for the color factors
of the complete set of diagrams arising from expanding
the exponential.

Over the past four years, non-Abelian exponentiation
has been shown to generalize to correlators of an
\emph{arbitrary} number of Wilson lines
\cite{Gardi:2010rn,Mitov:2010rp,Gardi:2011wa,Gardi:2011yz,Dukes:2013wa,Gardi:2013ita,Dukes:2013gea}.
In this section we will study the interplay of this remarkable theorem
with the formalism of section~\ref{sec:Im_of_L-loop_Wilson_lines}.

\begin{figure}[!h]
\begin{center}
\includegraphics[angle=0, width=0.7\textwidth]{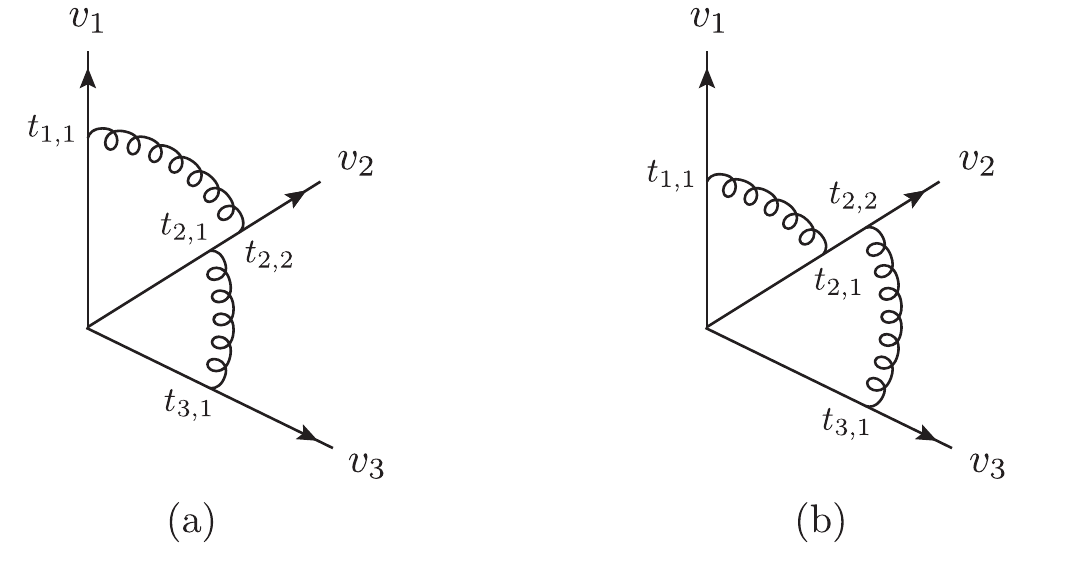}
{\vskip -3mm}
\caption{The two diagrams constituting the $(1,2,1)$ web. The
antisymmetric combination of the two diagrams is not captured
by exponentiation of the one-loop subdiagrams, and this linear
combination of the diagrams, defining the web, appears in the exponent.
Each separate diagram diverges as $\mathcal{O}(\frac{1}{\epsilon^2})$,
but the leading divergences conspire to cancel in the web, leaving
an $\mathcal{O}(\frac{1}{\epsilon})$ divergence.}
\label{fig:web_121}
\end{center}
\end{figure}

The statement of non-Abelian exponentiation in the multi-line
case requires a new classification of the set of diagrams which
appear in the exponent. In the multi-line case, a web
is defined as a \emph{collection} of diagrams which are mutually related by
permutations of the order of gluon attachments, acting on each
Wilson line separately. As an example, consider the $(1,2,1)$ web
in figure~\ref{fig:web_121}, whose imaginary part we will turn to shortly.

The labels $(1,2,1)$ refer to the number of gluon attachments on
each eikonal line, and the two diagrams are related by permutations
of the attachment points, where the permutations act on each line
separately. The individual diagrams in figure~\ref{fig:web_121}
are clearly not eikonal-line 2PI diagrams. However, the contributions
of these diagrams are not entirely reproduced by exponentiation
of their one-loop subdiagrams. Indeed, the sum of the one-loop
subdiagrams spanned by the pairs of lines $\{v_1 , v_2\}$ and $\{v_2 , v_3\}$
will appear in the exponent, but upon expanding the exponential
to second order, these subdiagrams only reproduce the symmetric
linear combination of the two-loop diagrams. To compensate, the
antisymmetric linear combination of the two-loop diagrams, which
defines the web, must be added to the exponent as a contribution.
Denoting the color and kinematical factors of the diagrams
in figure~\ref{fig:web_121} as $C_\mathrm{a,b}$ and
$\mathcal{F}_\mathrm{a,b}$, the web is accordingly defined as
\begin{equation}
\label{eq:Web-v1}
W^{(2)}_{(1,2,1)} \hspace{0.6mm}
= \hspace{0.6mm} \big(C_{\mathrm{a}}, \hspace{0.4mm} C_{\mathrm{b}}\big)
\left( \hspace{-1.2mm} \begin{array}{cc} \frac{1}{2} & -\frac{1}{2} \\[1.5mm]
-\frac{1}{2} & \frac{1}{2} \end{array}\hspace{-1.2mm}\right)
\colvec{{\cal F}_{\mathrm{a}} \\[1.5mm] {\cal F}_{\mathrm{b}}} \hspace{0.6mm}
= \hspace{0.6mm} \frac{1}{2} (C_{\mathrm{a}}-C_{\mathrm{b}})
({\cal F}_{\mathrm{a}} - {\cal F}_{\mathrm{b}}) \,.
\end{equation}
We can now state the general non-Abelian exponentiation theorem
\cite{Gardi:2010rn,Mitov:2010rp,Gardi:2011wa,Gardi:2011yz,Dukes:2013wa,Gardi:2013ita,Dukes:2013gea}.
Recalling the definition in eq.~(\ref{eq:def_of_Wilson_line})
of the Wilson line $\Phi_v$ spanned by the four-velocity $v^\mu$,
the statement is that the correlator
of an arbitrary number of Wilson lines is given as the exponential
of the sum of webs,
\begin{equation}
\big\langle \Phi_{v_1} \cdots \Phi_{v_n} \big\rangle
\hspace{0.7mm}=\hspace{0.7mm} \exp \bigg( \sum_{i \in \{ \mathrm{webs}\}}
C_i^T R_i \mathcal{F}_i \bigg) \,,
\end{equation}
where each web $i$ contributes to the exponent through
the color $C_i$ and kinematical $\mathcal{F}_i$
factors of its constituent diagrams, weighted by means
of the web mixing matrix $R_i$ in analogy with
eq.~(\ref{eq:Web-v1}). The web mixing matrices can be
computed systematically by means of the replica trick
of statistical mechanics \cite{Laenen:2008gt,Gardi:2010rn}. Among
several properties they satisfy the zero-sum-row condition
$\sum_b R_{ab} = 0$ which ensures that the symmetric
linear combination of the constituent diagrams is
projected out \cite{Gardi:2011wa}.

The mixing matrices satisfy an additional weighted
zero-sum column condition which ensures that the leading
divergence of the constituent diagrams of a web conspire to cancel
when the diagrams are added. This is a general feature of webs,
ultimately following from their renormalization properties
\cite{Gardi:2010rn,Mitov:2010rp,Gardi:2011yz}, leaving in many
cases a web with an $\mathcal{O}(\frac{1}{\eps})$ divergence.
As~a~result, webs are particularly amenable to the formalism
of section~\ref{sec:Im_of_L-loop_Wilson_lines}. It is here
important to keep in mind that the cutting prescription
should be applied to an entire web rather than its
constituent diagrams separately, as the separate imaginary
parts, computed to leading order in $\eps$, will cancel.

To illustrate the procedure in detail, we turn
to the web in figure~\ref{fig:web_121} and compute its imaginary part.
In analogy with sections~\ref{sec:2loopNPladder}~and~\ref{sec:3loopNPladder},
we take $\cosh \gamma_{12} \equiv v_1 \cdot v_2 > 0$ and
$\cosh \gamma_{23} \equiv v_2 \cdot v_3 > 0$, in order to have
a non-vanishing contribution to the imaginary part from both
kinematical channels, and set $\chi \equiv e^{-\gamma_{12}}$
and $\psi \equiv e^{-\gamma_{23}}$. Our first task will be to
show that the leading $\mathcal{O}(\frac{1}{\eps^2})$
divergences of the individual diagrams conspire to cancel,
leaving an $\mathcal{O}(\frac{1}{\eps})$ divergence. We will
then apply the cutting prescription of
eqs.~(\ref{eq:Dij})--(\ref{eq:Master Im F}) directly
to the web written in a form with a manifest
$\mathcal{O}(\frac{1}{\eps})$ divergence.

Let us start by considering diagram 6a in \refF{fig:web_121}.
Its kinematical factor is given by
\begin{align}
\label{eq:F6a-v1}
\Fau_{\mathrm{a}} &= C^{(2)}\,\mu^{4\epsilon}
\int_0^{\infty} \frac{\d t_{1,1} \,\d t_{2,1} \,\d t_{2,2} \,\d t_{3,1} ~\theta(t_{2,2}-t_{2,1})\, (v_1 \cdot v_2)\, (v_2 \cdot v_3)}
{\big[ {-} (t_{1,1} v_1 - t_{2,2} v_2)^2 +i\eta \big]^{1-\epsilon}
\big[ {-} (t_{2,1} v_2 - t_{3,1} v_3)^2 +i\eta \big]^{1-\epsilon}} \,,
\end{align}
where $C^{(2)}$ contains coupling constants etc., but no color factor.
In analogy with sections~\ref{sec:2loopNPladder}~and~\ref{sec:3loopNPladder}
our first task is to write this expression in the form of
eq.~(\ref{eq:L-loop_ladder_diagram_rep_2}). This is achieved
through the changes of variables in eq.~(\ref{eq:radial_coordinates})
(with $t_{\ell_j, m_j} = t_{2,3-j}$),
setting $(x_1, x_2) = (x,y)$ for convenience.

After these transformations, the kinematical factor takes the form
\begin{align}
\label{eq:F6a-v2}
\Fau_{\mathrm{a}} &= C^{(2)}
\int_0^1 \d x \hspace{0.7mm} \d y \hspace{0.7mm}
P_{12}^{[\eps]}(x) \,P_{23}^{[\eps]}(y) \, K(x,y)\,,
\end{align}
where the kernel is given by
\begin{align}
K(x,y)&=
\mu^{4\epsilon}
\int_0^{\infty} \frac{\d\rho_1 \hspace{0.6mm} \d\rho_2}{(\rho_1 \rho_2)^{1-2\eps}}
\,\theta(\rho_1 x - \rho_2 y)
\nn&=
\frac{\Gamma(4\eps)}{2\eps}\left(\frac{\mu}{\Lambda}\right)^{4\epsilon}
\big[ u^{2\eps} {}_2F_1(2\eps,4\eps;1+2\eps;-u) \big]^{u=\infty}_{u=y/x} \,.
\label{eq:kernel_of_121-web}
\end{align}
The result for $K(x,y)$ was obtained by applying the
substitution in eq.~(\ref{eq:tau_y_variables})
and performing the remaining integrations in complete
analogy with section~\ref{sec:2loopNPladder}.
The second diagram of the web, $\Fau_{\mathrm{b}}$, differs only in
the step function which reads $\theta(\rho_2 y - \rho_1 x)$,
changing the lower integration bound in
eq.~(\ref{eq:kernel_of_121-web}) from $y/x$ to $x/y$. After
expanding the gamma function and the hypergeometric function in $\eps$, we thus
find the kinematical factor of the $(1,2,1)$ web to take the form
\begin{align}
\label{eq:Web-v2}
\Fau_{\mathrm{a}}-\Fau_{\mathrm{b}} &=
\frac{C^{(2)}}{8\eps^2} \left(\frac{\mu}{\Lambda}\right)^{4\epsilon}
\int_0^1 \d x \hspace{0.7mm} \d y \hspace{0.7mm}
P_{12}^{[\eps]}(x)\, P_{23}^{[\eps]}(y)
\left[ \bigg(\frac{x}{y}\bigg)^{2\eps}\!\! - \bigg(\frac{y}{x}\bigg)^{2\eps} \right]
\big( 1+\Ord(\eps^2)\big) \,.
\end{align}
Upon expanding $\big[ \cdots \big]$ in $\eps$,
we observe that the leading poles of the separate diagrams cancel, leaving an
$\Ord(\tfrac{1}{\eps})$ divergence, in agreement with the
discussion above.

Factoring out the remaining pole, we can write
the web in the convenient form
\begin{align}
\label{eq:Web-v3}
W^{(2)}_{(1,2,1)} &=
\frac{C_{\mathrm{a}}-C_{\mathrm{b}}}{2}\,
\frac{C^{(2)}}{2\eps}\left(\frac{\mu}{\Lambda}\right)^{4\epsilon}
\Fau^{(2)}_{(1,2,1)} \,,
\end{align}
where $\Fau^{(2)}_{(1,2,1)}$ is finite and given to the leading order in $\eps$ by
\begin{align}
\label{eq:Wau}
\Fau^{(2)}_{(1,2,1)} &=
\int_0^1 \d x \hspace{0.7mm} \d y \hspace{0.7mm}
P_{12}^{[0]}(x)\, P_{23}^{[0]}(y) \,
\log \frac{x}{y} \,.
\end{align}
As the prefactor of $\Fau^{(2)}_{(1,2,1)}$ in
eq.~(\ref{eq:Web-v3}) is real, it factors out on
both sides of eq.~(\ref{eq:Master Im F}), yielding the formula
\begin{align}
\label{eq:ImWeb-v1}
\Im \Fau^{(2)}_{(1,2,1)} &=\Cut_{x} \Fau^{(2)}_{(1,2,1)} + \Cut_{y} \Fau^{(2)}_{(1,2,1)} \,.
\end{align}
More explicitly, by inserting the definition
of the operator $\Cut_{x_i}$ in \refE{eq:Cut_x F},
we have
\begin{align}
\label{eq:ImWeb-v2}
\Im \Fau^{(2)}_{(1,2,1)} &=
\int_0^1 \hspace{-0.9mm} \d x \hspace{0.3mm}
\PV \hspace{-0.6mm} \int_0^1 \hspace{-0.9mm} \d y
\hspace{0.7mm} \Delta_{12}(x) \hspace{0.3mm} P_{23}^{[0]}(y) \hspace{0.2mm}
\log \frac{x}{y} \hspace{0.3mm}
+ \hspace{0.3mm} \PV \hspace{-0.6mm} \int_0^1 \hspace{-0.9mm} \d x \hspace{-0.2mm}
\int_0^1 \hspace{-0.9mm} \d y \hspace{0.7mm}
P_{12}^{[0]}(x) \hspace{0.3mm} \Delta_{23}(y) \hspace{0.2mm}
\log \frac{x}{y} \,.
\end{align}
We observe that the second term equals minus
the first term with the two cusp angles and
integration variables interchanged, making
the imaginary part of the web antisymmetric under the interchange
$\gamma_{12} \longleftrightarrow \gamma_{23}$.
(The antisymmetry is of course inherited from
the full web which has this property by construction.)
This observation allows us to write the imaginary
part in the manifestly antisymmetric form
\begin{align}
\label{eq:ImWeb-v3}
\Im \Fau^{(2)}_{(1,2,1)} &=J(\chi,\psi) - J(\psi,\chi) \,,
\end{align}
where the auxiliary function is defined as the first
term of eq.~(\ref{eq:ImWeb-v2}),
\begin{align}
J(\chi,\psi) =
\int_0^1 \d x  \PV \int_0^1 \d y ~\Delta_{12}(x)\, P_{23}^{[0]}(y)\, \log \frac{x}{y} \,.
\label{eq:J_integral_def}
\end{align}
Evaluation of the imaginary part of the web
thus reduces to the evaluation of the integral~$J(\chi,\psi)$.
The latter can be computed by recalling the
partial-fractioned expressions for $\Delta_{12}$
and $P_{23}^{[0]}$ given in
\refE{eq:Pij(0)-partialfraction} which in the
present notation read
\begin{align}
\Delta_{12}(x) &= -\pi \frac{R(\chi)}{2}\Big(\delta\big(x-\rho_1(\chi)\big)+\delta\big(x-\rho_2(\chi)\big)\Big) \,,\nn
P_{23}^{[0]}(y) &= \frac{R(\psi)}{2}\left(\frac{1}{y-\rho_1(\psi)+i\eta}-\frac{1}{y-\rho_2(\psi)-i\eta}\right) \,,
\label{eq:Pij(0)-partialfraction_two_channels}
\end{align}
where we wrote out the expressions explicitly
to emphasize their dependence on the two distinct
kinematical invariants $\chi$ and $\psi$.

Integrating out the delta functions in
eq.~(\ref{eq:J_integral_def}) leaves one
principal-value integral to be evaluated. This
integral is computed as the corresponding full
integral minus its imaginary part, as explained in \refS{sec:2loopNPladder}.%
\footnote{To proceed we assume, without loss of generality,
that $\psi < \chi$. This fixes
$\rho_1(\chi) < \rho_1(\psi) < \rho_2(\psi) < \rho_2(\chi)$,
which allows the step functions in eqs.~(\ref{eq:Im_G_weight_1})--(\ref{eq:Im_G_weight_4}) to be resolved.
}
In this way we find
\begin{align}
J(\chi,\psi) &=
-\frac{\pi}{4}R(\chi)R(\psi)
\sum_{k=1}^2
\PV \int_0^1 \d y \left(\frac{1}{y-\rho_1(\psi)+i\eta}-\frac{1}{y-\rho_2(\psi)-i\eta}\right) \log\left(\frac{\rho_k(\chi)}{y}\right)
\nn&=
-\frac{\pi}{4}R(\chi)R(\psi)
\Big(
{-}4\text{Li}_2(-\psi)
+\log^2 \psi
-4\log \psi \big(\log(\psi+1)-\log(\chi+1) \big)
\nn&\hspace{30mm}
-2\log \chi \log \psi
-2\zeta_2
\Big) \,.
\label{eq:J_integral_result}
\end{align}
Upon the antisymmetrization in eq.~(\ref{eq:ImWeb-v3})
the terms on the last line of eq.~(\ref{eq:J_integral_result})
cancel, and we find the following result
for the imaginary part of the $(1,2,1)$ web,
\begin{align}
\label{eq:ImWeb-v4}
\Im \Fau^{(2)}_{(1,2,1)} &=
-\pi \,
R(\chi)R(\psi)
\Big(
\text{Li}_2(-\chi)-\text{Li}_2(-\psi)
-\tfrac{1}{4}\big(\log^2 \chi -\log^2 \psi \big)
\nn&\hspace{30mm}
+\big(\log \chi +\log \psi \big)\big(\log(\chi+1)-\log(\psi+1)\big)
\Big) \,.
\end{align}

\noindent As a crosscheck of this result, we can alternatively
compute the imaginary part of the web by
evaluating the diagram for space-like kinematics $v_1 \cdot v_2 < 0$
and $v_2 \cdot v_3 < 0$, in which case it will be purely real
(cf. the discussion at the end of section~\ref{sec:Im_of_L-loop_Wilson_lines}), and subsequently
perform the analytic continuation to time-like kinematics.

To the leading order in $\eps$, the web is given by
eq.~(\ref{eq:Wau}), although we must bear in mind that for
space-like kinematics the propagator roots $\rho_k$ are given
by the lower case of eq.~(\ref{eq:propagator_roots}). We now insert
into eq.~(\ref{eq:Wau}) the expressions for $P^{[0]}_{12}$
given in eq.~(\ref{eq:Pij(0)-partialfraction})
and perform the integrals in analogy with the calculations
in eqs.~(\ref{eq:F2-tch-v2})--(\ref{eq:two-loop_ladder_space-like_result})
for the case of the non-planar two-loop ladder. This leads
to the following result for the web in space-like kinematics,
\begin{align}
\label{eq:Web-tch-v1}
\widetilde{\Fau}^{(2)}_{(1,2,1)} &=
R(\chi)R(\psi)\, \big(L(\psi) \log \chi - L(\chi) \log \psi\big)
\,,
\end{align}
where we introduced the auxiliary function
\begin{align}
L(\chi)&=-\text{Li}_2(1-\chi)-\tfrac{1}{4}\log^2 \chi \,.
\end{align}
These expressions are consistent with results previously
obtained in the literature, see for example eq.~(3.11)
in ref.~\cite{Gardi:2013saa}, as well as references therein.
We can now obtain the result for the web in time-like kinematics
by performing the analytic continuations
$\chi \rightarrow -1/\chi-i\eta$ and $\psi \rightarrow -1/\psi-i\eta$
on eq.~(\ref{eq:Web-tch-v1}).
Under the analytic continuation, the functions
appearing in \refE{eq:Web-tch-v1} transform as
\begin{align}
\label{eq:AnalyticCont-Log-L}
R(z) &\rightarrow -R(z) \nn
\log z &\rightarrow
-\log z - \pi i \nn
L(z) &\rightarrow
-\text{Li}_2(-z)-\log z \log(z+1)+\tfrac{1}{4}\log^2 z -\tfrac{1}{2}\zeta_2
\nn&\hspace{5mm}
-\pi i \big( \log(z+1) - \tfrac{1}{2} \log z \big) \,.
\end{align}
Upon analytic continuation in $\chi$ and $\psi$ we thus
find the following result for the web with time-like kinematics,
\begin{align}
\label{eq:Web-tch-v2}
\Fau^{(2)}_{(1,2,1)} &=
R(\chi) R(\psi) \, \Big[
- i \pi \Big(
\text{Li}_2(-\chi)
-\tfrac{1}{4}\log^2 \chi
+\big(\log \chi +\log \psi \big)\log(\chi+1)
\Big)
\nn&\hspace{24mm}
+\log \chi \text{Li}_2(-\psi)
-\big(\log \chi \log \psi -6\zeta_2\big)\log(\chi+1)
\nn&\hspace{24mm}
+\tfrac{1}{4}\big(\log \chi \log \psi -10\zeta_2\big)\log \chi
\Big]
~ - \big(\chi \longleftrightarrow \psi\big)
\,.
\end{align}
We observe that the imaginary part of eq.~(\ref{eq:Web-tch-v2})
agrees with the result found in eq.~(\ref{eq:ImWeb-v4}),
as expected. We conclude that the cutting prescription
for the two-loop web stated in eq.~(\ref{eq:ImWeb-v1})
produces the correct imaginary part. The graphical representation
of the cutting prescription (\ref{eq:ImWeb-v1}) is similar
to that in figure~\ref{fig:Im_of_two-loop_ladder}, and we omit it here.
\\
\\
In the above we have computed the imaginary part of the
web with time-like kinematics. As the web depends on
two distinct angles, we may also consider the diagram
in the case of mixed time- and space-like kinematics,
for example $\cosh\gamma_{12} \equiv v_1 \cdot v_2 > 0$ and
$\cosh\gamma_{23} \equiv -v_2 \cdot v_3 > 0$.%
\footnote{The opposite-type kinematics $v_1 \cdot v_2 < 0$
and $v_2 \cdot v_3 > 0$ is of course equivalent by
the antisymmetry of the web under the interchange
$\gamma_{12} \longleftrightarrow \gamma_{23}$
of the cusp angles.}
A natural question is then whether also in this case
the imaginary part is computed correctly by the formalism
of section~\ref{sec:Im_of_L-loop_Wilson_lines}. As we
shall see shortly, the formalism readily applies, with
the one difference that the imaginary part has no
contribution from the $\psi$-channel, as propagators
stretched between mutually space-like eikonal lines
have vanishing cuts. Put differently, the discontinuity
in the $\psi$-channel does not contribute to
the imaginary part, cf. eq.~(\ref{eq:Im_and_discontinuities_relation}).

Returning to the formula for the imaginary part
in the explicit form (\ref{eq:ImWeb-v2}), we observe that
in the above case of mixed time- and space-like kinematics,
the roots of the propagator $P_{23}^{[0]}(y)$ lie outside
the range of integration, cf. the remarks below
eq.~(\ref{eq:propagator_roots}). As a result,
the second term in eq.~(\ref{eq:ImWeb-v2}) vanishes
and in the first term the principal-value prescription may be dropped,
\begin{align}
\label{eq:ImWeb-v5}
\Im \Fau^{(2)}_{(1,2,1)} &=
\int_0^1 \d x \hspace{0.7mm} \d y \hspace{0.7mm}
\Delta_{12}(x)\, P_{23}^{[0]}(y) \,
\log \frac{x}{y} \,.
\end{align}
After inserting eq.~(\ref{eq:Pij(0)-partialfraction_two_channels})
and integrating out the delta functions, the imaginary part is
readily expressed in terms of multiple polylogarithms;
these in turn can be simplified into classical polylogarithms,
yielding
\begin{align}
\label{eq:ImWeb-v6}
\Im \Fau^{(2)}_{(1,2,1)} &=
\frac{\pi}{4} R(\chi) R(\psi) \Big(
{-}2G(\rho_2(\psi),0;1)+2G(\rho_1(\psi),0;1)
\nn&\hspace{27mm}
+G(0;\rho_2(\chi))G(\rho_2(\psi);1)
+G(0;\rho_1(\chi))G(\rho_2(\psi);1)
\nn&\hspace{27mm}
-G(0;\rho_2(\chi))G(\rho_1(\psi);1)
-G(0;\rho_1(\chi))G(\rho_1(\psi);1)
\Big)
\nn&=
\pi R(\chi) R(\psi) \Big(
\text{Li}_2(\psi)-\log \psi \Big(
\tfrac{1}{4}\log \psi -\log(1-\psi)+\log(\chi+1)-\tfrac{1}{2}\log \chi
\Big)
-\zeta_2
\Big)
\,.
\end{align}
Using eq.~(\ref{eq:AnalyticCont-Log-L}) it is
straightforward to verify that this result agrees
with the imaginary part acquired by eq.~(\ref{eq:Web-tch-v1})
upon the analytic continuation $\chi \rightarrow -1/\chi-i\eta$.
We conclude that the formula (\ref{eq:Master Im F})
reproduces the correct imaginary part of the web also
for mixed time- and space-like kinematics, as expected.

\section{Position-space cuts of eikonal diagrams
with internal vertices}\label{sec:Im_of_diagrams_with_internal_vertices}

\begin{figure}[!h]
\begin{center}
\includegraphics[angle=0, width=0.35\textwidth]{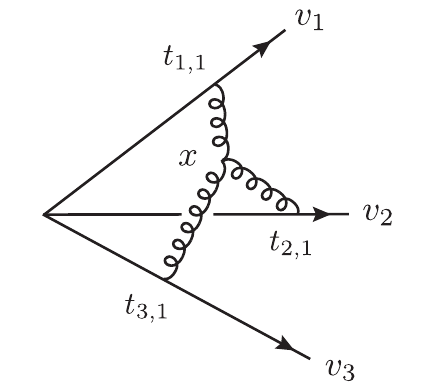}
{\vskip -2.5mm}
\caption{The non-planar two-loop three-gluon vertex diagram.}
\label{fig:non-planar_3g-vertex}
\end{center}
\end{figure}

In this section we turn to the application of the formalism
of ref.~\cite{Laenen:2014jga} to diagrams with internal (i.e., three-
and four-gluon) vertices. Here we provide details on
the calculation of the imaginary part of the diagram
involving a three-gluon vertex connected to three Wilson lines,
as illustrated in figure~\ref{fig:non-planar_3g-vertex}.

\begin{figure}[!h]
\begin{center}
\includegraphics[angle=0, width=0.66\textwidth]{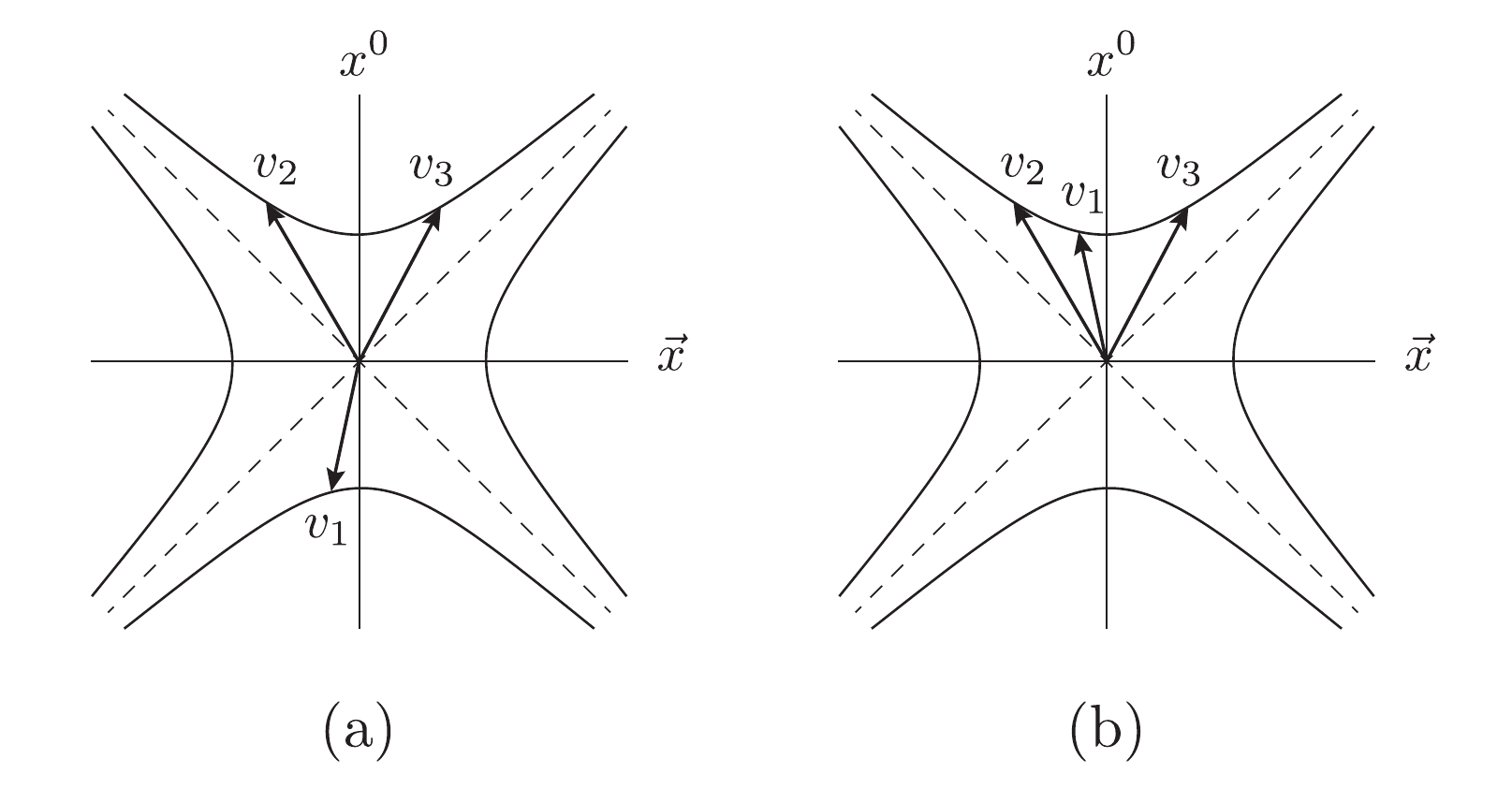}
{\vskip -2.5mm}
\caption{Physical configurations for three distinct Wilson-line velocities: (a) one incoming and two outgoing lines, or (b) three outgoing lines. Configurations related by time-reversal are omitted. In this section we compute the imaginary part of the three-gluon vertex diagram in the physical configuration (b).}
\label{fig:Kinematics_Three_Gluon_Vertex_Diagram}
\end{center}
\end{figure}

The integrated result for the diagram in figure~\ref{fig:non-planar_3g-vertex} was first obtained in refs.~\cite{Ferroglia:2009ii,Ferroglia:2009ep} using a Mellin-Barnes representation of the two loop-momentum integrals. In terms of the cusp angles $\gamma_{ij}$, defined through $\cosh \gamma_{ij} = - v_i \cdot v_j$, it is given by
\begin{align}
\widetilde{F}_{3g} &=
- i f^{abc}{\bf T}_1^a {\bf T}_2^b {\bf T}_3^c \,
\frac{2}{\eps} \left(\frac{\alpha_s}{4\pi}\right)^2 \sum_{i,j,k=1}^{3} \varepsilon_{ijk} ~\gamma_{ij}^2 \gamma_{ki} \coth \gamma_{ki} ~.
\label{eq:F3g-spacelike}
\end{align}
This expression is valid for an unphysical configuration with space-like kinematics for all pairs of Wilson lines, i.e. $v_i \cdot  v_j < 0$, as indicated by the tilde on $\widetilde{F}_{3g}$. In agreement with our observations in section~\ref{sec:causality_and_unitarity_of_WL}, $\widetilde{F}_{3g}$ has no imaginary part. In contrast, in a physical configuration of massive Wilson lines, each velocity is constrained to the unit three-hyperboloid, either inside the future light cone or inside the past light cone. There are two inequivalent physical configurations, shown in figure~\ref{fig:Kinematics_Three_Gluon_Vertex_Diagram}.
In both configurations at least one pair of Wilson lines is time-like separated, i.e. $v_i \cdot v_j >0$, leading to a non-vanishing imaginary part.

In the following we consider the configuration in figure~\ref{fig:Kinematics_Three_Gluon_Vertex_Diagram}(b), where all Wilson lines correspond to outgoing states, such that $v_i \cdot v_j > 0$ for each pair of Wilson lines.
The analytic result for such time-like kinematics, denoted by $F_{3g}$, is obtained from the space-like expression $\widetilde{F}_{3g}$ in eq.~(\ref{eq:F3g-spacelike}) by analytic continuation $\gamma_{ij} \rightarrow i \pi - \gamma_{ij}$ for all $i \neq j$ (cf. eq.~(\ref{eq:analytic_continuation_gamma})).
The imaginary part of the resulting expression is
\begin{align}
\Im F_{3g} &=
- f^{abc}{\bf T}_1^a {\bf T}_2^b {\bf T}_3^c \,
\left(\frac{\alpha_s}{4\pi}\right)^2 \, \frac{2}{\eps} \,
\sum_{i,j,k=1}^{3} \varepsilon_{ijk} ~
 \big(
\gamma_{ij}^2 \gamma_{ki}
-2\pi^2 \gamma_{ij}
\big) \coth\gamma_{ki} ~.
\label{eq:Im-F3g-timelike}
\end{align}

In the remainder of this section our task is to compute this imaginary part with our formalism. As in section~\ref{sec:Im_of_L-loop_Wilson_lines} we first need to extract the leading divergence of $F_{3g}$ in its position-space representation. The three-gluon vertex diagram has a only a simple pole in $\eps$. This divergence is extracted from the radial integral over the three-gluon vertex position. Having extracted the leading divergence, the diagram can be written as $\tfrac{1}{\eps} \times (\text{finite})$. We then apply position-space cuts to the finite function. The one-dimensional integrals along the Wilson lines are then trivially performed using the delta functions arising from the cut. The remaining integrations over the direction of the three-gluon vertex are performed numerically, after which the final result is compared to the analytic expression in eq.~(\ref{eq:Im-F3g-timelike}).
\newline

We start by writing down the position-space representation of $F_{3g}$. It reads\footnote{See ref.~\cite{Mitov:2009sv} for the corresponding position-space representation of this diagram in Euclidean space.}
\begin{align}
F_{3g} &=
- f^{abc}{\bf T}_1^a {\bf T}_2^b {\bf T}_3^c \,
\Big(\frac{\alpha_s}{4\pi}\Big)^2 \frac{4}{\pi^2}\, \mu^{4\eps}\,
\int \frac{\d^D x}{r^{4-6\eps}}
\sum_{i,j,k=1}^{3} \varepsilon_{ijk} \,
v_i \cdot v_j \, \zeta_i \, \zeta_k
\bigg( \! \frac{\partial}{\partial \zeta_i} \, g(\zeta_i,\eps) \! \bigg)
g(\zeta_j,\eps) \, g(\zeta_k,\eps) ~.
\label{eq:F3g_from_Feynmanrules}
\end{align}
Here the three-gluon vertex position $x$ is integrated over all of Minkowski space. In the integrand $x$ is decomposed into a radial distance $r$ and direction $u$, via $x^\mu = r \, u^\mu$, such that $u^2=1$ for time-like $x$ and $u^2=-1$ for space-like $x$. Dot products between $u$ and the Wilson line velocities are denoted by $\zeta_i = v_i \cdot u$. The one-dimensional integrals along the Wilson lines are contained in the functions $g(\zeta_i,\eps)$, which are defined as
\begin{align}
g(\zeta_i,\eps)&=\int_0^\infty \frac{\d x_i}{[-(u^2-2\,x_i\,\zeta_i+x_i^2)+i\eta]^{1-\eps}}~.
\end{align}

After a change of variables to hyperspherical coordinates in eq.~(\ref{eq:F3g_from_Feynmanrules}), the radial integral contains the overall divergence\footnote{This divergence is regulated by including an exponential damping factor in the radial integral
(cf. eq.~(\ref{eq:exponential_damping_factor})).} and may easily be performed, yielding a factor of $\tfrac{1}{4\eps}$. Restricting attention to the leading order in $\eps$ allows us to set $\eps=0$ in the finite function $\Fau_{3g}$, yielding
\begin{align}
F_{3g} &=
- f^{abc}{\bf T}_1^a {\bf T}_2^b {\bf T}_3^c \,
\Big(\frac{\alpha_s}{4\pi}\Big)^2 \frac{1}{\pi^2\,\eps} \, \Fau_{3g} ~,
\nn
\Fau_{3g} &= \int_{\RP} \d^3 u
\sum_{i,j,k=1}^{3} \varepsilon_{ijk} \,
v_i \cdot v_j \, \zeta_i \, \zeta_k
\bigg( \! \frac{\partial}{\partial \zeta_i} \, g(\zeta_i,0) \! \bigg)
g(\zeta_j,0) \, g(\zeta_k,0) ~.
\label{eq:F3g-in-cut-form}
\end{align}
The integration domain for the three-gluon vertex direction $u$ is $\RP \equiv \dH{+} \cup \dH{-} \cup \dS{+} \cup \dS{-}$, the union of the upper and lower sheets of the unit three-hyperboloid and three-dimensional de~Sitter space, defined by
\begin{align}
\begin{aligned}
\dH{\pm}   \hspace{0.8mm}&=\hspace{0.8mm}  \{  u \in \mathbb{R}^{1,3} : \hspace{0.7mm} u^2 = 1
\hspace{4.5mm} \mathrm{and} \hspace{1.5mm} u_0 \gtrless 0 \} ~,\\
\dS{\pm}  \hspace{0.8mm}&=\hspace{0.8mm}  \{  u \in \mathbb{R}^{1,3} : \hspace{0.7mm} u^2 = -1
\hspace{1.5mm} \mathrm{and} \hspace{1.5mm} u_0 \gtrless 0 \}~.
\end{aligned}
\end{align}
Having written the three-gluon vertex diagram in the form in eq.~(\ref{eq:F3g-in-cut-form}), we are ready to apply our formalism to obtain the imaginary part of $\Fau_{3g}$ from its cuts. This in turn gives the imaginary part of the full diagram $F_{3g}$, as they are proportional up to a real constant.
\newline

\begin{figure}[!t]
\begin{center}
\includegraphics[angle=0, width=\textwidth]{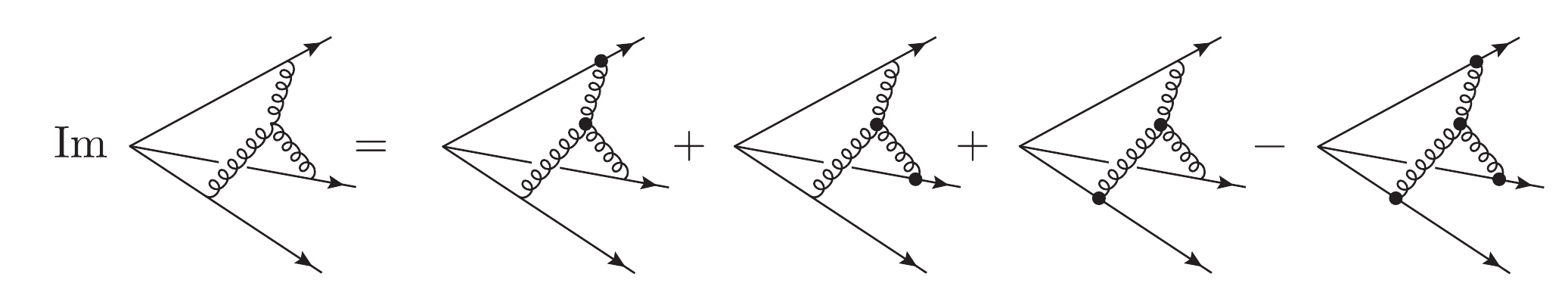}
{\vskip -2.5mm}
\caption{Graphical representation of position-space cuts applied to a diagram with an internal vertex.}
\label{fig:Three_Gluon_Vertex_Im_Part}
\end{center}
\end{figure}
\begin{figure}[!h]
\begin{center}
\includegraphics[angle=0, width=\textwidth]{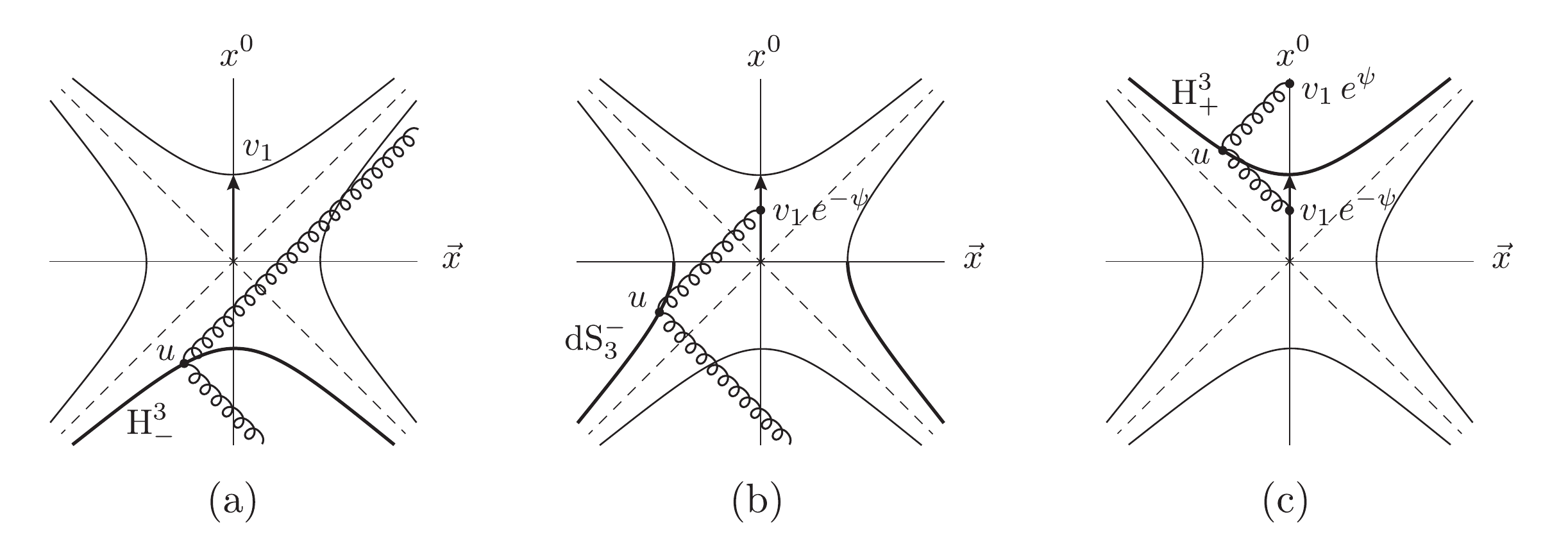}
{\vskip -2.5mm}
\caption{Spacetime pictures of a Wilson line along the positive $x^0$-axis (with normalized velocity $v_1$) and a three-gluon vertex $u$, located in the region (a) $\dH{-}$, (b) $\dS{-}$ or (c) $\dH{+}$. The region $\dS{+}$ is very similar to $\dS{-}$ and is omitted. Each figure shows two lightlike gluons emanating from $u$: one along the future light cone and the other along the past light cone. Of these two on-shell gluons respectively zero, one or two gluons are able to connect to the Wilson line. In other words, in case (a), (b) and (c) there are respectively zero, one and two solutions to the delta functions coming from the cut operators. This means that in the region $\dH{-}$ the operator $\Cut_{x_1}$ vanishes, thereby producing no imaginary part. Furthermore, as discussed in the main text, the imaginary parts from the space-like regions $\dS{+}$ and $\dS{-}$ cancel each other. The only contribution to the imaginary part of the diagram thus arises from the region $\dH{+}$.}
\label{fig:Cuts_Three_Gluon_Vertex_Diagram}
\end{center}
\end{figure}

The imaginary part of $\Fau_{3g}$ is computed from the formula in eq.~(\ref{eq:Master Im F}), as illustrated schematically in figure~\ref{fig:Three_Gluon_Vertex_Im_Part}. The cut propagators stretching between the three-gluon vertex and the Wilson lines take the obvious form $\Delta_{i}(x_i) = -\pi\,\delta(u^2 - 2\,x_i\,\zeta_i + x_i^2)$, rather than eq.~(\ref{eq:Dij}) for propagators connecting two Wilson lines. In order to resolve the support of these delta functions in the different subregions of $\RP$ it is convenient to introduce variables $y_i$ that are equal to $\zeta_i$, possibly up to a sign depending on the location of $u$. Explicitly, we let $y_i = \zeta_i$ for $u \in \dH{+} \cup \dS{+}$ and $y_i = -\zeta_i$ for $u \in \dH{-} \cup \dS{-}$. In this way $\zeta_i$ flips sign between the $\pm$ regions, but $y_i$ does not. In terms of these variables a cut operator acting on a function $g(\zeta_i,0)$ yields
\begin{align}
\Cut_{x_i} g(\zeta_i,0)
&= - \pi \int_0^\infty \d x_i ~ \delta(u^2-2\,x_i\,\zeta_i+x_i^2) =
\begin{cases}
- \tfrac{\pi}{2}\,(1\pm1)\, \big( y_i^2-1 \big)^{-1/2}
&~\text{in}~\dH{\pm} \\
- \tfrac{\pi}{2} \, \big( y_i^2+1 \big)^{-1/2}
&~\text{in}~\dS{\pm}~,
\end{cases}
\label{eq:Cut-g(zeta)}
\end{align}
while the principal-value part, due to $\Cut_{x_j} g(\zeta_i,0)$ with $i \neq j$, evaluates to
\begin{align}
\PV \int_0^\infty \frac{\d x_i}{-(u^2-2\,x_i\,\zeta_i+x_i^2)+i\eta} =
\begin{cases}
\pm \arcosh(y_i)\, \big( y_i^2-1 \big)^{-1/2}
&~\text{in}~\dH{\pm} \\
\mp \arsinh(y_i)\, \big( y_i^2+1 \big)^{-1/2}
&~\text{in}~\dS{\pm}~.
\end{cases}
\label{eq:PV-g(zeta)}
\end{align}
The right-hand side of eq.~(\ref{eq:Cut-g(zeta)}) shows that the cut vanishes for $u$ in the region $\dH{-}$. This is a consequence of the delta function having no support inside the domain of integration $[0,\infty)$. In the regions $\dS{\pm}$ and $\dH{+}$ there are respectively one and two solutions to the delta-function constraint, as can also be understood by inspection of figure~\ref{fig:Cuts_Three_Gluon_Vertex_Diagram}.
Focusing on the contributions to the imaginary part from $\dS{+}$ and $\dS{-}$, we see that both the single and the triple cuts acting on the product $g(\zeta_i,0)\,g(\zeta_j,0)\,g(\zeta_k,0)$ in $\Fau_{3g}$ yield the same results in both regions. But apart from this product of $g$'s, the sum in eq.~(\ref{eq:F3g-in-cut-form}) also contains $\zeta_i\,\zeta_k\,\frac{\partial}{\partial \zeta_i}$, which differs by a sign between the two regions. As a result, the imaginary part arising from the regions $\dS{+}$ and $\dS{-}$ cancel each other. The upshot is thus that the imaginary part of $\Fau_{3g}$ arises solely from the region $\dH{+}$.
\newline

The final step in the computation of $\Im \Fau_{3g}$ is now to perform the integration over $u$, the direction of the three-gluon vertex. We do not have analytic results for the integrals involved, but a numerical evaluation is sufficient to show agreement with the analytic formula in eq.~(\ref{eq:Im-F3g-timelike}). Let us give a few details regarding the setup of the numerical integration.

The three-gluon vertex direction $u$ may be parametrized explicitly in terms of Minkowski angles $\psi, \vartheta$ and $\phi$. As discussed above, the imaginary part arises solely from the region $\dH{+}$, which may be parametrized as
\begin{equation}
\dH{+}: \hspace{8mm} \left\{ \begin{array}{rll}
u^0 \hspace{0mm}&=&\hspace{0mm} \cosh \psi \\[0.3mm]
u^1 \hspace{0mm}&=&\hspace{0mm} \sinh \psi \sin \vartheta \cos \phi \\[0.3mm]
u^2 \hspace{0mm}&=&\hspace{0mm} \sinh \psi \sin \vartheta \sin \phi \\[0.3mm]
u^3 \hspace{0mm}&=&\hspace{0mm} \sinh \psi \cos \vartheta \,, \end{array}
\right. \hspace{10mm}
\begin{array}{rcccl}
0 \hspace{0mm}&\leq&\hspace{0mm} \psi      \hspace{0mm}&<&\hspace{0mm} \infty \\
0 \hspace{0mm}&\leq&\hspace{0mm} \vartheta \hspace{0mm}&\leq&\hspace{0mm} \pi \\
0 \hspace{0mm}&\leq&\hspace{0mm} \phi      \hspace{0mm}&\leq&\hspace{0mm} 2\pi \,.
\end{array}
\end{equation}
To facilitate the numerical integration over $\psi \in [0,\infty)$ we perform a further change of variables $z = \tanh\psi$, which has the effect of producing a finite integration domain $z \in [0,1]$. Explicit expressions for $\zeta_i = y_i = u \cdot v_i$ in terms of $z$, the angles $\vartheta,\phi$ and the cusp angles $\gamma_{ij}$ are obtained by choosing a convenient Lorentz frame. For example,
\begin{align}
\begin{aligned}
v_1^\mu &= (1 ,\, 0 ,\, 0 ,\, 0) ~,\\
v_2^\mu &= (\cosh\gamma_{12} ,\, 0 ,\, 0 ,\, \sinh\gamma_{12}) ~,\\
v_3^\mu &= (\cosh\gamma_{13} ,\, 0 ,\, \sin\theta_3\sinh\gamma_{13} ,\, \cos\theta_3\sinh\gamma_{13}) ~.
\end{aligned}
\label{eq:Lorentz_frame_for_Wilson_lines}
\end{align}
These velocities manifestly satisfy $v_i^2 = 1$ and $v_1 \cdot v_{k} = \cosh \gamma_{1j}$ for $j=2,3$. The remaining identity, $v_2 \cdot v_3 = \cosh \gamma_{23}$, fixes $\theta_3$ in terms of the cusp angles,
\begin{align}
\cos\theta_3 = \frac{\cosh\gamma_{12}\cosh\gamma_{13}-\cosh\gamma_{23}}{\sinh\gamma_{12}\sinh\gamma_{13}}~.
\end{align}
The explicit parametrization of the Wilson-line velocities in eq.~(\ref{eq:Lorentz_frame_for_Wilson_lines})  breaks the antisymmetry of $\Fau_{3g}$ under interchange of any pair of cusp angles at the integrand level. However, the antisymmetry must be recovered after integration (cf. eq.~(\ref{eq:F3g-spacelike})). At the level of numerical integration this indeed happens for small cusp angles, while for large cusp angles numerical instabilities arise from the integration near $z \approx 1$, i.e. very large $\psi$. Averaging over the cusp angles $\gamma_{ij}$ remedies those instabilities.

\begin{figure}[t!]
\begin{center}
\includegraphics[angle=0, width=0.49\textwidth]{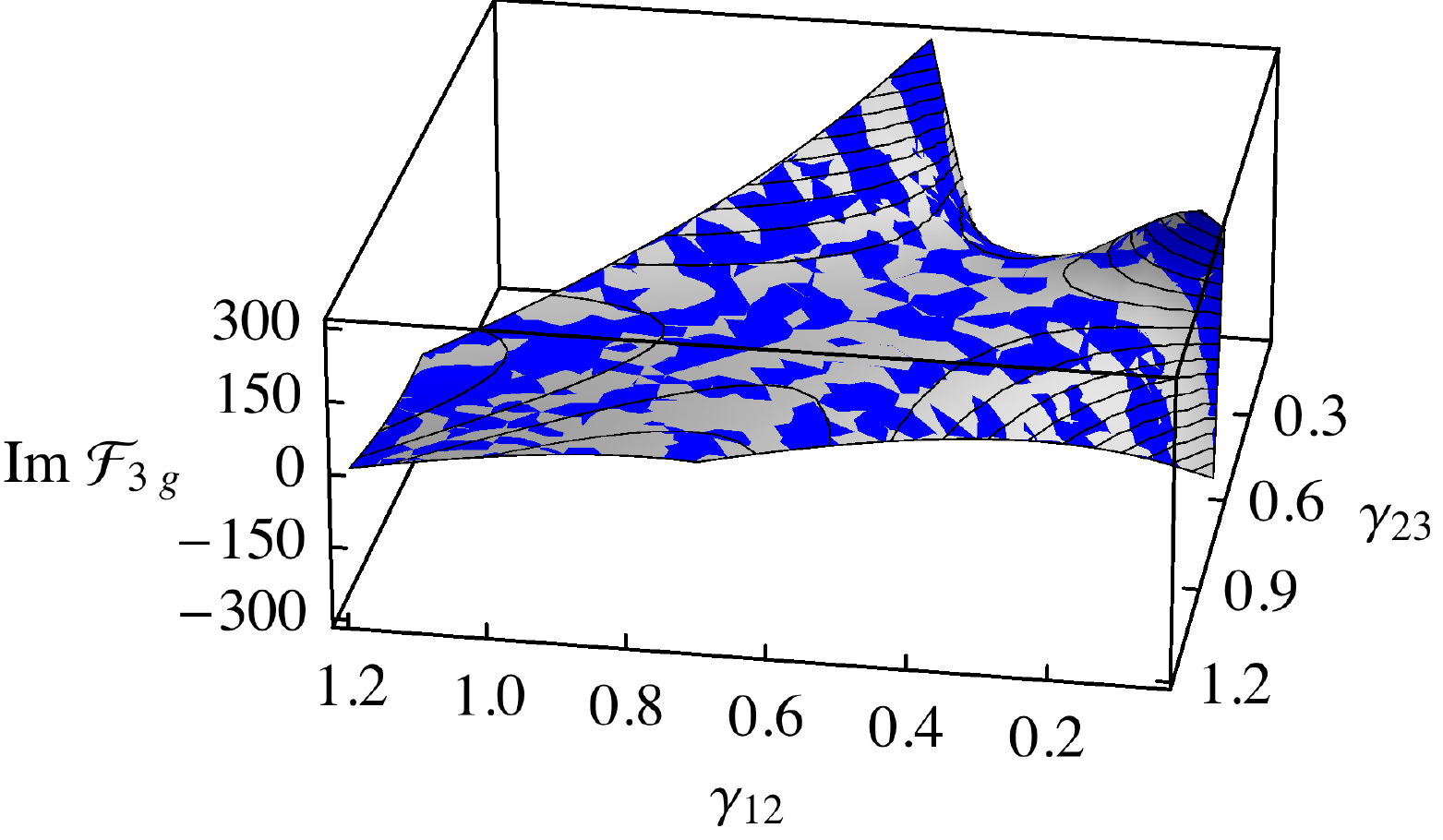}
\includegraphics[angle=0, width=0.49\textwidth]{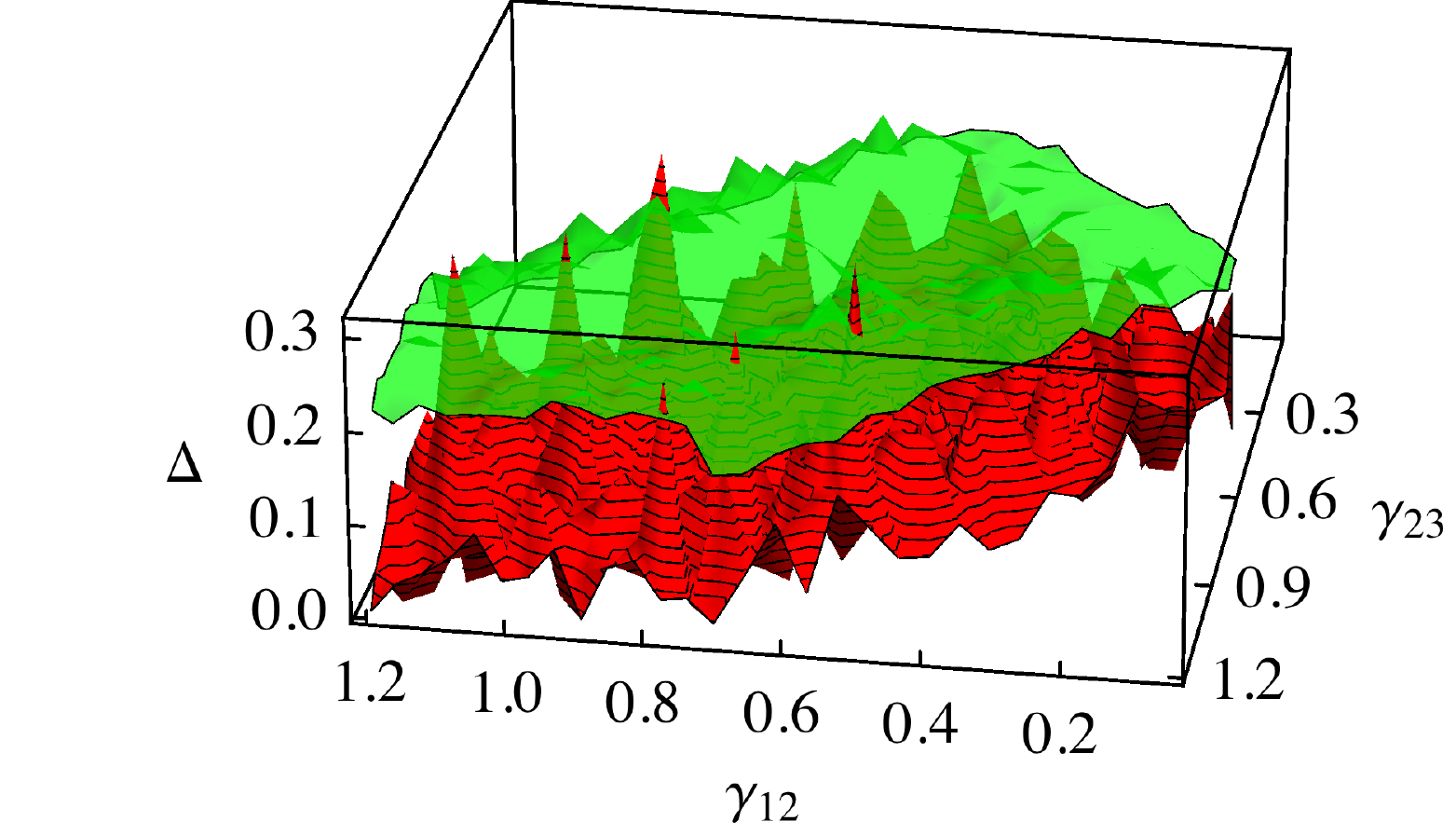}
{\vskip -2.5mm}
\caption{(Color online.) A comparison between the numerical and analytical results for the imaginary part of the three-gluon vertex diagram. The left pane shows $\Im \Fau_{3g}$ as a function of two cusp angles from numerical integration (in blue) and from the analytical result (in light gray) superimposed. The numerical result fits the analytic function rather well, with an overall scale factor deviating from one by about $2 \cdot 10^{-5}$. In the right pane we observe that $\Delta$, the absolute difference between the numerical and the analytical values, (in red) is below the numerical errors (in translucent green) for nearly all points ($\gamma_{12},\gamma_{23})$ and is on average about three times smaller. The relative difference with respect to the analytic formula, $\Delta / |\Im \Fau_{3g}|$, is on average of the order of $2$ percent, leading to the conclusion that there is excellent agreement between the numerical and analytical results. In both plots $\gamma_{13}=0.5$, while $\gamma_{12}$ and $\gamma_{23}$ vary between $0.0$ and $1.2$.
}
\label{fig:three_gluon_vertex_diagram_numerical_results}
\end{center}
\end{figure}

Having constructed expressions for $\zeta_i$ in terms of $z$, the angles $\vartheta,\phi$ and the cusp angles $\gamma_{ij}$, the imaginary part of the three-gluon vertex diagram is explicitly given by
\begin{align}
\Im \Fau_{3g} &=
\int_0^1 \frac{\d z \, z^2}{(1-z^2)^2} \int_0^\pi \d \vartheta \sin\vartheta \int_0^{2\pi} \d \phi
\sum_{i,j,k=1}^{3} \varepsilon_{ijk} \,
v_i \cdot v_j \, \zeta_i \, \zeta_k
\bigg(\! \frac{\partial}{\partial \zeta_i} \, {\cal G}\big(\{\zeta_i\}\big) \!\bigg) ~,
\label{eq:numerical_ImF3g}
\end{align}
where the cut operators are absorbed into the function ${\cal G}\big(\{\zeta_i\}\big)$, given by
\begin{align}
{\cal G}\big(\{\zeta_i\}\big)&\equiv
\big( \Cut_{x_i} + \Cut_{x_j} + \Cut_{x_k} - \Cut_{x_i,x_j,x_k} \big)\,
g(\zeta_i,0) g(\zeta_j,0) \, g(\zeta_k,0)
\nn
&=
- \frac{
\pi \arcosh\zeta_j \arcosh\zeta_k
+\pi \arcosh\zeta_i \arcosh\zeta_k
+\pi \arcosh\zeta_i \arcosh\zeta_j
-\pi^3
}{(\zeta_i^2-1)^{1/2}\,(\zeta_j^2-1)^{1/2}\,(\zeta_k^2-1)^{1/2}} ~.
\end{align}
For the numerical integration of eq.~(\ref{eq:numerical_ImF3g}) we have used GSL \cite{Galassi:2009}. A comparison between the numerical and analytical results for $\Im \Fau_{3g}$ is shown in figure~\ref{fig:three_gluon_vertex_diagram_numerical_results}. We find that the relative difference between the numerical and analytical results are at the percent level, with absolute differences smaller than numerical errors. We conclude that our formula for the imaginary part in eq.~(\ref{eq:numerical_ImF3g}) is in excellent agreement with the analytic expression in eq.~(\ref{eq:Im-F3g-timelike}). This suggests the applicability of the formalism introduced in ref.~\cite{Laenen:2014jga} to obtain the imaginary part of any eikonal diagram with internal vertices.

\section{Conclusions}\label{sec:Conclusions}

In this paper we have provided algorithms for the compution of the
position-space cuts of eikonal diagrams introduced in ref.~\cite{Laenen:2014jga}
and discussed the interplay of the cutting prescription with non-Abelian exponentiation.
The cutting prescription is applied directly to the position-space
representation of an eikonal diagram and computes its imaginary part to the
leading order in the dimensional regulator $\eps$. The prescription is stated in
eqs.~(\ref{eq:Dij})--(\ref{eq:Master Im F}). The relation of the imaginary part to the
branch cut discontinuity is given in
eq.~(\ref{eq:Im_and_discontinuities_relation}).

Momentum-space cuts of eikonal diagrams, analogous to the Cutkosky
rules for standard Feynman diagrams, were introduced in
ref.~\cite{Korchemsky:1987wg} where they were used to show
that the exchanges of Glauber-region gluons (i.e., maximally
transverse gluons) produce imaginary parts of the Wilson-line
correlator. Any given momentum-space cut separates the eikonal diagram
into two disjoint subdiagrams, putting the eikonal and, depending
on the cut, possibly also a number of standard Feynman propagators on shell
(see figure~\ref{fig:Cut_non-planar_three-loop_ladder} for an illustration).
As a result, momentum-space cuts have the conceptual advantage
of factoring eikonal diagrams into on-shell lower-loop and tree diagrams
which can be computed as independent objects. In practice, however,
the resulting cut diagrams involve integrations over two-, three-,
four-, $\ldots$ particle phase space. The evaluation of these
phase-space integrals poses a substantial computational challenge,
limiting the applicability of momentum-space cuts for computing
imaginary parts.

In contrast, position-space cuts do not factor the eikonal diagram
into disjoint subdiagrams, but rather constrain the gauge bosons
exchanged between the energetic partons to be lightlike. For
space-like external kinematics such exchanges are causally
impossible, and the imaginary part vanishes. For time-like
kinematics such exchanges are allowed and generate a nontrivial
evolution of the phases of the parton states, leading in turn
to a close relation of the imaginary part of the cusp anomalous
dimension to the static interquark potential. Position space thus
offers a causality viewpoint on the origin of the imaginary part
of the eikonal diagram. This is complementary to the unitarity
viewpoint provided by momentum space---i.e., that the imaginary
part arises from the hard partons going on shell
and exchanging Glauber-region gluons.  At the computational level,
the number of position-space cut diagrams contributing to the imaginary part
of a given eikonal diagram is in practice smaller than the number
of momentum-space cut contributions, and several of the position-space
cut diagrams can be seen to be equal a priori.

We have applied our formalism to several two- and three-loop
eikonal diagrams, finding agreement with results previously
obtained in the literature
\cite{Korchemsky:1987wg,Correa:2012nk,Falcioni:2014pka,Gardi:2013saa,Ferroglia:2009ii}.
These computations also serve to demonstrate that the
position-space cut diagrams contributing to the imaginary
part of a given eikonal diagram can be evaluated
in practice in nontrivial cases. In particular, for eikonal diagrams
without internal vertices---i.e., QED-like diagrams---the contributing
cut diagrams can be evaluated systematically by means of our
algorithm for computing the principal-value integrals involved
(supplemented with a slight generalization of the algorithm of that in
ref.~\cite{Anastasiou:2013srw} for expressing multiple polylogarithms
in terms of ones with constant indices).

The formalism developed in this paper allows us
to compute the imaginary part of eikonal diagrams only to the
leading order in $\eps$. This appears to limit the applicability
of the approach, but in practice Wilson line correlators can often be
expressed in terms of diagrams with simple poles in $\eps$
(once the diagrams are expressed in terms of the renormalized
coupling). This owes to the non-Abelian exponentiation theorem
\cite{Gatheral:1983cz,Frenkel:1984pz,Gardi:2010rn,Mitov:2010rp,Gardi:2011wa,Gardi:2011yz,Dukes:2013wa,Gardi:2013ita,Dukes:2013gea}
which states that the correlator can be expressed as the exponential
of specific linear combinations of diagrams mutually related
by permutations of the soft-gluon attachment points. These linear
combinations, called \emph{webs}, have the property that the leading
divergence of the constituent diagrams cancels, leaving in many cases
webs with simple poles in $\eps$. The organization of the exponent
of the Wilson line correlator in terms of webs is particularly
beneficial for the applicability of the present cutting prescription:
the cuts must be applied to an entire web rather than its
constituent diagrams separately, as the separate imaginary
parts, computed to leading order in $\eps$, will cancel. In this
sense the cutting prescription has a nontrivial interplay with
non-Abelian exponentiation.

It would be intriguing to investigate whether the position-space
cuts studied in this paper can be utilized, or serve as inspiration,
for developing (generalized) unitarity methods
\cite{Bern:1994zx,Bern:1994cg,Bern:1996je,Britto:2004nc,Britto:2005ha,Forde:2007mi,Mastrolia:2009dr,Kosower:2011ty,CaronHuot:2012ab}
for correlators of Wilson lines. Another
interesting direction for future research is the extension of
the present formalism to computations of imaginary parts of
Wilson line correlators to subleading orders in $\eps$.

\section*{Acknowledgments}

We thank Samuel Abreu, Simon Caron-Huot, Einan Gardi, Johannes Henn, Paul Hoyer,
Lorenzo Magnea, George Sterman, Iain Stewart, Ward Vleeshouwers, Andries Waelkens,
Chris White and especially Gregory Korchemsky for useful discussions.
We are grateful for the hospitality of the Higgs Centre of the University of Edinburgh.
KJL is grateful for the hospitality of
the Institute for Advanced Study in Princeton and
the Institut de Physique Th{\'e}orique, CEA Saclay,
where part of this work was carried out.
The research leading to these results has received
funding from the European Union Seventh Framework
Programme (FP7/2007-2013) under grant agreement no.~627521.
This work was supported by
the Foundation for Fundamental Research of Matter (FOM),
program 104 ``Theoretical Particle Physics in the Era of the LHC''
and by the Research Executive Agency (REA)
of the European Union under the Grant Agreement number
PITN-GA-2010-264564 (LHCPhenoNet) and PITN-GA-2012-316704
(HIGGSTOOLS).

\appendix

\section{Real and imaginary parts of multiple polylogarithms}
\label{App:ImG}

The position-space cut prescription in eq.~(\ref{eq:Master Im F})
produces principal-value integrals when applied to eikonal diagrams
beyond one loop. In practice, we compute such integrals as the corresponding
full integral (which evaluates into multiple
polylogarithms) minus its imaginary part, cf. eq.~(\ref{eq:PV-formula}).
In this appendix we describe how to construct the required real and
imaginary part of multiple polylogarithms to arbitrary weight
in a systematic way. Explicit formulas for imaginary parts
are given up to weight four, while the real parts are obtained
by subtracting the imaginary part from the original function.

Let us first introduce some notation.
Multiple polylogarithms are defined recursively by
\begin{align}
\label{eq:G definition recursion}
G(a_{1},\dotsc,a_{n}; x) = \int_{0}^{x} \frac{\d t}{t-a_{1}} G(a_{2},\dotsc,a_{n};t)
\hspace{8mm} \mathrm{for} \hspace{5mm} (a_1,\ldots,a_n) \neq \vec{0}_n \,,
\end{align}
starting from the special cases
\begin{align}
\label{eq:G definition}
&G( ; 0) \equiv 0 \,, \hspace{6mm}
G( ; x) \equiv 1 \,, \hspace{6mm} 	
G(\vec{0}_n; x) \equiv \frac{1}{n!} \log^{n} x \,,
\end{align}
where $\vec{a}_n=(a,\dots,a)$ denotes a vector with $n$ equal indices.
Multiple polylogarithms satisfy a variety of properties.
They form a shuffle algebra,
\begin{align}
\label{eq:G shuffle algebra}
G(a_1,\ldots,a_{n_1};x) \, G(a_{n_1+1},\ldots,a_{n_1+n_2};x)
&\,=\sum_{\sigma\in\Sigma(n_1, n_2)}\,G(a_{\sigma(1)},\ldots,a_{\sigma(n_1+n_2)};x) \,.
\end{align}
They are invariant under a common rescaling of
all arguments: setting $\vec a=(a_1, \ldots, a_n)$
we have
\begin{align}
\label{eq:G rescale relation}
G(k\,\vec a;k\, x) = G(\vec a;x) \qquad \mathrm{for} \hspace{4mm}
a_i\neq 0 \hspace{3mm} \mathrm{and} \hspace{3mm} k \in \mathbb{C}^* \,.
\end{align}
They reduce to classical polylogarithms in certain cases,
\begin{align}
\label{eq:GtoClassic}
G(\vec 0_n;x) = {1\over n!} \log^n x \,,
\qquad &
G(\vec a_n;x) = {1\over n!} \log^n\left(1-{x\over a}\right) \,,\nn
G(\vec 0_{n-1},a;x) = -\textrm{Li}_n\left({x\over a}\right) \,,
\qquad &
G(\vec 0_{n}, \vec{a}_{p};x) = (-1)^p\,S_{n,p}\left({x\over a}\right) \,,
\end{align}
or to harmonic polylogarithms, introduced in ref.~\cite{Remiddi:1999ew},
\begin{align}
\label{eq:GtoHPL}
G(\vec{a}; x) = (-1)^k \, H_{\vec{a}}( x) \hspace{8mm}
\text{if}\hspace{4mm} \forall a_i \in \vec{a} {:} \hspace{3mm} a_i \in \{\pm 1,0 \} \,,
\end{align}
where $k$ denotes the number of $+1$'s in $\vec{a}$.
We refer to ref.~\cite{Duhr:2011zq} for more details.

Having set the notation, we turn to the problem of constructing
the real and imaginary parts of multiple polylogarithms.
As it turns out, it is most convenient to obtain the real part
as the difference of the full function and its imaginary part,
\begin{align}
\label{eq:Re is Full minus Im}
\Re G(\vec{a};x) = G(\vec{a};x) - i \Im G(\vec{a};x) \,.
\end{align}
Thus, it suffices to determine the imaginary part.
In the remainder of this section we accordingly focus on
the construction of imaginary parts.

The imaginary part of a multiple polylogarithm arises when
one or more of its indices $a_i$ are located along the path
of integration of the corresponding iterated integral, that
is $a_i \in [0,x]$. By giving either the argument or the indices
an infinitesimal imaginary part, the imaginary part of any
multiple polylogarithm is fixed recursively in terms of the
imaginary part of the classical logarithm.
This is most easily seen in the special case of a polylogarithm
with all indices equal to zero by using its definition in
\refE{eq:G definition} in terms of logarithms,
\begin{align}
\label{eq:Im G(zeros)}
\Im G(\vec{0}_n; x \pm i \eta)
&= \frac{1}{n!} \Im \log^n (x \pm i\eta)
= \frac{1}{n!} \Im \Big[ \big(\log |x| \pm i \pi \theta(-x) \big)^n \Big]
\,.
\end{align}
The imaginary part on the right-hand side may be obtained by
simply expanding the product and collecting the terms proportional
to $i$. The other two special cases in \refE{eq:G definition} are
real constants and thus have a vanishing imaginary part.
This concludes the computation of the imaginary part of all the special cases
listed in \refE{eq:G definition}.

We thus turn to determining the imaginary part of a
multiple polylogarithm in the general case $G(\vec{a}; x)$
with at least one non-zero index, cf.~\refE{eq:G definition recursion}.
Since an imaginary part arises when some of the indices
$a_i$ are located along the path of integration, it will be necessary
to know the relative locations of the $a_i$'s in the complex plane. To this end,
we will make a few assumptions on the indices and arguments of
$G(\vec{a}; x)$. First, let us observe that all polylogarithms encountered
in this paper will have real indices $a_i$ (up to an infinitesimal
imaginary part whose sign is fixed by the Feynman rules). In addition,
we will assume, without loss of generality, that the last index is non-zero, and that
the endpoint of integration $x$ is real and positive.

The fact that the latter two assumptions may be imposed without
loss of generality follows from the properties of multiple polylogarithms.
Indeed, multiple polylogarithms with any number of trailing
zeros may be expressed, with the help of the shuffle algebra
in \refE{eq:G shuffle algebra}, in terms of multiple
polylogarithms with a non-zero last index, multiplied by pure
logarithms. For example,
\begin{align}
\label{eq:G shuffles}
G(a,0;x)&=G(a;x)G(0;x)-G(0,a;x) \,,\nn
G(a,0,0;x)&=G(a;x)G(0,0;x)-G(0,a,0;x)-G(0,0,a;x)\nn
&=G(a;x)G(0,0;x)-G(0,a;x)G(0;x)+G(0,0,a;x) \,.
\end{align}
Their imaginary part is thus given in terms of the (real and) imaginary parts of
multiple polylogarithms with a non-zero last index and pure logarithms.
The latter are known from \refE{eq:Im G(zeros)}. Thus, it suffices to
determine the imaginary part of polylogarithms with a non-zero last index.

Now, taking the last index to be non-zero, we may apply
the rescaling relation \refE{eq:G rescale relation} with $k=-1$,
\begin{align}
\label{eq:G rescale by -1}
G(\vec{a};x)=G(-\vec{a};-x)\hspace{8mm} \text{for} \hspace{4mm} a_n \neq 0 \,,
\end{align}
to map any negative argument $x$ to a positive argument
\cite{Vollinga:2004sn}. Likewise, a complex argument can be mapped to
a real number after rescaling by $k=1/x$, yielding $G(\vec{a}/x;1)$.

Our task is thus to determine the imaginary part of a
multiple polylogarithm in the general case $G(\vec{a}; x)$
with a non-zero last index, and with the endpoint of integration
$x$ being real and positive. Starting at weight one,
we let $a,x \in \mathbb{R}$, $a \neq 0$ and $x>0$.
Explicit computation of the imaginary part---using the fact
that the imaginary part of an integral equals the integral
over the imaginary part---yields
\begin{align}
\label{eq:Im G(a,x)}
\Im G(a \pm i \eta; x) &= \Im \int_{0}^{x}\frac{\d t}{t-(a \pm i\eta)}
= \pm\pi \int_{0}^{x}\d t ~\delta(t-a)
= \pm\pi \, \theta(a) \theta(x-a) \,.
\end{align}
This is the explicit formula for the imaginary part at lowest weight.

Moving on to higher-weight polylogarithms, let us consider a weight $n$
multiple polylogarithm. The integrand takes the form
\begin{align}
\label{eq:G integrand fractions}
\mathcal{I} =
\frac{1}{t_1-(a_1\pm i\eta)}
\frac{1}{t_2-(a_2\pm i\eta)}\dotsm
\frac{1}{t_n-(a_n\pm i\eta)} \,.
\end{align}
The imaginary part of $\mathcal{I}$ may be broken into real and imaginary
parts of each of the above fractions. The imaginary part of
a single fraction is simple, because it localizes the
corresponding integration variable to a point. In contrast, the real
parts do not simplify, so it is convenient to have as
few real-part evaluations as possible. To this end, we split the
imaginary part of the integrand (\ref{eq:G integrand fractions})
into products of imaginary parts and real parts of
products by recursively applying
\begin{align}
\label{eq:Split Im and Merge Re}
\Im (PQ) &= \Im P \Re Q + \Re P \Im Q \,,\nn
\Re P \Re Q &= \Re (PQ) + \Im P \Im Q \,,
\end{align}
where $P$ and $Q$ represent either a single fraction or products of fractions.
For example,
\begin{align}
\label{eq:Im(ab...)}
\Im (ab) &= \Im a \Re b + \Im b \Re a \,,
\nn
\Im (abc) &= \Im a \Re (bc)
+ \Im b \Re (ac)
+ \Im c \Re (ab)
+ 2 \Im a \Im b \Im c \,,
\nn
\Im (abcd) &= \Im a \Re (bcd)
+ \Im b \Re (acd)
+ \Im c \Re (abd)
+ \Im d \Re (abc)
+2\Im a \Im b \Im c \Re d
\nn&\quad
+2\Im a \Im b \Im d \Re c
+2\Im a \Im c \Im d \Re b
+2\Im b \Im c \Im d \Re a \,,
\end{align}
where $a,b,c$ and $d$ represent individual fractions in
the integrand $\mathcal{I}$ of a multiple polylogarithm.
Notice that each term on the right-hand side of eq.~(\ref{eq:Im(ab...)})
contains only one real part (of a product) and a product
of imaginary parts (of single factors).

The imaginary part of higher-weight multiple polylogarithms
are then computed by applying \refE{eq:Im(ab...)} to the
integrand and integrating out the delta functions arising from
eq.~(\ref{eq:PV-formula}). For example,
the imaginary part of a weight-two multiple polylogarithm for
$a,b,x \in \mathbb{R}$, $b \neq 0$ and $x>0$ is computed as follows,
\begin{align}
\label{eq:Im G(a,b,x)}
&\Im G(a \pm i \eta,b \pm i \eta; x)
\nn
&=\int_0^x \d t \int_0^t \d u \hspace{0.9mm} \mathrm{Im}
\left( \frac{1}{t-(a\pm i\eta)} \frac{1}{u-(b\pm i\eta)} \right)
\nn
&=\int_0^x \d t \int_0^t \d u \hspace{-0.2mm} \left[
\mathrm{Im} \left( \frac{1}{t-(a\pm i\eta)} \right)
\mathrm{Re} \left( \frac{1}{u-(b\pm i\eta)} \right) +
\mathrm{Re} \left( \frac{1}{t-(a\pm i\eta)} \right)
\mathrm{Im} \left( \frac{1}{u-(b\pm i\eta)} \right) \right]
\nn
&=\pm\pi\int_{0}^{x}\d t ~\delta(t-a) \int_{0}^{t} \d u \hspace{0.3mm}
\Re\left(\frac{1}{u-(b \pm i \eta)}\right)
\hspace{0.7mm} \pm \hspace{0.7mm}\pi\int_{0}^{x} \d t \hspace{0.3mm}
\Re \left( \frac{1}{t-(a \pm i \eta)}\right) \int_{0}^{t}\d u ~\delta(u-b)
\nn
&=\pm\pi\,\theta(a)\theta(x-a)\Re\int_{0}^{a}\frac{\d u}{u-(b \pm i \eta)}
\hspace{0.7mm} \pm \hspace{0.7mm}\pi \Re\int_{0}^{x}
\frac{\d t}{t-(a \pm i \eta)}\theta(b)\theta(t-b)
\nn
&=\pm\pi\,\theta(a)\theta(x-a)\Re G(b \pm i\eta;a)
\pm\pi\,\theta(b)\theta(x-b)\Re \big(G(a\pm i\eta;x)-G(a\pm i\eta;b) \big) \,.
\end{align}
The real parts of lower-weight functions on the right-hand
side are known inductively from expressions for lower-weight
imaginary parts, together with \refE{eq:Re is Full minus Im}
for the real part. We remark that when $a=0$, the first term
in \refE{eq:Im G(a,b,x)} vanishes because $G(b \pm i\eta;0)=0$.
The imaginary part of weight-three and -four multiple
polylogarithms have been computed along the same lines.
Before quoting the results, it is advantageous to introduce
some notation, in terms of which the formulas assume a nice form.

The imaginary parts are conveniently expressed in terms of two
new functions: a slightly different notation for multiple
polylogarithms, together with generalized step functions.
Let us first introduce the former, which is an iterated
integral where the base point of integration may be freely chosen,
\begin{align}
\label{eq:Iterated Integral definition}
I(a_0;a_1,\dotsc,a_n;a_{n+1}) = \int_{a_0}^{a_{n+1}} \frac{\d t}{t-a_n}I(a_0;a_1,\dotsc,a_{n-1};t)
\hspace{5mm} \mathrm{with} \hspace{5mm} I(a_0 ; \hspace{0.6mm} ; x) \equiv 1 \,.
\end{align}
Setting the base point to zero we obviously recover the
multiple polylogarithms defined in \refE{eq:G definition recursion},
up to a conventional reversal of the indices. Although
\refE{eq:Iterated Integral definition} appears to define
a larger class of integrals, it actually does not. Any $I$
can be written as a linear combination of (products of)
$G$'s. This is achieved by splitting the range of integration
into a difference of paths with basepoint zero, cf.
ref.~\cite{Duhr:2012fh}. For example,
\begin{align}
\label{eq:I-1 in terms of G}
I(a_0;a_1;a_2) = I(0;a_1;a_2) - I(0;a_1;a_0) = G(a_1;a_2) - G(a_1;a_0) \,.
\end{align}
At higher weight one splits the innermost integrals first. At weight two,
\begin{align}
\label{eq:I-2 in terms of G}
I(a_0;a_1,a_2;a_3)
&= \int_{a_0}^{a_3}\frac{\d t}{t-a_2}\big(G(a_1;t) - G(a_1;a_0)\big)\nn
&= G(a_2,a_1;a_3) - G(a_2,a_1;a_0) - G(a_1;a_0)\big(G(a_2;a_3)-G(a_2;a_0)\big) \,.
\end{align}
In this way any $I$ can be written in terms of $G$'s.

We also introduce generalized step functions $\theta(a_1,\dotsc,a_n)$,
which may be thought of as enforcing $a_1 \leq \dotsm \leq a_n$.
In terms of ordinary single-variable step functions,
\begin{align}
\label{eq:General theta}
\theta(a_1,\dotsc,a_n) &\equiv \prod_{i=1}^{n-1} \theta(a_{i+1}-a_{i}) \hspace{6mm}
\text{for}~n>1 \hspace{5mm} \mathrm{and} \hspace{4mm}
\forall~i {:} \hspace{3mm}a_i \neq a_{i+1} \hspace{2mm} \text{and} \hspace{2mm} a_i \in \mathbb{R}\,.
\end{align}
Equal adjacent arguments are dealt with using the following definition
\begin{align}
\label{eq:General theta equal adjacent indices}
\theta(\dotsc,a,\underbrace{b,\dotsc,b}_{n\,\text{times}},c,\dotsc)
&= \frac{1}{n!} \theta(\dotsc,a,b,c,\dotsc)
\qquad \text{for real indices}~a\neq b \neq c \,.
\end{align}
Infinitesimal imaginary parts produce an overall sign,
\begin{align}
\label{eq:General theta small im parts}
\theta(\dotsc,a \pm i \eta,\dotsc)
&= \pm \theta(\dotsc,a,\dotsc) \hspace{6mm} \text{for} \hspace{3mm}
a \in \mathbb{R} \,.
\end{align}
In this notation, the imaginary part of multiple polylogarithms
up to weight four are
\begin{align}
\Im G(a;x) &= \pi \,\theta(0,a,x) \,,
\label{eq:Im_G_weight_1}\\
\Im G(a,b;x) &=
	\pi\,\theta(0,a,x) \Re G(b;a) 				
	+\pi\,\theta(0,b,x) \Re I(b;a;x) \,,
\label{eq:Im_G_weight_2}\\
\Im G(a,b,c;x) &=
	\pi\,\theta(0,a,x) \Re G(b,c;a) 				
	+\pi\,\theta(0,b,x) \Re \big[I(b;a;x)G(c;b)\big]
	\nn&\quad
	+\pi\,\theta(0,c,x) \Re I(c;b,a;x)				
	+2\pi^3\theta(0,c,b,a,x) \,,
\label{eq:Im_G_weight_3}\\
\Im G(a,b,c,d;x) &=
	\pi\,\theta(0,a,x) \Re G(b,c,d;a) 				\nn&\quad
	+\pi\,\theta(0,b,x) \Re \big[I(b;a;x)G(c,d;b)\big]
	\nn&\quad
	+\pi\,\theta(0,c,x) \Re \big[I(c;b,a;x)G(d;c)\big] 	\nn&\quad
	+\pi\,\theta(0,d,x) \Re I(d;c,b,a;x)
	\nn&\quad
	+2\pi^3\theta(0,c,b,a,x) \Re G(d;c)			\nn&\quad
	+2\pi^3\theta(0,d,b,a,x) \Re I(d;c;b)
	\nn&\quad
	+2\pi^3\theta(0,d,c,a,x) \Re I(c;b;a)			\nn&\quad
	+2\pi^3\theta(0,d,c,b,x) \Re I(b;a;x) \,,
\label{eq:Im_G_weight_4}
\end{align}
where $a,b,c,d,x \in \mathbb{R}$, $x$ is positive, and in each case
the last index is non-zero. The suppressed Feynman $i\eta$'s may
be reinstated by replacing the indices according to $a \rightarrow a \pm i \eta$.

We conclude this section by providing proofs of eqs.~(\ref{eq:Im_G_weight_2})
and (\ref{eq:Im_G_weight_3}).

\begin{proof}[Proof of eq.~\textup{(}\ref{eq:Im_G_weight_2}\textup{)}]
We proceed by direct computation, using eq.~(\ref{eq:Im(ab...)})
and the identity $\Im \frac{1}{\xi \pm i0} = \mp \pi \delta(\xi)$,
\begin{align}
\Im G(a,b;x)&=\Im\left(\int_{0}^{x} \tfrac{\d t}{t-a-i0} \int_{0}^{t} \tfrac{\d u}{u-b-i0}\right)
\nn
&=\int_{0}^{x} \d t \int_{0}^{t} \d u \Im\left(\tfrac{1}{t-a-i0}\tfrac{1}{u-b-i0}\right)
\nn
&=\int_{0}^{x} \d t \int_{0}^{t} \d u
\left[
\Im\left(\tfrac{1}{t-a-i0}\right)\Re\left(\tfrac{1}{u-b-i0}\right)
+\Re\left(\tfrac{1}{t-a-i0}\right)\Im\left(\tfrac{1}{u-b-i0}\right)
\right]
\nn
&=\int_{0}^{x} \d t \int_{0}^{t} \d u
\left[
\pi \delta(t-a) \Re\left(\tfrac{1}{u-b-i0}\right)
+\Re\left(\tfrac{1}{t-a-i0}\right) \pi \delta(u-b)
\right]
\nn
&=\pi \int_{0}^{x} \d t ~ \delta(t-a)
\int_{0}^{t} \d u \Re\left(\tfrac{1}{u-b-i0}\right)
+\pi \int_{0}^{x} \d t \Re\left(\tfrac{1}{t-a-i0}\right)
\int_{0}^{t} \d u ~ \delta(u-b)
\nn
&=\pi \theta(0,a,x)\int_{0}^{a} \d u \Re\left(\tfrac{1}{u-b-i0}\right)
+\pi \int_{0}^{x} \d t \Re\left(\tfrac{1}{t-a-i0}\right) \theta(0,b,t)
\nn
&=\pi \theta(0,a,x)\Re\left(\int_{0}^{a} \tfrac{\d u}{u-b-i0}\right)
+\pi \theta(0,b,x) \Re\left(\int_{b}^{x} \tfrac{\d t}{t-a-i0} \right)
\nn
&=\pi \theta(0,a,x)\Re G(b;a)
+\pi \theta(0,b,x) \Re I(b;a;x) \,.
\end{align}
This completes the proof of eq.~(\ref{eq:Im_G_weight_2}).
\end{proof}

\begin{proof}[Proof of eq.~\textup{(}\ref{eq:Im_G_weight_3}\textup{)}]
We proceed by direct computation, using eq.~(\ref{eq:Im(ab...)})
and the identity $\Im \frac{1}{\xi \pm i0} = \mp \pi \delta(\xi)$,
\begin{align}
\Im G(a,b,c;x)&=\Im\left(\int_{0}^{x} \tfrac{\d t}{t-a-i0} \int_{0}^{t} \tfrac{\d u}{u-b-i0} \int_{0}^{u} \tfrac{\d v}{v-c-i0}\right)
\nn
&=\int_{0}^{x} \d t \int_{0}^{t} \d u \int_{0}^{u} \d v \Im\left(\tfrac{1}{t-a-i0}\tfrac{1}{u-b-i0}\tfrac{1}{v-c-i0}\right)
\nn
&=\int_{0}^{x} \d t \int_{0}^{t} \d u \int_{0}^{u} \d v
\left[
\Im\left(\tfrac{1}{t-a-i0}\right)\Re\left(\tfrac{1}{u-b-i0}\tfrac{1}{v-c-i0}\right)
\right. \nn &\hspace{21mm} +\Im\left(\tfrac{1}{u-b-i0}\right)\Re\left(\tfrac{1}{t-a-i0}\tfrac{1}{v-c-i0}\right)
+\Im\left(\tfrac{1}{v-c-i0}\right)\Re\left(\tfrac{1}{t-a-i0}\tfrac{1}{u-b-i0}\right) \nn
&\hspace{21mm}\left. +2\Im\left(\tfrac{1}{t-a-i0}\right)\Im\left(\tfrac{1}{u-b-i0}\right)\Im\left(\tfrac{1}{v-c-i0}\right)
\right]
\nn
&=\int_{0}^{x} \d t \int_{0}^{t} \d u \int_{0}^{u} \d v
\left[
\pi \delta(t-a)\Re\left(\tfrac{1}{u-b-i0}\tfrac{1}{v-c-i0}\right)
+\pi \delta(u-b)\Re\left(\tfrac{1}{t-a-i0}\tfrac{1}{v-c-i0}\right)
\right. \nn &\hspace{34mm} \left.
+\pi \delta(v-c)\Re\left(\tfrac{1}{t-a-i0}\tfrac{1}{u-b-i0}\right)
+2 \pi^3 \delta(t-a) \delta(u-b) \delta(v-c)
\right]
\nn
&\equiv I_{(3,1)} + I_{(3,2)} + I_{(3,3)} + I_{(3,4)} \,.
\label{eq:Im_G(a,b,c;x)_four_terms}
\end{align}
We continue by evaluating each term in the last line of
eq.~(\ref{eq:Im_G(a,b,c;x)_four_terms}) separately.
The first term evaluates to,
\begin{align}
I_{(3,1)}
&=\pi \int_{0}^{x} \d t ~ \delta(t-a) \int_{0}^{t} \d u \int_{0}^{u} \d v \Re\left(\tfrac{1}{u-b-i0}\tfrac{1}{v-c-i0}\right)
\nn
&=\pi \theta(0,a,x) \int_{0}^{a} \d u \int_{0}^{u} \d v \Re\left(\tfrac{1}{u-b-i0}\tfrac{1}{v-c-i0}\right)
\nn
&=\pi \theta(0,a,x) \Re\left( \int_{0}^{a} \d u \int_{0}^{u} \d v \tfrac{1}{u-b-i0}\tfrac{1}{v-c-i0}\right)
\nn
&=\pi \theta(0,a,x) \Re G(b,c;a) \,.
\label{eq:I_{(3,1)}}
\end{align}
The second term in eq.~(\ref{eq:Im_G(a,b,c;x)_four_terms}) evaluates to,
\begin{align}
I_{(3,2)}
&=\pi \int_{0}^{x} \d t \int_{0}^{t} \d u ~ \delta(u-b) \int_{0}^{u} \d v \Re\left(\tfrac{1}{t-a-i0}\tfrac{1}{v-c-i0}\right)
\nn
&=\pi \int_{0}^{x} \d t ~\theta(0,b,t) \int_{0}^{b} \d v \Re\left(\tfrac{1}{t-a-i0}\tfrac{1}{v-c-i0}\right)
\nn
&=\pi \theta(0,b,x) \int_{b}^{x} \d t \int_{0}^{b} \d v \Re\left(\tfrac{1}{t-a-i0}\tfrac{1}{v-c-i0}\right)
\nn
&=\pi \theta(0,b,x) \Re\left(\int_{b}^{x} \d t \tfrac{1}{t-a-i0} \int_{0}^{b} \d v \tfrac{1}{v-c-i0}\right)
\nn
&=\pi \theta(0,b,x) \Re\big( I(b;a;x) G(c;b) \big) \,.
\label{eq:I_{(3,2)}}
\end{align}
The third term in eq.~(\ref{eq:Im_G(a,b,c;x)_four_terms}) evaluates to,
\begin{align}
I_{(3,3)}
&=\pi \int_{0}^{x} \d t \int_{0}^{t} \d u \int_{0}^{u} \d v ~ \delta(v-c) \Re\left(\tfrac{1}{t-a-i0}\tfrac{1}{u-b-i0}\right)
\nn
&=\pi \int_{0}^{x} \d t \int_{0}^{t} \d u ~\theta(0,c,u) \Re\left(\tfrac{1}{t-a-i0}\tfrac{1}{u-b-i0}\right)
\nn
&=\pi \int_{0}^{x} \d t ~\theta(0,c,t) \int_{c}^{t} \d u \Re\left(\tfrac{1}{t-a-i0}\tfrac{1}{u-b-i0}\right)
\nn
&=\pi \theta(0,c,x) \int_{c}^{x} \d t \int_{c}^{t} \d u \Re\left(\tfrac{1}{t-a-i0}\tfrac{1}{u-b-i0}\right)
\nn
&=\pi \theta(0,c,x) \Re\left( \int_{c}^{x} \d t \int_{c}^{t} \d u \tfrac{1}{t-a-i0}\tfrac{1}{u-b-i0}\right)
\nn
&=\pi \theta(0,c,x) \Re I(c;b,a;x) \,.
\label{eq:I_{(3,3)}}
\end{align}
The fourth term in eq.~(\ref{eq:Im_G(a,b,c;x)_four_terms}) evaluates to,
\begin{align}
I_{(3,4)}
&=2 \pi^3 \int_{0}^{x} \d t \int_{0}^{t} \d u \int_{0}^{u} \d v ~\delta(t-a) \delta(u-b) \delta(v-c)
\nn
&=2 \pi^3 \int_{0}^{x} \d t  ~\delta(t-a) \int_{0}^{t} \d u ~\delta(u-b) \int_{0}^{u} \d v ~\delta(v-c)
\nn
&=2 \pi^3 \int_{0}^{x} \d t  ~\delta(t-a) \int_{0}^{t} \d u ~\delta(u-b) \theta(0,c,u)
\nn
&=2 \pi^3 \int_{0}^{x} \d t  ~\delta(t-a) \theta(0,c,b,t)
\nn
&=2 \pi^3 \theta(0,c,b,a,x) \,.
\label{eq:I_{(3,4)}}
\end{align}
Adding up the four contributions in eqs.~(\ref{eq:I_{(3,1)}})--(\ref{eq:I_{(3,4)}})
according to eq.~(\ref{eq:Im_G(a,b,c;x)_four_terms}) we obtain the result given
in eq.~(\ref{eq:Im_G_weight_3}). This completes the proof.
\end{proof}

\noindent The identity (\ref{eq:Im_G_weight_4}) may be shown
by completely analogous steps, and we therefore omit its proof here.

\section{Algorithm for achieving canonical-form polylogarithms}
\label{App:algorithm}

In their position-space representation, eikonal diagrams without
internal vertices, and the corresponding cut diagrams, take the form of iterated integrals.
A first step in the evaluation of these diagrams therefore amounts to
recognizing the definition of multiple polylogarithms in
terms of iterated integrals, cf. \refE{eq:G definition recursion}.
The resulting multiple polylogarithms depend in general on the
kinematical variables of the problem through both their indices
and their argument. They may be rewritten in terms of polylogarithms
with constant indices, in so-called canonical form, by
exploiting the Hopf algebra structure of multiple polylogarithms,
which encodes the plethora of functional relations within
this class of functions. In this appendix we describe the
algorithm to cast multiple polylogarithms in canonical form,
which is extensively used in the computations in section~\ref{sec:examples}.

The algorithm to cast multiple polylogarithms in canonical form
relies on the Hopf algebra structure of multiple polylogarithms.
In particular, we make use of the notions of the symbol and
coproduct of multiple polylogarithms, see refs.~\cite{Duhr:2011zq,Duhr:2012fh}
and references therein, as well as a procedure from
ref.~\cite{Anastasiou:2013srw} to map symbols to polylogarithms.
In order to describe the algorithm, we start by setting up
notation. Let us denote an eikonal diagram, or a cut eikonal diagram
(or a partially integrated result thereof), by the function
$g(\vec{x})$, depending on $n$ variables, $\vec{x} = \{x_1,\dotsc,x_n\}$.
(This set of variables typically contains the cusp angles and
possibly some remaining integration variables.) After recognizing
the definition of multiple polylogarithms, the function $g(\vec{x})$
is given in terms of functions $G(\vec{a}(\vec{x}); z(\vec{x}))$
depending on $\vec{x}$ through both their indices $\vec{a}(\vec{x})$
and their argument $z(\vec{x})$. To simplify the following presentation,
we will assume\footnote{If this assumption fails, the algorithm may be
applied to each subexpressions of uniform weight.} that $g(\vec{x})$
has uniform transcendental weight $w$.

The algorithm follows three steps. Let us first state the algorithm
and subsequently elaborate on each of the steps separately.

\begin{algorithm}[Multiple polylogarithms in canonical form]
\begin{enumerate}
\item
Compute the symbol $S[g(\vec{x})]$ of the function $g(\vec{x})$.
\item
Apply a map $M_{\vec{x}}$ to the symbol $S[g(\vec{x})]$,
whose purpose is to construct a polylogarithm in canonical form
with the same symbol as the original function. The resulting
expression differs only from the original function by terms
proportional to transcendental constants (which are in the
kernel of the symbol map).
\item
Compute subsequently the coproducts $\Delta_{2,1,\dotsc,1},
\, \Delta_{3,1,\dotsc,1}, \, \dotsc, \, \Delta_{w-1,1}$
to reconstruct any missing terms proportional to constants
with transcendental weight $2,3,\dotsc,w-1$ respectively.
\end{enumerate}
\end{algorithm}
The output of this algorithm is a new function $h(\vec{x})$
in canonical form, which is numerically equal to the original
function $g(\vec{x})$. In the remainder of this appendix we
shall give the definitions of the symbol $S$, the map $M_{\vec{x}}$
and the coproduct $\Delta$. We conclude by illustrating
the application of this algorithm to the non-planar two-loop
ladder diagram considered in section~\ref{sec:2loopNPladder}.

The first step of the algorithm involves computing the symbol
of multiple polylogarithms, first introduced in
ref.~\cite{Goncharov.A.B.:2009tja}. The idea of the symbol map
is to encode the functional relations among multiple polylogarithms
as simple algebraic identities in the corresponding tensor algebra.

One way to define the symbol is by considering the total differential,
\begin{align}
\d G(a_{n-1},\dotsc,a_{1};a_{n}) = \sum_{i=1}^{n-1}
G(a_{n-1},\dotsc,a_{i+1},a_{i-1},\dotsc,a_{1};a_{n}) ~
\d\log \left(\frac{a_{i} - a_{i+1}}{a_{i}-a_{i-1}}\right) \,,
\end{align}
and to define the symbol of a multiple polylogarithm analogously,
cf. ref.~\cite{Goncharov:2010jf},
\begin{align}
\label{eq:definition symbol}
S\big[G(a_{n-1},\dotsc,a_{1};a_{n})\big] = \sum_{i=1}^{n-1}
\Tensor{ S\big[G(a_{n-1},\dotsc,a_{i+1},a_{i-1},\dotsc,a_{1};a_{n})\big] }{\left(\frac{a_{i} - a_{i+1}}{a_{i}-a_{i-1}}\right)}\,,
\end{align}
in the case of generic indices $a_i$; i.e., non-zero and mutually
different. The formula for the symbol in \refE{eq:definition symbol},
augmented with formulas for special cases and the rules of symbol
calculus (for which we refer the reader to
refs.~\cite{Duhr:2011zq,Duhr:2012fh}), allows the symbol $S[g(\vec{x})]$
to be computed. This completes the first step of the algorithm.

The second step of the algorithm takes the resulting symbol
as input and maps it to an expression of multiple polylogarithms
in canonical form, whose symbol is the same as the symbol of the
original function. A procedure that achieves this goal was given
in appendix D of ref.~\cite{Anastasiou:2013srw}. We cast their
procedure in the form of an explicit map and furthermore make
a slight generalization in order to deal with functions of
more than two variables which have a sufficiently factorized form.
Let us first define the map and then point out wherein the
slight generalization resides.

The map $M_{\vec{x}}$ is defined recursively in the number of
variables. Starting with the case of a single variable, we
define the map $M_x$ which takes tensors to functions,
\begin{align}
\label{eq:mapMx}
M_x(T) =
\begin{cases}
G\left( -\tfrac{b_1}{a_1}, \dotsc, -\tfrac{b_w}{a_w} \, ; \, x \right)
&\text{if}~ T = (a_w x + b_w) \,\otimes\, \dotsm \,\otimes\, (a_1 x + b_1)~, \\
0
&\text{otherwise}~,
\end{cases}
\end{align}
where $a_i$ and $b_i$ are independent of $x$.
The map $M_x$ is linear in the space of tensors: given a symbol
$S = \sum_{i} c_i T_i$, with rational numbers $c_i$ and
tensors $T_i$, one has $M_x(S) = \sum_{i} c_i \, M_x(T_i)$.
The map $M_x(T)$ is designed to construct a function in
canonical form, such that its symbol is given by $T$ plus
possibly other tensors which have at least one entry independent
of $x$. The proof of this statement was given in ref.~\cite{Anastasiou:2013srw}.

Generalizing to the multivariate case, we let
$\vec{x} = \{x_1,\dotsc,x_n\}$ denote a collection of
at least two variables, and define the multivariate map
\begin{align}
\label{eq:Multivariate-map}
M_{\vec{x}}(S) &=
\left(
P_{\vec{x},S}^{(w)} \circ
P_{\vec{x},S}^{(w-1)} \circ
\dotsm \circ
P_{\vec{x},S}^{(1)} \circ
M_{\vec{x}}^{(0)}
\right) (S)
\nn
&\equiv
P_{\vec{x},S}^{(w)} \Big(
P_{\vec{x},S}^{(w-1)} \Big(
\dotsm
P_{\vec{x},S}^{(1)} \Big(
M_{\vec{x}}^{(0)} \big(S \big)
\Big) \dotsm \Big)\Big)~,
\end{align}
where the projectors $P$ map functions to functions according to
\begin{align}
\label{eq:Projectors}
P_{\vec{x},S}^{(r)}(h) = h + M_{\vec{x}}^{(r)}(S - S[h])~.
\end{align}
The maps $M_{\vec{x}}^{(r)}(S)$ which occur on the right-hand sides
of eqs.~(\ref{eq:Multivariate-map})--(\ref{eq:Projectors}) are
linear in the space of tensors (as in the case of a single
variable) and are defined to act on elementary tensors $T$
by recursion in the number of variables,
\begin{align}
\label{eq:M(r,x)(T)}
M_{\vec{x}}^{(r)}(T) =
\begin{cases}
M_{x_1}(T)
& \text{for}~ r=0 ~,\\
M_{x_2,\dotsc,x_n}(\Tensor{ T^{(1)} }{ \dotsm }{ T^{(r)} })
~M_{x_1}(\Tensor{ T^{(r+1)} }{ \dotsm }{ T^{(w)} })
& \text{for}~ r=1,2,\dotsc,w-1 ~,\\
M_{x_2,\dotsc,x_n}(T)
& \text{for}~ r=w~.
\end{cases}
\end{align}
For the map $M_{\vec{x}}^{(r)}(T)$ to be non-vanishing,
the last $w-r$ indices must depend on $x_1$. Its output
is then given by a canonical function with argument $x_1$
and weight $w-r$, multiplied by an $x_1$-independent function
of weight $r$. The projectors $P_{\vec{x},S}^{(r)}(h)$
add such functions to their input, thus gradually constructing
a function in canonical form, starting with functions of $x_1$
with weight $w$ down to weight $1$ and repeating the process
for the remaining variables $x_2, x_3, \dotsc, x_n$.
As a result, the multivariate map $M_{\vec{x}}(S)$ generates
a function of the form
\begin{align}
\label{eq:Canonical-form-multivariate}
\sum_{(i_1,\dotsc,i_n)} c_{i_1,\dotsc,i_n}
G(\vec{a}_{i_n}; x_n) \dotsm G(\vec{a}_{i_1}; x_1) \,,
\end{align}
where the $\vec{a}_{i_k}$ are independent of $x_1, \dotsc, x_k$.
This expression is by definition in canonical form.
For a single variable, the indices are in fact constants
and as slight abuse of terminology this is what we occasionally
refer to as canonical form, bearing in mind that
\refE{eq:Canonical-form-multivariate} is the proper
definition of canonical form in the multivariate case.

Having defined the multivariate map, let us point out
the difference with respect to the procedure described
in ref.~\cite{Anastasiou:2013srw}. A slight generalization
resides in the definition of $M_{\vec{x}}^{(r)}(T)$.
In particular, two of the maps on the right-hand side
of \refE{eq:M(r,x)(T)} depend on all remaining variables
$x_2,\dotsc,x_n$, rather than just the next variable $x_2$.
This alteration allows us to reconstruct functions of
more than two variables which have a sufficiently factorized form.

A simple example illustrates the point. Consider
$g(x,y,z) = \log x \log y \log z$. This function is
already in canonical form, but the algorithm should
nonetheless be able to reconstruct this function from
its symbol. In this case the weight is $w=3$, and we set
$\vec{x} = \{x,y,z\}$. The symbol of $g(x,y,z)$ is given by
\begin{align}
S \,=\,
\Tensor{x}{y}{z} \,+\,
\Tensor{x}{z}{y} \,+\,
\Tensor{y}{x}{z} \,+\,
\Tensor{y}{z}{x} \,+\,
\Tensor{z}{x}{y} \,+\,
\Tensor{z}{y}{x}~.
\end{align}
Let us apply the map to this symbol,
\begin{align}
M_{\vec{x}}(S) = \left(
P_{\vec{x},S}^{(3)} \circ
P_{\vec{x},S}^{(2)} \circ
P_{\vec{x},S}^{(1)} \circ
M_{\vec{x}}^{(0)}
\right) (S) ~.
\end{align}
Since $S$ contains no tensors with all three entries
depending on $x$, the first map gives zero,
\begin{align}
M_{\vec{x}}^{(0)}(S) = M_{x}(S) = 0 ~.
\end{align}
Subsequently, the first projector acts on this result.
Inserting its definition from \refE{eq:Projectors},
it reduces to applying the map $M_{\vec{x}}^{(1)}(S)$,
which is non-zero only for tensors whose last two entries
depend on $x$. Since $S$ does not contain such tensors,
the result is zero,
\begin{align}
P_{\vec{x}}^{(1)}(0) = M_{\vec{x}}^{(1)}(S) = 0 ~.
\end{align}
A non-vanishing contribution is found in the next step,
coming from the two tensors which contain $x$ in the last entry.
\begin{align}
\label{eq:Projector2}
P_{\vec{x}}^{(2)}(0)
\,=\, M_{\vec{x}}^{(2)}(S)
\,=\, M_{y,z}(\Tensor{y}{z}) \, M_{x}(x) \, +\,  M_{y,z}(\Tensor{z}{y}) \, M_{x}(x) ~.
\end{align}
From the definition in \refE{eq:mapMx} we have
$M_{x}(x) = G(0;x) = \log x$. The other map $M_{y,z}$
is computed along the very same lines, with the results
$M_{y,z}(\Tensor{y}{z}) = 0$ and $M_{y,z}(\Tensor{z}{y}) = \log y \log z$.
By now we have that $M_{\vec{x}}(S) = P_{\vec{x},S}^{(3)}(h)$,
with $h=P_{\vec{x}}^{(2)}(0) = \log x \log y \log z$.
Because $S=S[h]$, the last projector $P_{\vec{x}}^{(3)}$ becomes the identity, resulting in
\begin{align}
M_{\vec{x}}(S) \,=\, \log x \log y \log z \,=\, g(x,y,z) ~,
\end{align}
as required, because our algorithm should reconstruct
the original function from its symbol. This example
shows that it is crucial to have both variables $y$
and $z$ as parameters in the maps $M_{y,z}$ in \refE{eq:Projector2}.
If one would use a single variable only, then the vanishing
maps $M_{y}(\Tensor{z}{y}) = M_{y}(\Tensor{y}{z}) = 0$,
as one can easily verify from \refE{eq:mapMx},
would lead to a vanishing, and hence incorrect, result.
This illustrates the purpose of the slight generalization of the map.

In conclusion, the second step of the algorithm constructs
from the symbol $S[g(\vec{x})]$ an expression in canonical
form, which has the same symbol as $g(\vec{x})$,
by applying the map $M_{\vec{x}}(S)$ in \refE{eq:Multivariate-map}.

The resulting expression differs from the original function
$g(\vec{x})$ only by terms proportional to transcendental
constants, because such terms are in the kernel of the symbol
map. Finding these missing terms is the task of the next step
in the algorithm.

The third step in the algorithm revolves around the coproduct
of multiple polylogarithms, which generalizes the concept
of the symbol. Before describing how it may be used to
construct missing terms proportional to transcendental constants,
let us first define the required coproducts of the form
$\Delta_{p,q,\dotsc,r}$. They are derived from the general
coproduct $\Delta$, which is defined by its action on iterated
integrals with a freely specified base point of integration,
see \refE{eq:Iterated Integral definition}, cf. ref.~\cite{Goncharov:2005sla}
\begin{align}
\label{eq:coproduct-definition}
&\Delta\big[I(a_0; a_1,\dotsc,a_n;a_{n+1})\big]
=
\nn
&\hspace{-2mm}\sum_{k=0}^{n} ~
\sum_{0=i_0<i_1<\dotsm<i_k<i_{k+1}=n+1}
\Tensor{
I(a_{0}; a_{i_1},\dotsc,a_{i_k};a_{n+1})
}{
\prod_{p=0}^{k}
I(a_{i_p}; a_{i_p+1}, \dotsc, a_{i_{p+1}-1}; a_{i_{p+1}})
}\,.
\end{align}
The right-hand side of \refE{eq:coproduct-definition}
consists of tensors with two entries, each entry having a weight
between $0$ and $n$, such that the two weights add up to $n$,
the weight of the original function.%
\footnote{The weight of $I(a_0; a_1,\dotsc,a_n;a_{n+1})$ is
equal to its number of indices, $n$. Likewise, a pair of
weights $(p,q)$ is attributed to a tensor $\Tensor{T_p}{T_q}$
where the weights of $T_p$ and $T_q$ are $p$ and $q$, respectively.}
All possible pairs of weights are thus $(0,n),(1,n-1),\dotsc,(n,0)$.
Grouping the tensors by those pairs of weights decomposes the coproduct into
\begin{align}
\label{eq:decomposed-coproduct}
\Delta = \sum_{p+q=n} \Delta_{p,q} ~.
\end{align}
In other words, the action of $\Delta_{p,q}$ on
$I(a_0; a_1,\dotsc,a_n;a_{n+1})$ yields the subset
of terms in \refE{eq:coproduct-definition} of tensors
with weight $(p,q)$. Besides the operator $\Delta_{p,q}$, the
third step of our algorithm also uses operators with multiple
indices $\Delta_{p,q,\dotsc,r}$. Those are defined recursively
in terms of $\Delta_{p,q}$. For example, $\Delta_{p,q,r}$ is
defined by application of $\Delta_{q,r}$ to the second entry
of all tensors obtained from $\Delta_{p,q+r}$. These are
the definitions of the coproducts which are needed in the
third step of the algorithm. For more details, see ref.~\cite{Duhr:2012fh}.

The coproducts thus defined may be employed to construct
missing terms proportional to transcendental constants,
starting with the lowest-weight constants, cf. ref.~\cite{Anastasiou:2013srw}.
To be specific, in the previous step of the algorithm
we constructed a function $h(\vec{x})$ such that
$S[g(\vec{x})-h(\vec{x})] = 0$. This means that the
difference $g(\vec{x})-h(\vec{x})$ must be proportional
to transcendental constants. Following the recipe in the
third step of the algorithm we act on this difference
with the coproduct $\Delta_{2,1,\dotsc,1}$ to first
find terms proportional to $\zeta_2$. The coproduct takes the form
\begin{align}
\label{eq:Cop-2,1,...,1}
\Delta_{2,1,\dotsc,1}\big[ g(\vec{x})-h(\vec{x}) \big] \,=\,
\sum_{(i_3,\dotsc,i_{w})}
\Tensor{L_{i_3,\dotsc,i_w}(\vec{x})}{\log R_{i_3}(\vec{x})}{\dotsm}{\log R_{i_w}(\vec{x})} \,,
\end{align}
where $L_{i_3,\dotsc,i_w}$ is a linear combination of
weight-two multiple polylogarithms, and the
$R_{i_3}, \dotsc, R_{i_w}$ are rational functions.
Since the weight-two object $L_{i_3,\dotsc,i_w}$ must
be proportional to $\zeta_2$, we write
$L_{i_3,\dotsc,i_w} = k \, \zeta_2$ for some rational
number $k$. This constant of proportionality can be
determined by numerical evaluation at some specific
values for $\vec{x}$ using {\tt Ginac} \cite{Bauer:2000cp}
and running the PSLQ algorithm \cite{Ferguson:1992,Ferguson:1999}.
The hereby obtained transcendental constant $k \, \zeta_2$
multiplies a polylogarithmic function, whose symbol
is given by $\Tensor{R_{i_3}}{\dotsm}{R_{i_w}}$, arising
from the tail of the arguments of \refE{eq:Cop-2,1,...,1}.
Feeding this symbol back into the first step of
this algorithm and collecting the resulting multiple
polylogarithms from the output of step two produces
a function $h_{2}(\vec{x})$ in canonical form, which
is to multiply the constant $k \, \zeta_2$.
As a consequence we have that
\begin{align}
\Delta_{2,1,\dotsc,1}\big[
g(\vec{x})-h(\vec{x})-k\, \zeta_2 \, h_{2}(\vec{x})
\big] =0 \,,
\end{align}
and we conclude that the difference
$g(\vec{x})-h(\vec{x})-k\, \zeta_2 \, h_{2}(\vec{x})$
is equal to terms proportional to transcendental constants
of weight three and higher, which are in the kernel of
$\Delta_{2,1,\dotsc,1}$. Iterating this procedure with
the coproducts $\Delta_{3,1,\dotsc,1}, \, \dotsc, \, \Delta_{w-1,1}$
allows us to moreover reconstruct the other missing constants with
transcendental weight $3,\dotsc,w-1$, respectively.
The final output of the third step in the algorithm
is thus a rewritten version of the original function in canonical form.

Let us conclude this appendix by demonstrating an
explicit application of the algorithm, involving in
particular the use of the coproduct. To this end,
we consider the expression in the first line of
\refE{eq:F2-tch-v2}, $\widetilde{\Fau}^{(2)} = 2 R(\chi)^2 g(\chi)$,
where the interesting part is given by the weight-three function
\begin{align}
g(\chi)= \frac{1}{4}
\Big(
G(\rho_1,0,\rho_1;1)
-G(\rho_1,0,\rho_2;1)
-G(\rho_2,0,\rho_1;1)
+G(\rho_2,0,\rho_2;1)
\Big) ~.
\end{align}
It depends on a single variable $\chi$ through its
indices, $\rho_1=\tfrac{\chi}{\chi-1}$
and $\rho_2=\tfrac{1}{1-\chi}$. We wish to
express this in terms of multiple polylogarithms with
constant indices and argument $\chi$. Following the
algorithm, we start by computing the symbol of $g(\chi)$,
\begin{align}
S\big[g(\chi)\big] =
\Tensor{\chi}{(1-\chi)}{\chi}
~ - ~ \Tensor{\chi}{\chi}{\chi}
~ + ~ \Tensor{\chi}{(1+\chi)}{\chi} ~.
\end{align}
Application of the map from step two yields a multiple
polylogarithm in canonical form with the same symbol,
\begin{align}
h(\chi) =
M_{\chi}\big(S \big[g(\chi) \big] \big) =
G(0,-1,0;\chi) - G(0,0,0;\chi) + G(0,1,0;\chi) ~.
\end{align}
Indeed, the symbol of the difference vanishes,
$S\big(g(\chi)-h(\chi)\big)=0$. Yet the functions
$g(\chi)$ and $h(\chi)$ are not equal, because they
differ by terms proportional to zeta values up to
weight three, in this case $\zeta_2$ and $\zeta_3$.
According to the third step of the algorithm, the missing
terms proportional to $\zeta_2$ are obtained first by acting
with the coproduct $\Delta_{2,1}$ on the difference,
\begin{align}
\label{eq:Cop2,1}
\Delta_{2,1}\big[g(\chi)-h(\chi)\big]
= \Tensor{c_2}{G(0;\chi)}
= \Delta_{2,1}\big[c_2 \, G(0;\chi)\big] ~,
\end{align}
where we expect that $c_2 = k \, \zeta_2$ for some rational
number $k$. Explicitly, we find
\begin{align}
c_2 &=
-\tfrac{1}{4}G(\tfrac{\chi}{\chi-1},\tfrac{1}{1-\chi};1)
+\tfrac{1}{4}G(\tfrac{1}{1-\chi},\tfrac{\chi}{\chi-1};1)
-\text{Li}_2(\chi)-\text{Li}_2(-\chi)
+\tfrac{1}{2}\text{Li}_2(\tfrac{\chi-1}{\chi})
\nn&\quad
-\tfrac{1}{2}\text{Li}_2(1-\chi)
-\log(\chi)\log(1+\chi)
+\tfrac{1}{2}\log^2(\chi)
-\log(1-\chi)\log(\chi) ~.
\end{align}
Evaluating this expression at any value of $\chi$ with {\tt Ginac} yields
\begin{align}
c_2 &=-0.822467033424113218... ~
= - \tfrac{1}{2} \zeta_2 ~.
\end{align}
Inserting this result for $c_2$ into \refE{eq:Cop2,1},
we conclude that $\Delta_{2,1}\big[ g(\chi)- \big( h(\chi) - \tfrac{1}{2}\zeta_2 G(0;\chi) \big)\big] = 0$.
To find the last missing contribution proportional to $\zeta_3$,
it suffices to evaluate the remaining difference numerically
at a single point,
\begin{align}
g(\chi)-\big(h(\chi)-\tfrac{1}{2}\zeta_2 G(0;\chi)\big)
= -0.601028451579797142... ~
= -\tfrac{1}{2} \zeta_3 ~.
\end{align}
We finally conclude that
\begin{align}
g(\chi)&= h(\chi)-\tfrac{1}{2}\zeta_2 G(0;\chi)-\tfrac{1}{2}\zeta_3
\nn
&=
G(0,-1,0;\chi)
- G(0,0,0;\chi)
+ G(0,1,0;\chi)
- \tfrac{1}{2} \zeta_2 G(0;\chi)
- \tfrac{1}{2} \zeta_3 ~.
\end{align}
We have thus succeeded in expressing $g(\chi)$ in terms
of multiple polylogarithms with constant indices and argument $\chi$.
Inserting this result into $\widetilde{\Fau}^{(2)} = 2 R(\chi)^2 g(\chi)$
reproduces the second line of \refE{eq:F2-tch-v2}. It is
now a simple matter of applying \refE{eq:GtoClassic}
for multiple polylogarithms with constant indices,
to rewrite $g(\chi)$ in terms of classical polylogarithms,
thus reducing the expression to the form
given in the last line of \refE{eq:F2-tch-v2}.

This completes the illustration of our algorithm in a
practical example and thereby also completes our
description of each of the three steps in the algorithm
to rewrite multiple polylogarithms in canonical form.

\bibliography{Wilson_line_cuts}
\end{document}